\documentclass[twocolumn,traditabstract]{aa}

\usepackage{graphicx}
\usepackage{txfonts}
\usepackage{multirow}
\usepackage{tabulary}
\usepackage{tabularx}
\usepackage{multicol}
\usepackage{booktabs} 
\usepackage{array} 
\usepackage{tablefootnote}
\usepackage{xcolor,colortbl}
\usepackage{enumitem}

\def\farcs{\hbox{$.\!\!^{\prime\prime}$}}

\newcommand{\gp}{\mbox{$g^\prime$}}
\newcommand{\rp}{\mbox{$r^\prime$}}
\newcommand{\ip}{\mbox{$i^\prime$}}
\newcommand{\zp}{\mbox{$z^\prime$}}
\newcommand{\griz}{\gp\rp\ip\zp}

\newcommand*\nuc{$\nu_{\rm{c}} $}
\newcommand*\numm{$\nu_{\rm{m}} $}
\newcommand*\nusa{$\nu_{\rm{sa}} $}

\newcommand*\gammam{$\gamma_{\rm{m}} $}

\newcommand*\gammac{$\gamma_{\rm{c}} $}

\newcommand*\epse{$\epsilon_{\rm{e}} $}
\newcommand*\bepse{$\bar{\epsilon}_{\rm{e}} $}
\newcommand*\epsb{$\epsilon_{\rm{B}} $}
\newcommand*\den{$A_{\rm{*}} $}
\newcommand*\thh{$\theta_{\rm{0}} $}
\newcommand*\mloss{$\dot{M}_{\rm{W}} $}
\newcommand*\msun{$M_{\odot} $}

\newcommand*\eiso{$E_{\rm{K,iso}} $}
\newcommand*\eisog{$E_{\rm{iso}}^{\gamma} $}

\newcommand*\ejet{$E_{\rm{jet}} $}

\newcommand*\wind{stellar wind-like }
\newcommand*\chidof{$\chi^2/\rm{d.o.f}$}

\newcommand*\dusth{$A^{\rm{host}}_{\rm{v}} $}
\newcommand*\gash{$N_{\rm{H}}^{\rm{host}} $}
\newcommand*\dustg{$A^{\rm{Gal}}_{\rm{v}} $}
\newcommand*\gasg{$N_{\rm{H}}^{\rm{Gal}} $}

\newcommand*\al{$\alpha$}

\newcommand*\aax{$\alpha_{\rm{XRT}} $}
\newcommand*\apre{$\alpha_{\rm{pre}} $}
\newcommand*\apos{$\alpha_{\rm{pos}} $}
\newcommand*\aei{$\alpha_{\rm{EI}} $}

\newcommand*\aprex{$\alpha^{\rm{XRT}}_{\rm{pre}} $}

\newcommand*\apreo{$\alpha^{\rm{opt}}_{\rm{pre}} $}
\newcommand*\aposo{$\alpha^{\rm{opt}}_{\rm{pos}} $}

\newcommand*\tb{$t_{\rm{b}}$}
\newcommand*\tba{$t_{\rm{b_1}}$}
\newcommand*\tbb{$t_{\rm{b_2}}$}

\newcommand*\ebv{$E\rm{(B-V)} $}
\newcommand*\av{$A_{\rm{V}} $}

\newcommand*\be{$\beta$}
\newcommand*\bpre{$\beta_{\rm{pre}}$}
\newcommand*\bpos{$\beta_{\rm{pos}}$}
\newcommand*\bei{$\beta_{\rm{EI}}$}
\newcommand*\bo{$\beta_{\rm{opt}} $}
\newcommand*\bx{$\beta_{\rm{XRT}} $}
\newcommand*\nux{$\nu_{\rm{XRT}} $}
\newcommand*\nuo{$\nu_{\rm{opt}} $}

\newcommand*\sm{$sm$}
\newcommand*\sma{$sm_1$}

\newcommand*\swift{\textit{Swift}}

\newcommand*\q{$q$}

\newcommand*\s{$\,\,$}

\newcommand*\fref{Fig. \ref}
\newcommand*\tref{Table \ref}
\newcommand*\cref{Chap. \ref}
\newcommand*\sref{Sec. \ref}
\newcommand*\eref{Eq. \ref}

\newcommand*\tit{\textit}
\newcommand*\tbf{\textbf}

\newcommand*\pmm{$\pm$}

\newcommand*\pl{power-law}

\newcommand*\brpl{broken power-law}
\newcommand*\brplh{broken power-law with host}
\newcommand*\trbrpl{double broken power-law}

\newcommand*\eg{e.g., }
\newcommand*\ie{i.e., }
\def\fdg{\hbox{$.\!\!^\circ$}}

\begin{document}

\title{The GROND GRB sample: \\II. Fireball parameters for four GRB afterglows
\thanks{Based on data acquired with the Atacama Pathfinder Experiment (APEX) under ESO programme 091.D-0131(A).}}

   \author{K. Varela\inst{1}\thanks{\emph{Present address:} Influur Corporation, 1111 Brickell Ave, Miami, FL 33131, U.S.A.}  \and 
   J. Greiner\inst{1} \and
   P. Schady\inst{2} \and
   H. van Eerten\inst{2}
   }

    \institute{Max-Planck Institut f\"ur extraterrestrische Physik,
              Giessenbachstr. 1, 85748 Garching, Germany\\
         \email{jcg@mpe.mpg.de}
    \and
    Department of Physics, University of Bath, Building 3 West, Bath BA2 7AY, United Kingdom\\
         \email{ps2018@bath.ac.uk, hjve20@bath.ac.uk} 
   }

\date{Received  March 24, 2025; accepted July 7, 2025}

  \abstract{
  Afterglows of GRBs are, in general, well described by the fireball model. Yet, deducing the full set of model parameters from observations without prior assumptions has been possible for only a handful of GRBs.
   With GROND, a 7-channel simultaneous optical and near-infrared imager at the 2.2m telescope of the Max-Planck Society at ESO/La Silla, a dedicated gamma-ray burst (GRB) afterglow observing program was performed between 2007 and 2016.
   Here, we combine GROND observations of four particularly well-sampled GRBs with public Swift/XRT data and partly own sub-mm and radio data to determine the basic fireball afterglow parameters.
   We find that all four bursts exploded into a wind environment. We are able to infer the evolution of the magnetic field strength from our data, and find evidence for its origin through shock amplification of the magnetic field of the circumburst medium.
   }
   
   \keywords{Gamma rays: bursts -- gamma-ray burst: individual: GRB 100418A, 110715A, 121024A, 130418A --
                Techniques: photometric
               }
           
\maketitle

\section{Introduction}

In the standard GRB afterglow model, the dominant process during the afterglow phase is synchrotron emission from shock-accelerated electrons in a collimated relativistic blast wave interacting with the external medium \citep{PaczynskiRhoads1993, MeszarosRees1997, Waxman1997}.
Under the implicit assumption that the electrons are ``Fermi'' accelerated at the relativistic shocks to a power law distribution with an index $p$ ($p>2$ is usually assumed to keep the energy of the electrons finite), their dynamics can be expressed in terms of 4 main parameters (in addition to $p$):
(1) the total internal energy $E$ in the shocked region as released in the explosion, 
(2) the density and radial profile of the  surrounding medium,
(3) the fraction of the shock energy going into electrons $\epsilon_e$,
(4) the fraction of the energy density in the  magnetic field behind the shock $\epsilon_B$. Both, $\epsilon_e$ and $\epsilon_B$ are typically assumed to be constant throughout a burst.
The emission spectrum of the forward shock of such an electron population then consists of three segments above the self-absorption frequency \nusa\ \citep{Sari+1998, GS2002}: a low-energy tail where $F_\nu \sim \nu^{1/3}$, a high-energy power-law segment with $F_\nu \sim \nu^{-p/2}$, and an intermediate segment, the slope of which depends on the relative ordering of the cooling frequency $\nu_c$, and the characteristic frequency $\nu_m$.
 
With the advent of Swift's rapid slewing capabilities \citep{Gehrels+2004}, the community has collected a wealth of data to trace the temporally and spectrally changing afterglow emission over a wide range of frequencies, from the radio to the $\gamma$-rays. This allows us, in principle, to put constraints on the underlying physical model, and its boundary conditions.

In practice, this comes with various complications:
(i) Bright and/or nearby events allow us to detect additional features on top of the generic afterglow scenario, making it difficult to extract its basic parameters.
(ii) In canonical, mostly fainter afterglows, detections at multiple times at multiple frequencies are challenging, thus rarely covering sufficient spectral and temporal range to infer all basic parameters without extra assumptions.
(iii) A substantial fraction of X-ray afterglows shows a plateau phase, inconsistent with the standard afterglow model. Whether explained from energy injection, as is commonly assumed, or from other processes, such plateaus require additional parameters.
(iv) Finally, even in the perfect case of the observational data being consistent with the generic model, we may be surprised about best-fit parameters falling in regions unexpected in our favourite model scenario.

Examples for the latter case are jet breaks and the external density profile.
While the standard GRB afterglow model describes the majority of observations, the identification of clear jet breaks in only about 10\%--15\% of their light curves has been puzzling since the early times \citep{2009racusin}. Several factors may contribute, from the lack of multi-wavelength data (to prove the achromatic behaviour) over not sufficiently accurate measurements (to identify small decay slope changes) to diverse internal jet structure (changing the light curve profile around the break) or jet orientation relative to the observer \citep{2010ApJ...722..235V}. Similarly, the prevalence of a constant radial density profile around GRBs \citep{2001ApJ...554..667P, Schulze+2011} is at odds with the idea that massive stars with strong winds are the progenitors of long-duration GRBs \citep{ChevalierLi2000, 2006A&A...460..105V}.

The aim of the present study is to help form an unbiased picture of the physics of GRB afterglows by adding cases for which all parameters of the standard model can be deduced without any assumption (apart from a synchrotron origin from a decelerating blast wave). We do not include pre-GROND GRBs in our selection process, simply because of the spotty multi-wavelength (in particular NIR) data. From our sample of GROND-observed GRBs \citep{Greiner+2024} we picked those four (100418A, 110715A, 121024A, 130418A) for which our multi-epoch, multi-wavelength (incl. sub-mm/radio) observations cover the full range from radio to X-rays, so that we can measure the three break frequencies of the synchrotron spectrum, constrain the dust extinction, the X-ray absorption and the microphysical parameters, incl. tests of their constancy.  

The paper is structured as follows: Section 2 describes the observational data, Sect. 3 presents the modelling approach, separated into phenomenological analysis and derivation of the physical parameters, and Sec. 4 discusses the results in the context of previously published results.

\section{Observations and data reduction}

GROND \citep{gbc08}, a simultaneous 7-channel optical/near-infrared imager covering the wavelength range 0.4--2.4 $\mu$m (\griz$JHK_s$) and mounted at  the 2.2\,m  MPI/ESO telescope at La Silla (ESO, Chile), was designed and developed to rapidly identify GRB afterglows and measure their redshift via the drop-out technique.
GROND has observed all well-localized gamma-ray burst visible from Chile between GRB 070521 and GRB 161001A. For further details, the observing strategy and some statistics (see \citealt{Greiner+2024}).

We take the GROND-GRB sample as the parent sample because of the importance for well flux-calibrated and simultaneous optical+NIR afterglow data. While we have well-sampled optical/NIR light curves of 54 GRBs with GROND, only 8 of those have radio detections; as we will show below, these are crucial for the proper modelling of the afterglow. In addition, we only consider those GROND GRBs that have (i) radio afterglow data in at least 2 radio frequencies over at least 2 epochs (this excludes GRBs 141026A \citep{Corsi2014} and 141109A \citep{CorsiBhakta2014}), and (ii) a broadband spectral energy distribution (SED) that can be well described by a single synchrotron emission model (this excludes GRBs 081007 \citep{Jin+2013} and 090313 \citep{Melandri+2010}). 
Thus, we are left with 4 GRBs.

Here, we analyse observations of GRBs 100418A, 110715A and 130418A, based on public Swift/XRT data, our own extensive GROND monitoring, and the results of our radio (ATCA) and sub-millimeter (APEX) observing programs. Where appropriate, data from other radio/sub-mm observations as published in the literature is incorporated. In the final discussion, the analysis results of similar multi-epoch, multi-frequency data of GRB 121024A \citep{Varela2016A&A...589A..37V} are included.

GROND data were  reduced in the standard manner \citep{2008kruehler} using pyraf/IRAF \citep{1993tody, kkg08b}. The optical imaging data ($g^\prime r^\prime i^\prime z^\prime$) was calibrated against the Sloan Digital Sky Survey (SDSS)\footnote{http://www.sdss.org} \citep{Eisenstein+2011}, or Pan-STARRS1\footnote{http://pan-starrs.ifa.hawaii.edu} \citep{Chambers+2016},
and the NIR data ($JHK_{\rm s}$) against the 2MASS catalogue \citep{2006Skrutskie}. This results in typical absolute accuracies  of $\pm$0.03~mag in $g^\prime r^\prime i^\prime z^\prime$ and $\pm$0.05~mag in $JHK_{\rm s}$.
Finding charts and optical/NIR calibration of the GROND data for our three GRBs are given in Appendix A. 
Observational details of all multi-wavelength data and their corresponding analysis are described in Appendix B.

\section{Fireball modelling}

\subsection{Procedural notes}

We separate our analysis into two parts, 
(i) a model-independent phenomenological analysis where we determine the temporal and spectral slope(s) of the observed flux (using the convention $F\sim t^{-\alpha}\nu^{-\beta}$, with \al\s and \be\s the temporal and spectral slope, respectively),   
and also discuss particular features like flares, breaks in the light curve, flattening, or any behaviour different from that expected for a canonical afterglow light curve \citep{GS2002}, and 
(ii) a model-dependent analysis where we determine the break frequencies and the microphysical parameters by adopting the \cite{GS2002} formulation. Details of both steps are given below.

\subsubsection{Model-independent analysis}

\paragraph{Light curve fitting:}
\label{sedsec_grb130418a}
 
The main functions used here for the light curve fitting (and similarly for the SED, using $\nu$ instead of $t$)  are:
\begin{equation}
F_{\nu}(t) = 
\begin{cases}
F_0 \times \left( \frac{t}{t_0}\right)^{-\alpha}  \\
F_1 \times \left[ \left( \frac{t}{t_{\rm{1}}} \right)^{-\alpha_1 \rm{sm}}+\left( \frac{t}{t_{\rm{1}}} \right)^{-\alpha_2 \rm{sm}}  \right]^{-1/\rm{sm}} + host \\
F_1 \times \left[ F'_{\nu}(t_1)^{-\rm{sm}} + F'_{\nu}(t_2)^{-\rm{sm}}  \times \left( \frac{t}{t_0}\right)^{-\alpha_3}  \right]^{-1/\rm{sm}} + host 
\end{cases}
\label{smoothfit}
\end{equation}
for a simple \pl, a smoothly broken and \trbrpl, respectively \citep{1999AA...352L..26B}. $F_i$ are the normalisation factors at the time $t_i$, $\rm{sm}$ is the smoothness of the break $i$, $\alpha_i$ are the slopes for each power-law segment and $host$ refers to the contribution of the host galaxy (if detected in late-time observations). The analysis on the temporal evolution provides information on \al\s and possible features like flares, breaks in the light curve, plateau phases and information on the host galaxy (e.g. optical/NIR magnitudes). 
 
\paragraph{SED fitting:}
\label{sedsec}

The fitting of the spectral energy distribution is performed using XSPEC v12.7.1 \citep{1996ASPC..101...17A}. 
The SED analysis incorporates two steps:
First, an analysis of the optical/NIR and X-ray data is performed, including
estimates of the dust (UV/optical) and gas (X-ray) attenuation effects along the line of sight due to both the local environment and the host galaxy. 
Galactic reddening \dustg\s is taken from \cite{2011ApJ...737..103S} and a Milky Way extinction law with $R_V = 3.08$ is adopted \citep{1992ApJ...395..130P}. For the host galaxy, templates based on the Small and Large Magellanic Cloud are used \citep{1992ApJ...395..130P} and the values for extinction \dusth are derived from our afterglow SED fits.

The second step after the derivation of \dusth, \gash\s and \be\s is
the inclusion of sub-mm and radio data
in order to measure the three break frequencies. 
The slope \be\s of the GROND and XRT bands is  allowed to vary within its previous 3$\sigma$ uncertainty intervals. The smoothness of each break, depending on the temporal slopes in the optical/NIR and the X-rays, is taken from \citep{GS2002}. The break frequencies are left free to vary. 
We fit the SED with three breaks, but also allow for the possibility of fewer breaks if not all are covered by our observational range -- therefore, the different fit profiles described in \eref{smoothfit} are tested.

\subsubsection{Model-dependent analysis}
\label{modeldepana}

\paragraph{The standard afterglow model:}
This analysis is explicitly based on the canonical afterglow model as described analytically in \cite{GS2002}, where the dominant emission is associated with synchrotron radiation from shock-accelerated electrons. These electrons are assumed to have a power-law energy distribution with slope $p$ and minimum energy \gammam. The observed synchrotron spectrum is characterised by three main break frequencies (\nuc,\numm, \nusa) and a peak flux. The synchrotron injection frequency \numm\s is defined by \gammam. The cooling frequency \nuc\s is defined by the critical value \gammac, above which electrons radiate their energy on smaller timescales than the explosion timescale. The self-absorption frequency \nusa\s marks the frequency below which the optical depth to synchrotron-self absorption is $>1$. In this model, two main cooling regimes are defined by the relative position of the break frequencies: a fast cooling regime where \numm>\nuc\s and most of the electron are cooling fast, and a slow cooling regime where $\nu_{\rm{m}}<\nu_{\rm{c}}$
\citep{GS2002}.

The number of combinations of \al\s and \be\s is limited when a specific dynamical model and the synchrotron spectrum are given, though these are different for wind ($k$=2) and ISM ($k$=0) density profiles $r^{-k}$. This gives rise to a unique set of relations between \al\s and \be\s known as "closure relations" \citep{1994ApJ...430L..93R,1997MNRAS288wijers,Sari+1998,DC2001,zhang_meszaros:2004, 2009racusin, Gao2013141}. These relations constrain the cooling regime, the circumburst environment, the jet geometry and the electron energy distribution $p$. The fit results of the multi-power law segment spectrum to the multi-wavelength data is used to both determine all the break frequencies \nusa, \numm\ and \nuc, and to identify the correct (in most cases unique) closure relation. With this solution, we derive the fireball parameters using the relations of  \cite{2005ApJ...618..413G} (their Appendix C) for post-jet break SEDs.

\paragraph{Handling of plateau phases:}
The frequently observed plateaus in the light curve require an additional model ingredient \citep[e.g.][]{FanPiran2006}. One possibility is to assume a non-adiabatic (coasting) phase in the jet dynamics 
where the contact discontinuity between forward and reverse shock, separating ejecta from circumburst medium, is Rayleigh-Taylor unstable. The corresponding cooling of the forward shock will prevent the turn-on of the magnetic field until an observer time later than the deceleration time, thus leading to a plateau phase \citep{DuffellMacFadyen2014}. While several scenarios were put forward (see sect. \ref{sect:plateaus_all}) the more frequently used explanation is prolonged energy injection \citep{1998ApJ...496L...1R,2000ApJ...535L..33S} with the temporal luminosity evolution described via $L(t) = L_o\, t^{-q}$, where $q$ is the injection parameter and $L_o$ the initial luminosity. Phenomenologically, the injection parameter can be easily inferred from the light curve analysis.
Besides the many observed X-ray and optical plateau phases, prolonged energy injection phases have also been detected in sub-mm and radio data (\eg \citealt{2006ApJ...647.1238J, 2013ApJ...779..105M}).
A simultaneous detection of the plateau phase at all wavelengths implies a dynamical origin for the change in the temporal evolution of the afterglow emission.  
Physically, the energy injection mechanism depends on the type of the progenitor and the properties of the central engine, and three main mechanisms have been proposed \citep{2000ApJ...535L..33S, 2005ApJ...628..315Z}:
\begin{itemize}\vspace{-0.18cm}
\item A Poynting flux dominated outflow: a magnetar progenitor acts as a long-lived central engine with a constant luminosity, implying q = 0 \citep{1998A&A...333L..87D, 2000ApJ...537..803D}. 
\item Mass stratification: while all material is ejected promptly, different outflow velocities lead to the stratification of shells, \ie $M(\gamma) \propto \gamma^{-s}$. The subsequent collision of the shells (after the first one has decelerated) cause additional injection of energy during the afterglow evolution \citep{1998ApJ...496L...1R}. The slope $s$ is related to the injection parameter $q$:  $s = (10-7q)/(2+q)$ and  $s = (4-3q)/q$ for ISM or wind case, respectively \citep{2006ApJ...642..354Z, 2006ApJ...643.1036P}. For $s > 1$, corresponding to $q < 1$, the dynamics of the outflow is altered such that energy injection occurs.
\item Relativistic reverse shock:  
If the (usually sub-dominant) reverse shock is strong and relativistic, it could mimic an energy injection phase \citep{2000ApJ...545..807K, 2013ApJ...776..119L, 2014MNRAS.442.3495V}.
\vspace{-0.2cm}\end{itemize}

\noindent After the energy injection phase, only a decelerating forward shock remains and the standard afterglow emission model applies. However, the end of a plateau phase is connected with a break in the light curve, which needs to be distinguished from a canonical jet break. 
Using the flux and frequency equations for a radial flow from \citet{2009MNRAS.394.2164V} and \citet{2012MNRAS.427.1329L}, we derive the closure relations for a general density profile with an arbitrary $k$ during the deceleration stage following energy injection as $k=\frac{4(2\alpha-3\beta)}{1+2\alpha-3\beta}$ (valid for \numm<$\nu$<\nuc).

\paragraph{Additional emission component:}
\label{SSC}
Inverse Compton scattering of synchrotron photons from the relativistic electrons in the afterglow (in the following Synchrotron Self-Compton, SSC) can change the time evolution of the cooling break of the synchrotron component, and thus the optical and X-ray afterglow light curves, since SSC is an additional (and even more efficient) cooling process than synchrotron. The effect depends on the Comptonisation parameter Y = $\gamma^2\tau_e$: If $Y > 1$, which corresponds to $\epsilon_e > \epsilon_B$, a large fraction of the low energy synchrotron radiation will be up scattered by SSC. We test the effect of SSC via the $C$ parameter \citep[which is a combination of break frequencies and peak flux, and $C < 1/4$ holds; see eq. 4.9 in][]{sari...esin2001}, once all parameters have been determined. If SSC dominates, we use the appropriate equations for the physical parameters as given in \cite{sari...esin2001} instead of the canonical \cite{GS2002}.

\subsection{Phenomenological analysis}
\label{analysis}

\subsubsection{GRB 100418A}

\paragraph{Afterglow light curve fitting:}
\label{sect:lcsec_grb100418a}

\begin{figure}[bht]
\includegraphics[width=0.49\textwidth]{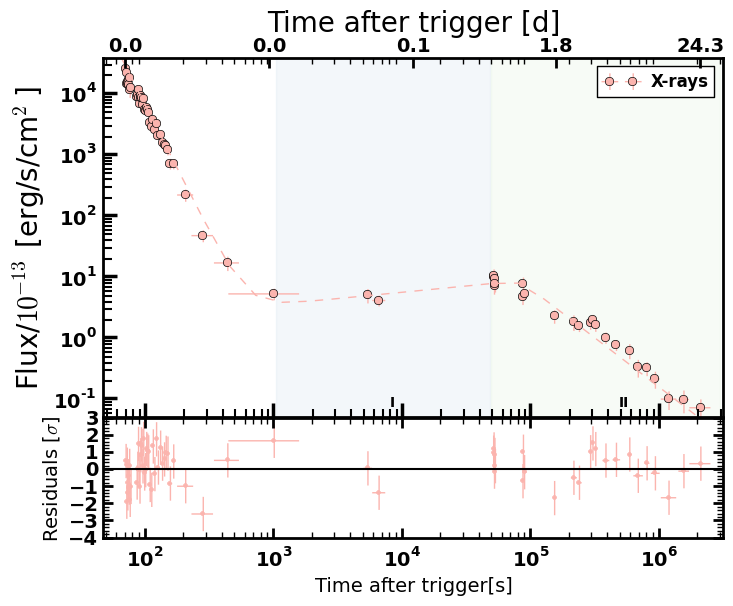}
\vspace{-0.2cm}
\caption{X-ray light curve of GRB 100418A from the XRT repository. The best fit is a smoothly \trbrpl\s shown in dashed lines. The epochs used are marked by vertical shaded regions: 
the steep decay phase (white),
the plateau phase (blue), and the
post-energy injection phase after a jet break (light green).}
\label{Fig:LC_xray_fit_100418A}
\end{figure}

\begin{figure}[ht]
\includegraphics[width=0.49\textwidth]{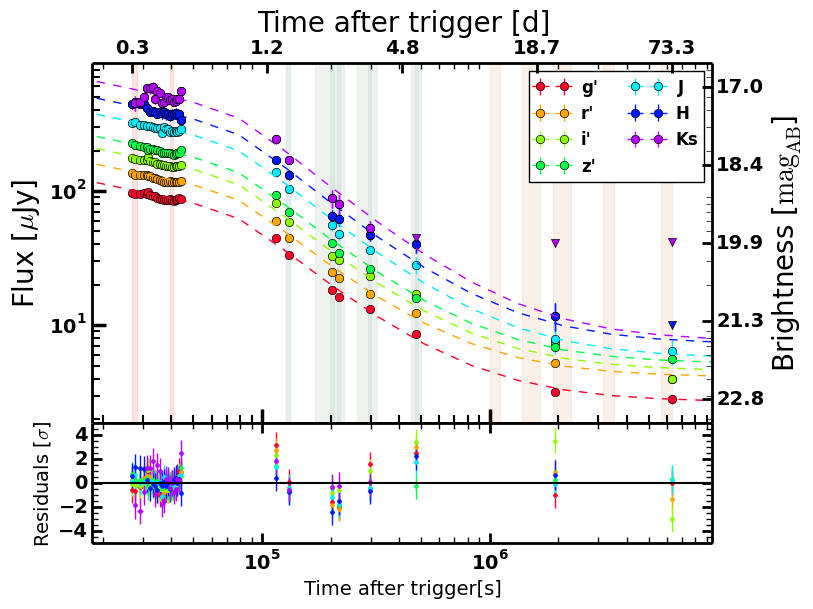}
\caption{GROND light curve $g'r'i'z'JHK_s$ of the afterglow of GRB 100418A. The best fit of the combined optical/NIR and X-ray data is a smooth \brplh\s contribution shown in dash lines. The epochs used for the spectral analysis are highlighted with the vertical bars. The first two epochs highlighted in light red correspond to the energy injection phase. 
The last five epochs in orange correspond to the slow cooling regime, the earlier epochs to fast cooling (see sect. \ref{sect:theory_grb100418a}). For the blue epochs joint GROND-XRT SEDs were produced (see Fig. ~\ref{Fig:SED_fit_grb100418A} and \sref{sect:sedsec_grb100418a}). The green and orange epochs correspond to the SED analysis that includes radio data (\sref{theory}).}
\label{Fig:LC_opt_fit_100418A}
\end{figure}

The X-ray temporal evolution for the afterglow of GRB 100418A (Fig. \ref{Fig:LC_xray_fit_100418A}) is well described by a \trbrpl\s with smooth breaks (see \eref{smoothfit}). 
The plateau phase (\aei=-0.21\pmm0.12 up to \tbb=82\pmm29 ks) after the initial steep decay (with \apre=4.16\pmm0.08) may be associated to an ongoing energy injection phase \citep{marshall2011}. 
The late decay (\apos=1.61\pmm0.19) can be associated to the normal afterglow decay where the break time is associated to either the end of the ongoing energy injection
or a jet break.

\begin{figure}[hb]
\hspace{-0.1cm}\includegraphics[width=0.49\textwidth]{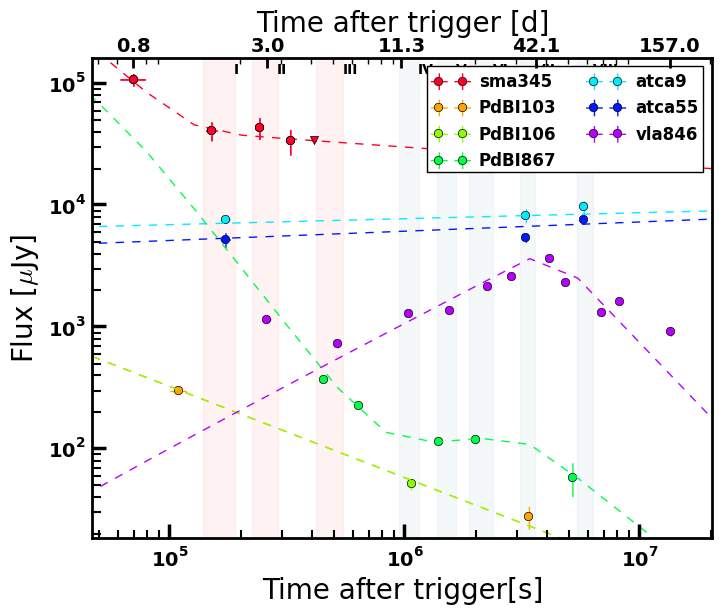}
\vspace{-0.2cm}
\caption{Sub-mm and radio light curves of the GRB 100418A afterglow at different frequencies, with dashed lines showing the best fit. The eight highlighted vertical regions correspond to the epochs used in the SED analysis. The orange (blue) regions correspond to the fast (slow) cooling regime. The light curves are scaled by an arbitrary factor for clarity.}
\label{Fig:LC_radio_fit_100418A}
\end{figure}

The optical/NIR light curve (\fref{Fig:LC_opt_fit_100418A} and \tref{Table:grb100418A_mags_opt}) in all 7 bands ($g'r'i'z'JHK_s$) starts with a shallow decay around 10$^4$\,s, followed by a steeper decay. The best fit describing the observations is a smooth \brplh\s contribution,
with best fit parameters \apre = 0.32\pmm0.04, \apos=1.41\pmm0.04, \tb=73.6\pmm2.5 ks, \sm=15\pmm11 (\chidof=181/184). The Swift/UVOT observations in the $white$ band \citep{2010GCN..10625...1S} and the observations in the R$_c$ band \citep{2010GCN..10635...1B,2010GCN..10794...1H} show a fast increase in flux between 2000 s and 7000 s, which could be the result of a flare on top of the plateau phase \citep{marshall2011} or a refreshed shock. However, since it is not covered by either Swift/XRT or GROND, it is difficult to determine the real cause.
After this time, the behaviour is consistent with GROND; the possible flare contribution is not taken into account in our study. 

Given that after T+10~ks the XRT and GROND data are both described well by a smooth \brpl\s with consistent best-fit parameters, a better constraint on the break time is obtained by a combined fit of both the XRT and GROND data. The main difference between the combined and the individual data sets are the values of the pre-break slopes. 
We performed three different fits (all of them with the break time linked):
the best fitting model is that with only the post-break slopes linked, resulting in fit parameters of  \aprex=0.11\pmm0.05 and \apreo=0.36\pmm0.04, \tb=76.4\pmm2.7 ks, \sm=6.9\pmm1.3 and  \apos=1.46\pmm0.04, but the fit with also the pre-break slopes linked is only marginally worse.

\begin{figure}[thb]
\includegraphics[width=0.49\textwidth]{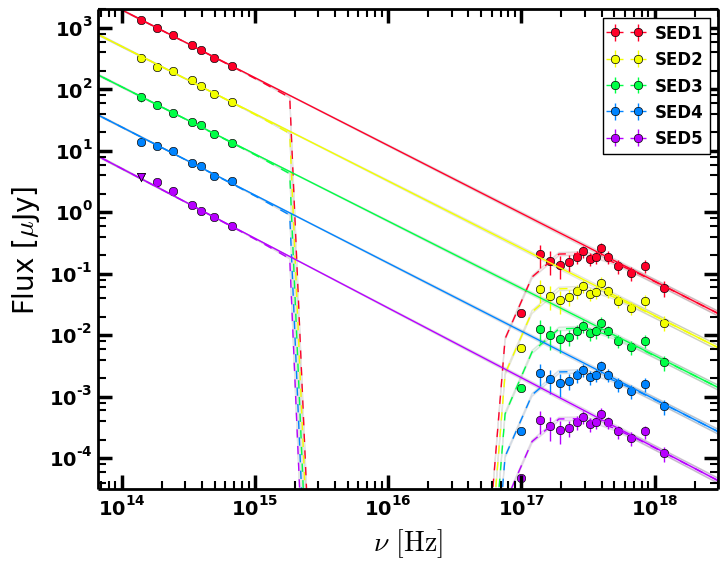}
\vspace*{-0.5cm}
\caption{SEDs of the afterglow of GRB 100418A using GROND and XRT data at five epochs:  SED1 ($t$=130.9 ks),  SED2 ($t$=202.1 ks), SED3 ($t$=217.8 ks), SED4 ($t$=296.8 ks), SED5 ($t$=476.4 ks). For clarity, the SEDs are scaled arbitrarily (GROND  magnitudes are given in \tref{Table:grb100418A_mags_opt}). The best-fit SED slope is \be=1.11\pmm0.02.
}
\label{Fig:SED_fit_grb100418A}
\end{figure}

Observations with SMA at 340 GHz begin at $\sim 0.8$~days post trigger and are well described by an initial decay with \apreo$\sim1.61$ up to \tb$\sim126$ ks, followed by a plateau phase of \aposo$\sim0.15$ (Fig.~\ref{Fig:LC_radio_fit_100418A}). Further observations at three epochs with PdBI at 106 GHz and 103 GHz are well described by a single \pl\s with a slope \al$\sim$0.75. In contrast, PdBI observations at 867 GHz  
are described by an initial slope with \al$\sim$2.1 up to \tba$\sim8.2\times10^{5}$ s followed by a plateau phase with \al$\sim$0.23 up to \tbb$\sim3.1\times10^{6}$ s and a final decay phase with \al$\sim$1.5 (Fig.~\ref{Fig:LC_radio_fit_100418A}). The observations with ATCA at 9.0 GHz and 5.5 GHz show a constant flux from $10^5$~s to $10^6$~s. However, it is possible that the first observations are affected by interstellar scintillation effects and therefore the actual flux might be lower.  
The VLA data show an increase in the flux with \al$\sim-1.0$ up to \tb$\sim4\times10^6$ s 
and then a decay phase with \al$\sim$2.1 (\fref{Fig:LC_radio_fit_100418A}). The scintillation effects on the observations are included as an additional error on the individual observations.

\begin{figure}[!ht]
\includegraphics[width=0.49\textwidth]{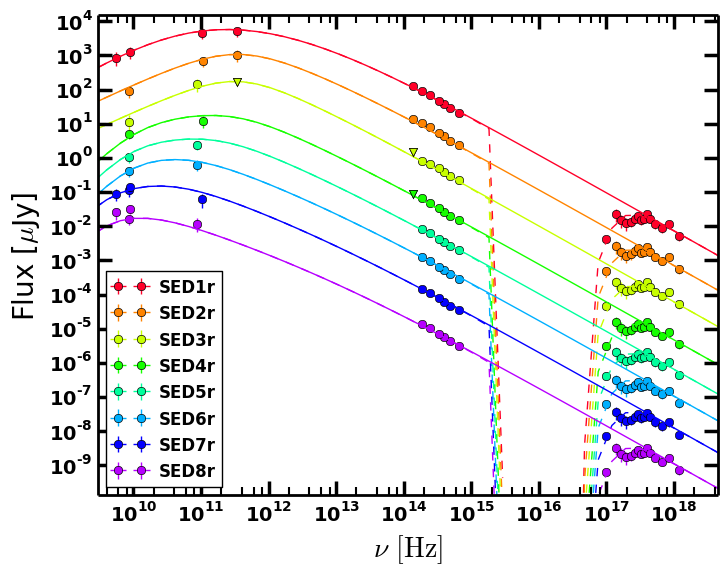}
\vspace{-0.6cm}
\caption{Broadband SED analysis for GRB 100418A, using eight epochs analysed using multi-epoch broad-band observations. The first three epochs correspond to the fast cooling regime (SED1r - SED3r); the last five epochs (SED4r - SED8r) correspond to the slow cooling regime. The best fit model for all SEDs is a \trbrpl\s with smooth breaks. 
}
\label{Fig:SED_fit_grb100418A_evolution}
\end{figure}

\paragraph{Afterglow SED fitting:}
\label{sect:sedsec_grb100418a}

We used seven epochs with combined XRT and GROND data; two epochs during the plateau phase and five epochs after the break in the light curve (Fig.~\ref{Fig:SED_fit_grb100418A}). 
The best-fit host galaxy magnitudes from the last epoch were subtracted from the optical/NIR data. In our fits, the host galaxy dust extinction and gas absorption were linked between all SED epochs.

\begin{table}[ht]
\caption{Break frequencies for the 8 epochs of the GRB 100418A afterglow, using a \trbrpl\s. 
\label{Table:grb100418A_sedfit_freq}}
\vspace{-0.15cm}
\begin{tabular}{c c c c c}
\toprule
SED & mid-time [ks] & $\nu_{\rm{c},{12}}$ [Hz] & $\nu_{\rm{m},{11}}$  [Hz] & $\nu_{\rm{sa},{10}}$  [Hz]   \\
\midrule
I    & 173   &  $0.22^{+0.02}_{-0.04}$ &  $93.3^{+14.7}_{-26.8}$ & $4.88^{+2.54}_{-2.34}$ \\
II   & 259   &  $0.49^{+0.26}_{-0.23}$ &  $22.1^{+3.3}_{-3.6}$ 	 & $3.31^{+2.01}_{-0.89}$ \\
III  & 450   &  $0.61^{+0.09}_{-0.35}$ &  $9.12^{+2.08}_{-2.35}$ & $2.26^{+0.31}_{-0.12}$ \\
IV  &1065  &  $1.24^{+0.43}_{-0.10}$ &  $4.96^{+3.27}_{-0.95}$ & $0.91^{+0.86}_{-0.51}$ \\
V   &1555  &  $1.34^{+0.36}_{-0.31}$ &  $2.25^{+1.21}_{-0.14}$ & $0.85^{+0.91}_{-0.28}$ \\
VI  & 2246 &  $2.06^{+0.31}_{-0.22}$ &  $0.79^{+0.41}_{-0.11}$ & $0.76^{+1.03}_{-0.29}$ \\
VII & 3283 &  $2.20^{+0.19}_{-0.70}$ &  $0.44^{+0.42}_{-0.18}$ & $0.62^{+0.35}_{-0.28}$ \\
VIII& 5788 &  $4.40^{+1.46}_{-0.98}$ &  $0.17^{+0.05}_{-0.03}$ & $0.44^{+0.61}_{-0.03}$ \\
\bottomrule
\end{tabular}
\vspace{-0.1cm}
\tablefoot{The goodness of the fit is \chidof=184/159.}
\end{table}

\begin{figure}[ht]
\includegraphics[width=0.49\textwidth]{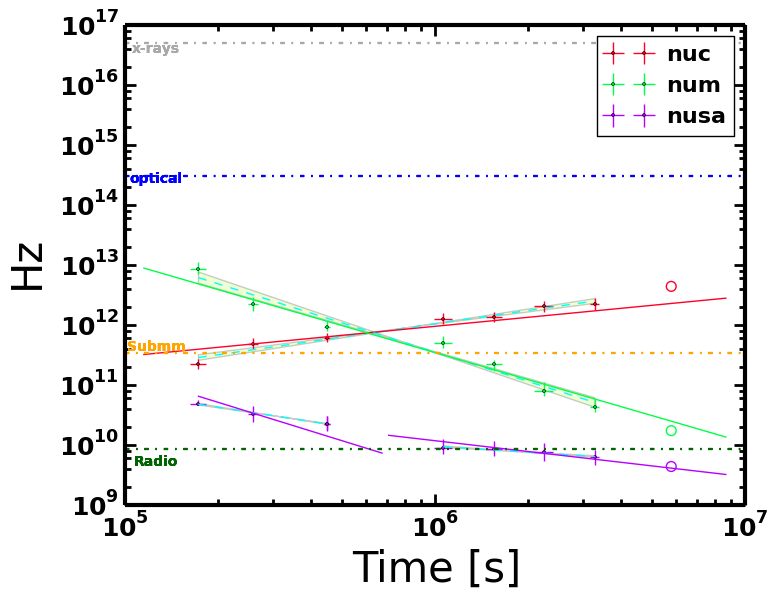}
\vspace{-0.5cm}
\caption{Evolution of the measured break frequencies for the eight optical to radio SEDs for the GRB 100418A afterglow. The solid lines represent the expected theoretical evolution for a wind environment during fast ($<$600 ks) and slow ($>$600 ks) cooling. The coloured dashed lines represent the best power-law fits to the measured temporal evolution (open data points were left out from the fit).
The horizontal dashed lines show the mean frequencies of our data (X-ray, optical, sub-mm, radio). 
}
\label{Fig:frec_evol_grb100418A}
\end{figure}

For the two epochs during the plateau phase (red regions in Fig. \ref{Fig:LC_opt_fit_100418A}), the SED slopes were initially left free to vary; the fits consistently have the same slope in both cases (\be$=1.14^{+0.08}_{-0.19}$ and \be$=1.11^{+0.08}_{-0.20}$, respectively).  When we tied the SED slopes, 
the fit returns a spectral index \be$=1.12^{+0.10}_{-0.18}$ and host dust extinction given by a Small Magellanic Cloud (SMC) reddening law \citep{1992ApJ...395..130P} \dusth$=0.06^{+0.19}_{-0.06}$ mag. In the case of the X-ray SEDs, the observations during the three analysed epochs are well described by a \pl\s with \be=0.94\pmm0.12, \gash = $3^{+1}_{-3}$ $\times$ 10$^{20}$ cm$^{-2}$ and a goodness of the fit of \chidof$=7.4/9$.

Optical and X-ray segments at later times have the same spectral slopes and can be described by a \pl. The XRT SED epoch between $t$=100 ks to $t$=300 ks is described by a mean spectral slope \be$=0.98^{+0.24}_{-0.20}$ with a goodness of the fit of \chidof$=9.0/12$ and \gash$=0.42^{+0.22}_{-0.08}\times10^{22}$ cm$^{-2}$. The five GROND epochs have a spectral slope \be$=1.01^{+0.11}_{-0.12}$. All the slopes are linked between the five SEDs; this is completely consistent with the fit with unlinked slopes and, therefore, it is evident that there is no spectral evolution. 
The combined fit results in a best-fit \pl\s with  slope \be=1.11\pmm0.02. The fact that this is slightly steeper than the separate GROND and Swift/XRT fits is likely due to the inaccuracy of the absolute flux calibrations between optical/NIR and X-ray bands.  
The host dust extinction is given by a Small Magellanic Cloud (SMC) reddening law \citep{1992ApJ...395..130P} with a value \dusth$=0.01^{+0.03}_{-0.01}$ mag. The gas column density is \gash$=5.7^{+0.9}_{-0.8}\times10^{21}$ cm$^{-2}$ and the goodness of the fit is \chidof=85/101.  
These constraints are going to be used as base values for the analysis in the following sections.

\paragraph{Broadband SED analysis:}
\label{BBSED:grb100418a}

When including the sub-mm and radio data 
a spectrum with three breaks is necessary to describe the observations; the gently curved SED in the 10$^{10}$--10$^{12}$ Hz range is too broad to be fit with two breaks (\fref{Fig:SED_fit_grb100418A_evolution}).
In total, eight epochs are used (\tref{Table:grb100418A_sedfit_freq} and \tref{Table:grbmags_sedr_GRB100418A}); these epochs are selected for the availability of radio data, and are distinct from the GROND-SED epochs shown in Fig.~\ref{Fig:SED_fit_grb100418A}. All the epochs were fitted simultaneously,  
with the only constraints being \dusth\s and \gash\s as derived previously. The slope of the GROND and XRT bands is allowed to vary within a 3$\sigma$ uncertainty interval around their previous best fit. The smoothness of each break depends on the temporal slopes in the optical/NIR and the X-ray \citep{GS2002}. 
\fref{Fig:SED_fit_grb100418A_evolution} shows the final fit for each SED,  the best-fit frequencies are given in \tref{Table:grb100418A_sedfit_freq}, 
and their movement with time is shown in \fref{Fig:frec_evol_grb100418A}.
The assignment of the nature of the breaks and the determination of the physical parameters is done in sect. \ref{sect:theory_grb100418a} based on the \citet{GS2002} framework.

\subsubsection{GRB 110715A}

\paragraph{Afterglow light curve fitting:}
\label{sect:lcsec_grb110715A}

\begin{figure}[!ht]
\hspace{-0.1cm}\includegraphics[width=0.49\textwidth]{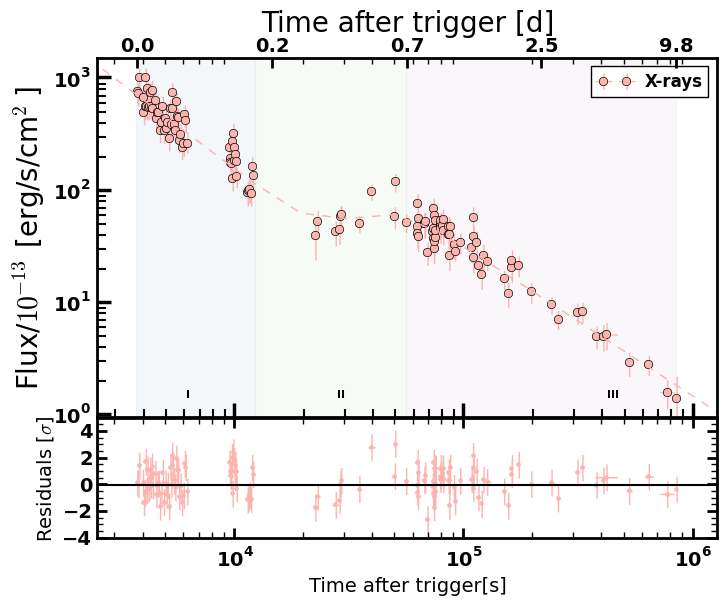}
\caption{X-ray light curve of the GRB 110715A afterglow, described by a smooth \trbrpl\s shown in dashed lines. The 3 regions used in the SED analysis are shown as shaded areas, corresponding to the GRB tail, the plateau and the final decay phase, respectively. }
\label{Fig:LC_xray_fit_110715A}
\end{figure}

\begin{figure}[ht]
\includegraphics[width=0.49\textwidth]{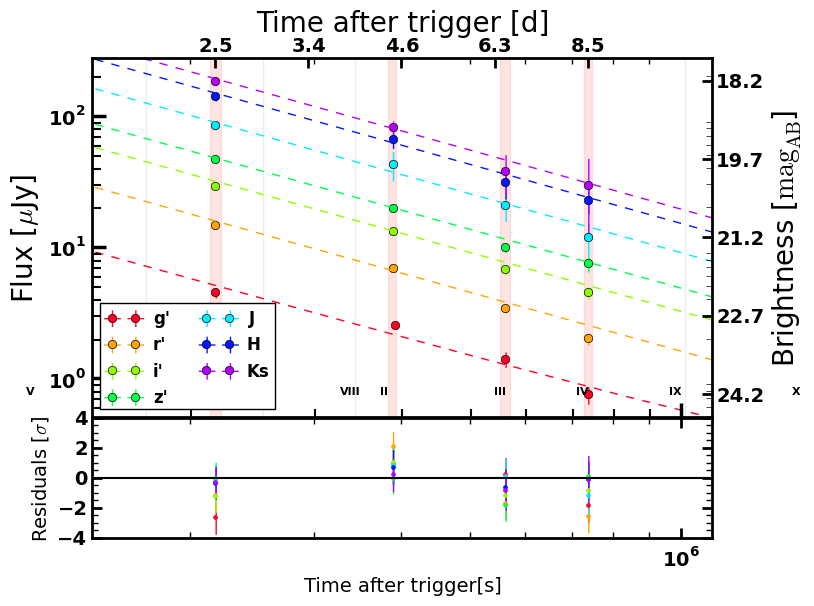}
\caption{GROND $g'r'i'z'JHK_{\rm{s}}$ light curve of the GRB 110715A afterglow. The best fit is a single \pl\s with \al = 1.51\pmm0.03 as shown with the dashed lines. The epochs used for the spectral analysis are highlighted with the vertical bars. All four epochs are after the plateau phase.}
\label{Fig:LC_opt_fit_110715A}
\end{figure}

The analysis follows the same methodology as for GRB 100418A: After individual X-ray (final decay with slope \apos = 1.34\pmm0.07; Fig. \ref{Fig:LC_xray_fit_110715A}) and GROND (single power law with slope \al = 1.51\pmm0.03; Fig. \ref{Fig:LC_opt_fit_110715A}) fits, a combined fit results in \apos = 1.48\pmm0.05 for both, optical/NIR and X-ray data (\chidof = 192/143).

We note that the time before the first GROND epoch was covered by Swift/UVOT and $R_c$ band observations  \citep{2011GCN..12162...1K, Nelson2011} and show a plateau phase like that seen in the X-ray light curve. 

\begin{figure}[th]
\includegraphics[width=0.49\textwidth]{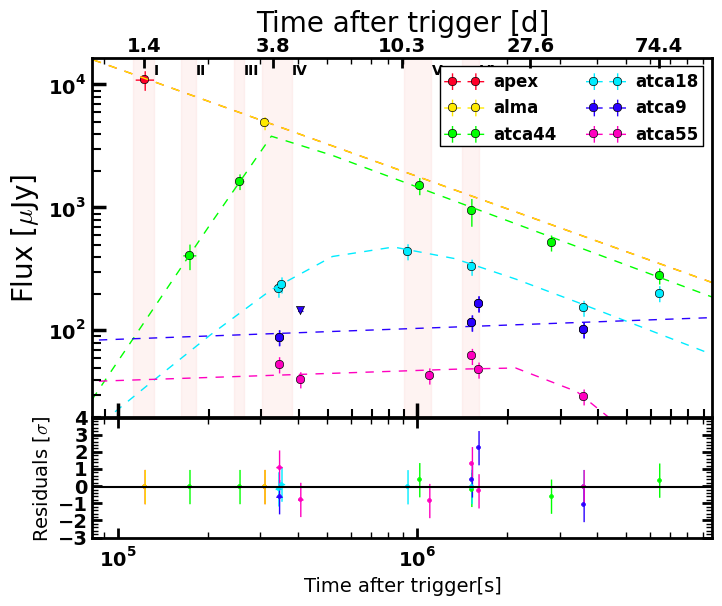}
\vspace{-0.5cm}
\caption{Sub-mm and radio light curves of the GRB 110715A afterglow. The best fit for each of the bands is represented by dashed lines. The six highlighted vertical regions correspond to the epochs used in the broadband multi-epoch SED analysis. The light curves are scaled by an arbitrary factor for clarity.}
\label{Fig:LC_radio_fit_110715A}
\end{figure}

Observations with APEX and ALMA at 345 GHz show a decaying flux between the two epochs, with a slope of $\alpha_{\rm{sub}}$ = 0.87\pmm0.23 \citep{2017MNRAS.464.4624S}. The six ATCA epochs at 44.0 GHz are described by a smoothly \brpl,
with slopes of \apre = -3.61\pmm0.71 and \apos = 0.91\pmm0.12 before and after
the break at \tb = 325.2\pmm28.2 ks, respectively.
The decay slope is consistent with the observations at 345 GHz. A similar behaviour is observed at 18 GHz, but with a late break time  
of \tbb = 613\pmm102 ks.
Finally, at frequencies of 9.0 GHz and 5.5 GHz the flux remains almost constant throughout the observations, with \al = 0.09\pmm0.07 and \al = 0.08\pmm0.11, respectively. At 5.5 GHz there is a change in the temporal evolution just before the last epoch where there is a steep decrease in flux with slope \al$\sim$1.8. The sub-mm and radio light curve fits are shown in \fref{Fig:LC_radio_fit_110715A}.

\paragraph{Afterglow SED fitting:}
\label{sect:sedsec_grb110715a}

\begin{figure}[ht]
\includegraphics[width=0.49\textwidth]{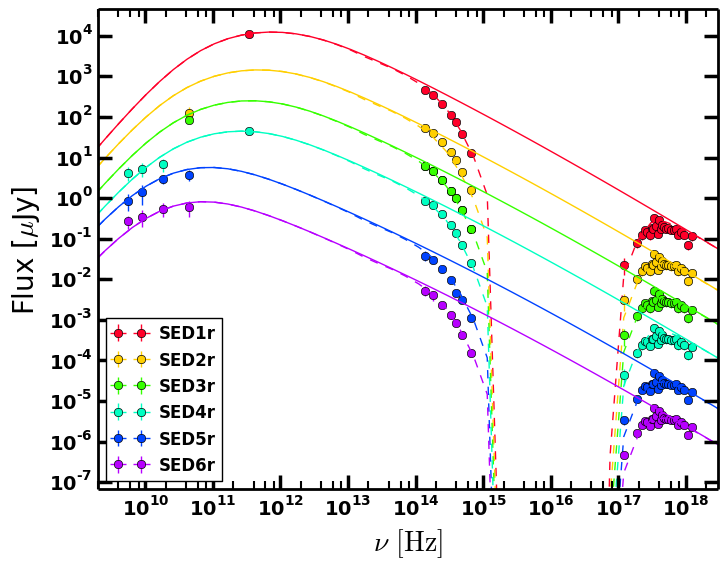}
\caption{Broad-band analysis of the GRB 110715A afterglow. Six epochs are presented with all the breaks measured.}
\label{Fig:SED_RADIO_GRB110715A}
\end{figure}

Three X-ray SEDs were analysed: before 
(3.7–12.1 ks), during (22.3–56.4 ks) and
after (62.3–849.1 ks) the plateau phase (see \fref{Fig:LC_xray_fit_110715A}).
The best fitting profile is a simple \pl\s with \gash = 0.55\pmm0.11$\times10^{22}$ cm$^{-2}$ and slopes \bpre = 1.01\pmm0.15, \bei = 0.85\pmm0.09 and \bpos = 1.06\pmm0.13.
For the analysis of the optical/NIR SEDs, the four slopes are very similar, and a fit with all four epochs linked
results in \dusth = 0.21\pmm0.05 mag and \be = 0.35\pmm0.12 
(\chidof = 14/22). 
A combined analysis of the XRT and optical/NIR observations was performed in order to check if a simple \pl\s can successfully describe the observations or if the suggested 0.65 slope difference  
is real. 
The XRT SEDs are renormalised to match the mid-time X-ray flux at the time of each of the optical SEDs. 
It turns out that a single \pl\s is indeed sufficient, 
leading to best fitting parameters of \be=1.05\pmm0.01, \gash=0.16$^{+0.03}_{-0.04}$, \dusth$=0.05\pm0.01$ (see \fref{Fig:SED_RADIO_GRB110715A}).

\begin{table}[th]
\caption{Break frequencies for the six epochs of GRB 110715A using broad-band observations. 
  \label{Table:grb110715A_sedfit_freqs}}
\vspace{-0.2cm}
\begin{tabular}{c c c c c}
\toprule
SED & mid-time [ks] & $\nu_{c,{13}}$ [Hz]  & $\nu_{m,{12}}$ [Hz] & $\nu_{{sa},{10}}$ [Hz]  \\
\midrule
I   & 122.7  & $2.49^{+0.83}_{-0.38}$ & $2.78^{+0.82}_{-0.25}$ & $5.17^{+1.05}_{-1.10}$ \\ 
II  & 173.2  & $3.94^{+0.92}_{-0.39}$ & $2.15^{+0.49}_{-0.23}$ & $2.85^{+0.77}_{-0.84}$ \\ 
III & 254.5  & $4.96^{+0.75}_{-0.44}$ & $1.26^{+0.23}_{-0.17}$ & $2.36^{+0.62}_{-0.36}$ \\ 
IV& 344.9   & $6.12^{+0.81}_{-0.36}$ & $0.82^{+0.02}_{-0.02}$ & $1.87^{+0.55}_{-0.38}$ \\ 
V & 1014.2 & $8.78^{+0.94}_{-0.93}$ & $0.18^{+0.02}_{-0.02}$ & $0.99^{+0.23}_{-0.16}$ \\ 
VI& 1513.8 & $9.70^{+0.86}_{-0.82}$ & $0.11^{+0.02}_{-0.03}$ & $0.76^{+0.25}_{-0.18}$ \\ 
\bottomrule
\end{tabular}
\vspace{-0.1cm}
\tablefoot{The epochs are highlighted in Fig. \ref{Fig:LC_radio_fit_110715A} and the final SED fitting is presented in Fig. \ref{Fig:SED_RADIO_GRB110715A}}
\end{table}
\smallskip

\begin{figure}[!ht]
\hspace{-0.3cm}\includegraphics[width=0.5\textwidth]{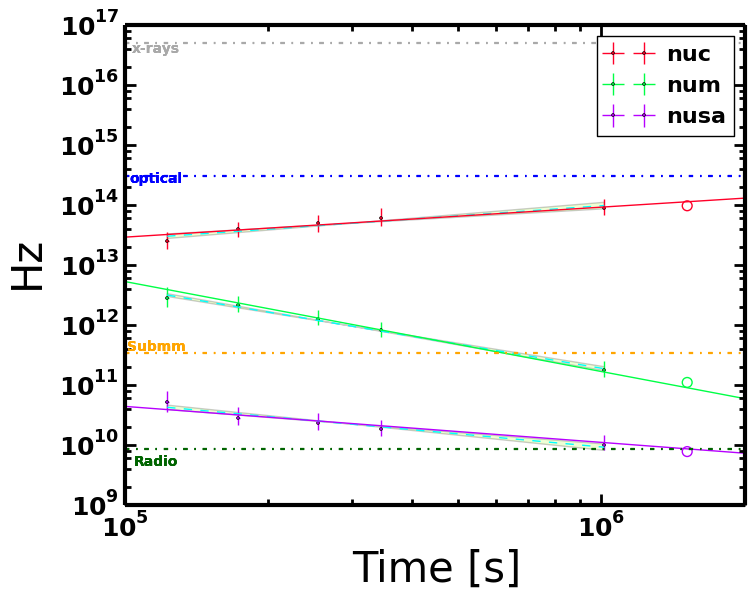}
\caption{Evolution of the break frequencies of the afterglow of GRB 110715A. The last SED is not included in the fit because the values for the optical/NIR bands were extrapolated. Lines are as in \fref{Fig:frec_evol_grb100418A}. 
}
\label{Fig:GRB110715A_freq_evol}
\end{figure}

\paragraph{Broadband SED fitting:}
 
Six epochs of SEDs including the sub-mm and radio data were fitted using a \trbrpl\s with smooth breaks.  
\tref{Table:grb110715A_sedfit_freqs} summarizes the break locations for all epochs, and \fref{Fig:GRB110715A_freq_evol} shows their movement with time.

\subsubsection{GRB 130418A}

\paragraph{Afterglow light curve fitting:}
\label{sect:lcsec_grb130418a}

\begin{figure}[!ht]
\includegraphics[width=0.49\textwidth]{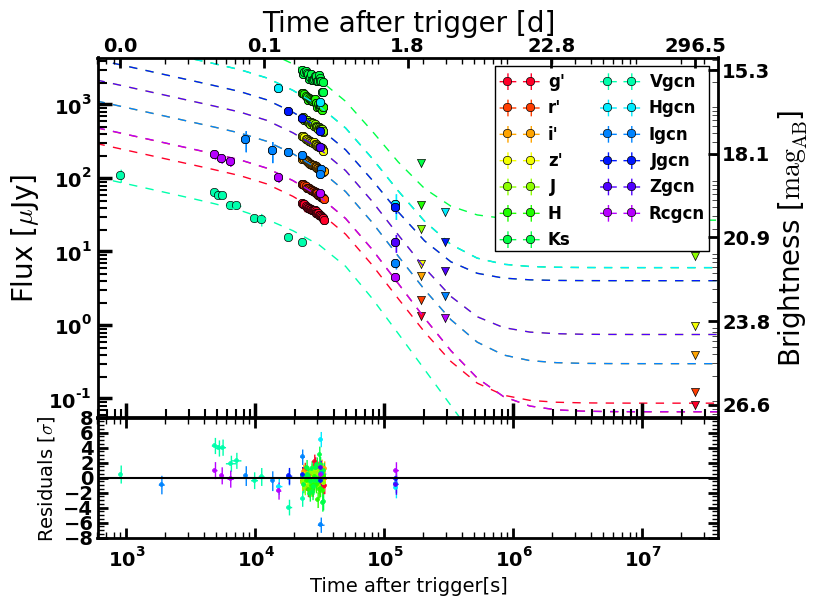}
\caption{Optical/NIR light curve of the afterglow of GRB 130418A observed with GROND, including data from the literature. The best fit model is a \trbrpl\s with smooth breaks, and includes a  host detection at r$^\prime$(AB) = 25.4$\pm$0.3 mag.}
\label{Fig:LCfit_grb130418a}
\end{figure}

While a single \pl\s fit to the X-ray data would be acceptable (\al = 1.47\pmm0.06 with  \chidof = 28/18),
The optical/NIR light curves from GROND (\fref{Fig:LCfit_grb130418a} and \tref{Table:GRONDmags_130418A})  
and data from the literature \citep{2013GCN..14378....1G, 2013GCN..14379....1Q, 2013GCN..14382....1K, 2013GCN..14388....1B} require two breaks. 
The best fitting parameters of the combined observations are: 
\aprex = 1.11\pmm0.14 and \apreo = 0.31\pmm0.08, a break time \tba = 18.8\pmm3.5 ks with smoothness \sm = 5.4\pmm1.3 followed by a decay with slopes \aei = 1.11\pmm0.14 up to \tbb = 61.7\pmm8.1 ks with smoothness \sma = 3.3\pmm0.8 and a final decay slope of \apos = 2.40\pmm0.19 (\chidof = 242/195).

The observations in the sub-mm and radio wavelength range show a rising,
and then fading flux at 340/345 GHz (SMA and APEX).

\begin{figure}[ht]
\includegraphics[width=0.49\textwidth]{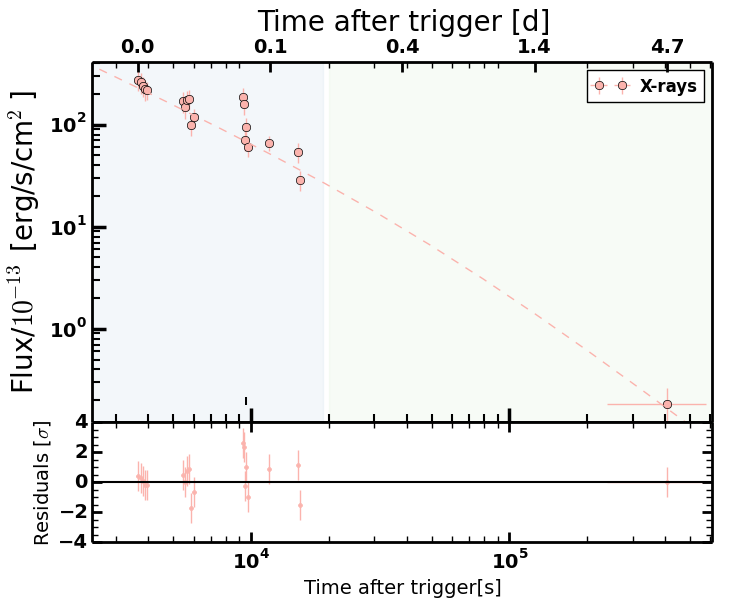}
\caption{X-ray light curve of the afterglow of GRB 130418A, with the early WT data omitted. 
The high-lighted vertical regions corresponds to the two main epochs of the SED analysis. 
}
\label{Fig:LC_xray_fit_130418A}
\end{figure}

\begin{figure}[th]
  \includegraphics[width=0.45\textwidth, height=6cm]{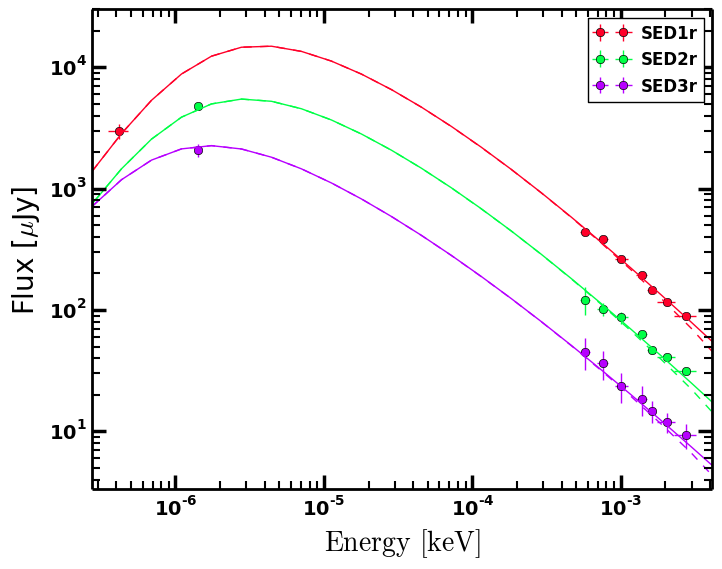}
  \vspace{-0.1cm}
\caption{Broad-band SEDs of the GRB 130418A afterglow at the three epochs of \tref{Table:grb130418A_sedfit_freqs}. 
Data are given in \tref{Table:GRONDmags_130418A}.}
\label{Fig:GRB130418A_radio_SED}
\end{figure}

\paragraph{Afterglow SED fitting:}
\label{sect:sedsec_grb130418a}
 
The X-ray data are best described by a single \pl\s with slope \be = 0.58\pmm0.11 with \gash = 8.6\pmm8.4$\times10^{20}$ cm$^{-2}$.
Three GROND SEDs are used, two before the break in the light curve at \tb = 45.4 ks and one after. 
The combined fit linking the individual slopes and the host \dusth
 provide best fitting parameters of \dusth\s = 0 and spectral slope \be=1.16\pmm0.07. 
A combined XRT/GROND SED was not possible given the non-matching coverage.

\paragraph{Broadband SED analysis:}
\label{BBSED:grb130418a}

\begin{figure}[!hb]
\includegraphics[width=0.47\textwidth, height=6.5cm]{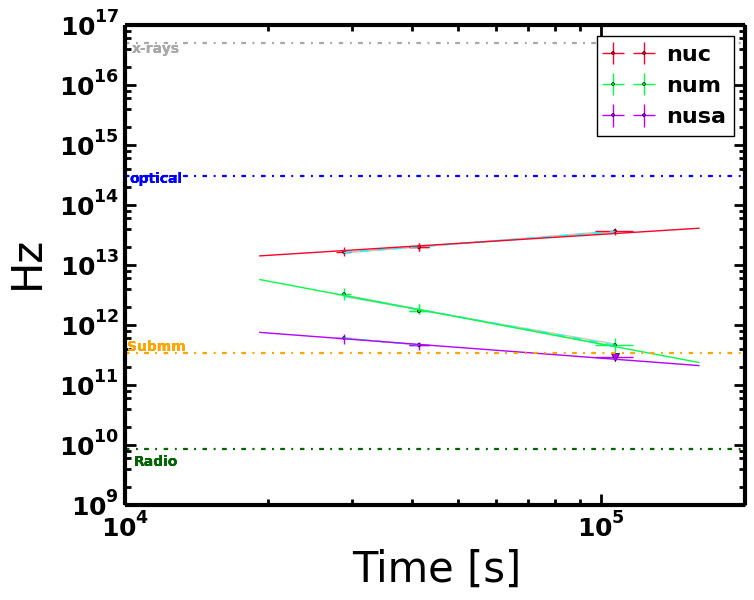}
\vspace{-0.2cm}
\caption{Evolution of the break frequencies of the GRB 130418A afterglow.  Lines are as in \fref{Fig:frec_evol_grb100418A}.
}
\label{Fig:frec_evol_grb130418A}
\vspace{-0.3cm}
\end{figure}

For the fits including the radio and sub-mm data, the values for the dust and gas attenuation effects \dusth, \dustg, \gash, \gasg\s along the line of sight
are set to the values obtained in the previous section. 
Three spectral breaks are required, and
the results are shown in \tref{Table:grb130418A_sedfit_freqs} and in \fref{Fig:GRB130418A_radio_SED}. The evolution of the break frequencies is shown in \fref{Fig:frec_evol_grb130418A}.

\begin{table}[!ht]
\caption{Break frequencies for the three epochs of the GRB 130418A afterglow, using a \trbrpl\ (\fref{Fig:GRB130418A_radio_SED}).}
\label{Table:grb130418A_sedfit_freqs}
\vspace{-0.2cm}
\begin{tabular}{c c c c c}
\toprule
SED & mid-time [ks] & $\nu_{c,{13}}$ [Hz]  & $\nu_{m,{12}}$  [Hz] & $\nu_{{sa},{11}}$  [Hz]  \\
\midrule
I   & 28.8 & $1.66^{+0.18}_{-0.23} $  & $3.26^{+0.33}_{-0.21} $  & $6.04^{+0.63}_{-0.48} $  \\ 
II  & 41.5 & $1.98^{+0.12}_{-0.19} $  & $1.73^{+0.23}_{-0.18} $  & $4.64^{+0.61}_{-0.46} $  \\ 
III & 106.8 & $3.65^{+2.48}_{-2.16} $  & $0.47^{+0.03}_{-0.02} $  & $2.93$ UL  \\ 
\bottomrule
\end{tabular}
\vspace{-0.2cm}
\end{table}

\subsection{Physical parameters of the standard afterglow model}
\label{theory}

We now proceed with the derivation of the microphysical and dynamical parameters of the GRB afterglow, based on the standard afterglow model
and including the effect of inverse Compton scattering as an additional possible way for the Fermi accelerated electrons to cool down.
For GRB afterglows with a plateau phase, we use the concept of energy injection as a valid interpretation,
which affects the pre-break slopes. 
During the plateau phase, the cooling can be either fast (\nuc < \numm) or slow (\numm < \nuc), 
with the blast wave moving into a \wind\s or an ISM external density profiles.  The end of the plateau phase can be associated with the end of an ongoing energy injection phase, a jet break, or both.

\noindent We follow two main steps to analyse the afterglow data: 
\begin{enumerate}
\vspace{-0.2cm}
\item Spectral regime: Considering the movement of the break frequencies, we apply the closure relations \citep{2009racusin} to determine the external density profile, the spectral regime and the electron index $p$ of the distribution of the accelerated electrons. We test these for each possible option of the observing frequency (\tref{Table:grb100418A_sedfit_freq}) being in any of the synchrotron spectral regimes. 
\item Microphysical and dynamical parameters: We compare the fits of the full multi-wavelength data to
the standard formalism for a spherical blast wave propagating into an external cold medium during the fast  and slow cooling regime 
\citep{GS2002,DC2001,2012MNRAS.427.1329L}, and subsequently check for consistency with the slow/fast cooling transition times.  
The half-opening angle of the jet is derived using Eq.(4) from \cite{2005ApJ...618..413G}.
For the efficiency of the conversion of the kinetic energy in the outflow to gamma-rays during the prompt emission we use $\eta$=\eisog/(\eisog+\eiso), where \eisog\s is the isotropic energy released in the prompt gamma-ray emission calculated using \eisog=$4\pi d_{\rm{L}}^2\rm{F}/(1+z)$, with $F$ the fluence in the rest-frame energy range $1-10^{4}$ keV.
For the radial profile of the external density, we follow the canonical $k=0$ (constant, ISM-like) and $k=2$ (wind profile) description. For the $k=2$ case, we report the density in terms of $A_* = A/(5\times10^{11} {\rm g\ cm^{-1}})$ \citep{ChevalierLi2000}.
\end{enumerate}

\begin{table}[bht]
  \caption{Derived microphysical and dynamical parameters for the GRB 100418A afterglow from the five epochs of slow cooling.\label{Table:params_grb100418a} }
\vspace{-0.2cm}
\begin{tabular}{cccccc}
\toprule
SED & $\!\!$mid-time$\!\!$ & \bepse$_{-2}$ & \epsb$_{,-1}$ & \den & $\!\!$\eiso$_{,52}$$\!\!$  \\
     &   [ks] & & & & [erg] \\
\midrule
IV & 1065   &  $6.8^{+2.4}_{-1.4}$ & $1.51^{+0.27}_{-0.05}$ 			& $2.28^{+1.74}_{-0.68}$     & $2.3^{+0.1}_{-0.1}$ \\
V &  1555 	  &  $6.8^{+2.2}_{-1.2}$ & $1.40^{+0.36}_{-0.06}$ 			& $2.33^{+1.89}_{-0.66}$     & $1.6^{+0.1}_{-0.1}$ \\
VI  & 2246  &  $5.6^{+1.8}_{-1.0}$ & $1.30^{+0.29}_{-0.21}$ 			& $2.1^{+1.8}_{-0.6}$     & $1.4^{+0.1}_{-0.1}$ \\
VII & 3283  &  $5.2^{+1.5}_{-0.9}$ & $1.53^{+2.06}_{-0.03}$ 			& $2.30^{+1.19}_{-0.52}$     & $1.5^{+0.1}_{-0.1}$ \\
$\!$VIII & 5788 &  $5.6^{+1.7}_{-1.0}$ & $0.97^{+0.15}_{-0.18}$ 			& $1.57^{+1.89}_{-0.71}$     & $1.5^{+0.1}_{-0.1}$  \\
\bottomrule
\end{tabular}
\vspace{-0.1cm}
\tablefoot{\bepse=\epse * $(|p-2|)/(p-1)$. 
The subscripts denote $C_x=C\times10^{x}$.}
\end{table}

\begin{table*}[th]
\caption{Secondary parameters derived using the values of the GRB~100418A afterglow parameters reported in \tref{Table:params_grb100418a}. 
\label{Table:params_sec_grb100418a}}
\vspace{-0.5mm}
\begin{tabular}{ccccccc}
\toprule
SED & mid-time [ks] & \thh[rad] & $\eta$ & $B$ [G] & \mloss$_{,-5}$ & \ejet$_{,\rm{tot},50}$ [erg] \\
\midrule
IV & 1065   & $0.20^{+0.02}_{-0.03}$ & $0.02^{+0.01}_{-0.01}$ & $0.35^{+0.02}_{-0.01}$ & $2.28^{+1.74}_{-0.69}$ & $4.78^{+0.12}_{-0.08}$ \\
V &  1555   & $0.22^{+0.03}_{-0.02}$ & $0.06^{+0.01}_{-0.02}$ & $0.28^{+0.01}_{-0.01}$ & $2.34^{+1.89}_{-0.66}$ & $3.98^{+0.14}_{-0.07}$  \\
VI  & 2246  & $0.21^{+0.03}_{-0.03}$ & $0.07^{+0.01}_{-0.01}$ & $0.19^{+0.02}_{-0.01}$ & $2.12^{+1.76}_{-0.58}$ & $3.67^{+0.10}_{-0.04}$   \\
VII & 3283  & $0.22^{+0.04}_{-0.02}$ & $0.06^{+0.01}_{-0.01}$ & $0.15^{+0.01}_{-0.01}$ & $2.02^{+1.19}_{-0.52}$ & $3.66^{+0.09}_{-0.05}$   \\
VIII & 5788 & $0.23^{+0.02}_{-0.02}$ & $0.06^{+0.02}_{-0.01}$ & $0.09^{+0.01}_{-0.02}$ & $2.31^{+1.89}_{-0.71}$ & $3.92^{+0.13}_{-0.07}$   \\
\bottomrule
\end{tabular}
\vspace{-0.1cm}
\tablefoot{The subscript denotes $C_x=C\times10^{-x}$. \mloss\ for a wind velocity of 1000 km/s. \ejet=\eiso$\times$\thh$^2$/2. \ejet$,_{\gamma}$=\eisog$\times$\thh$^2$/2. \eisog=$9.9^{+6.3}_{-3.4}\times10^{50}$ erg. \ejet$,_{\rm{tot}}$=\ejet+\ejet$,_{\gamma}$ .}
\end{table*}

\subsubsection{GRB 100418A}
\label{sect:theory_grb100418a}

\paragraph{Closure relations}
Applying the closure relations to the X-ray/optical/NIR data and based on the measured post-break temporal slope, and because no spectral evolution is detected between the observations before and after the break in the X-ray or optical bands (which excludes the passage of a break frequency), three possible scenarios are in agreement with the observations:

\begin{enumerate}
\vspace{-0.22cm}
\item An afterglow with the plateau phase associated with an ongoing energy injection into the outflow; the implied injection parameter is \q=0.23\pmm0.05 (from the fit with linked pre-break slopes). 
This is followed by a normal decay phase associated to a radial outflow with no energy injection. The outflow is evolving into an ISM external medium and the optical and X-ray data is on the spectral segment between \numm\s and \nuc. 
\item A break in the light curve at the end of the plateau phase is associated with a uniform non-spreading jet. In this case the outflow is propagating into a \wind\s medium. The X-ray/optical/NIR observing frequencies are above \nuc\s and \numm. The cooling regime can be either fast or slow, since both have the same temporal and spectral indices. 
\item  
The end of the plateau phase is associated to the end of the prolonged energy injection and a uniform spreading jet. Within a 3$\sigma$ uncertainty error bars, the external medium is consistent with both ISM or \wind\s medium.
\end{enumerate}

\noindent 
Including the temporal evolution of the radio data is then used to determine which of the above three scenarios is preferred.
First, the SMA flux is constant or slowly decreasing, thus the optical and SMA bands have to be in different segments of the synchrotron spectrum. The SMA data would be consistent with an ISM or \wind\s external medium in the fast cooling regime for \nuc<$\nu_{\rm{SMA}}$<\numm, but also with a \wind\s external density profile in the slow cooling regime with the SMA wavelength between \nusa\s and \numm. This is only consistent for a \wind\s density profile with \nuc\s below the optical data and the end of the plateau phase being associated to both the end of the energy injection and a (non-spreading) jet break. The PdBI data at 106 GHz, 103 GHz and 86.7 GHz are consistent with this model where the fast cooling regime continues until $t\sim600$ ks, and the PdBI wavelengths lie between \nusa\s and \nuc. 
The ATCA and VLA radio data then fall in the slow cooling regime, and are consistent with the expected temporal and spectral slope. 
The favoured scenario is therefore a plateau phase due to energy injection, and the end of the plateau phase is associated to the end of the energy injection together with a uniform non-spreading jet break expanding into a \wind density profile (above scenario 2). We recall that GROND and XRT data are above \nuc\s and \numm\s and therefore the electron index is $p$=2.22\pmm0.04.

\begin{figure}[ht]
\hspace{0.1cm}\includegraphics[width=0.47\textwidth]{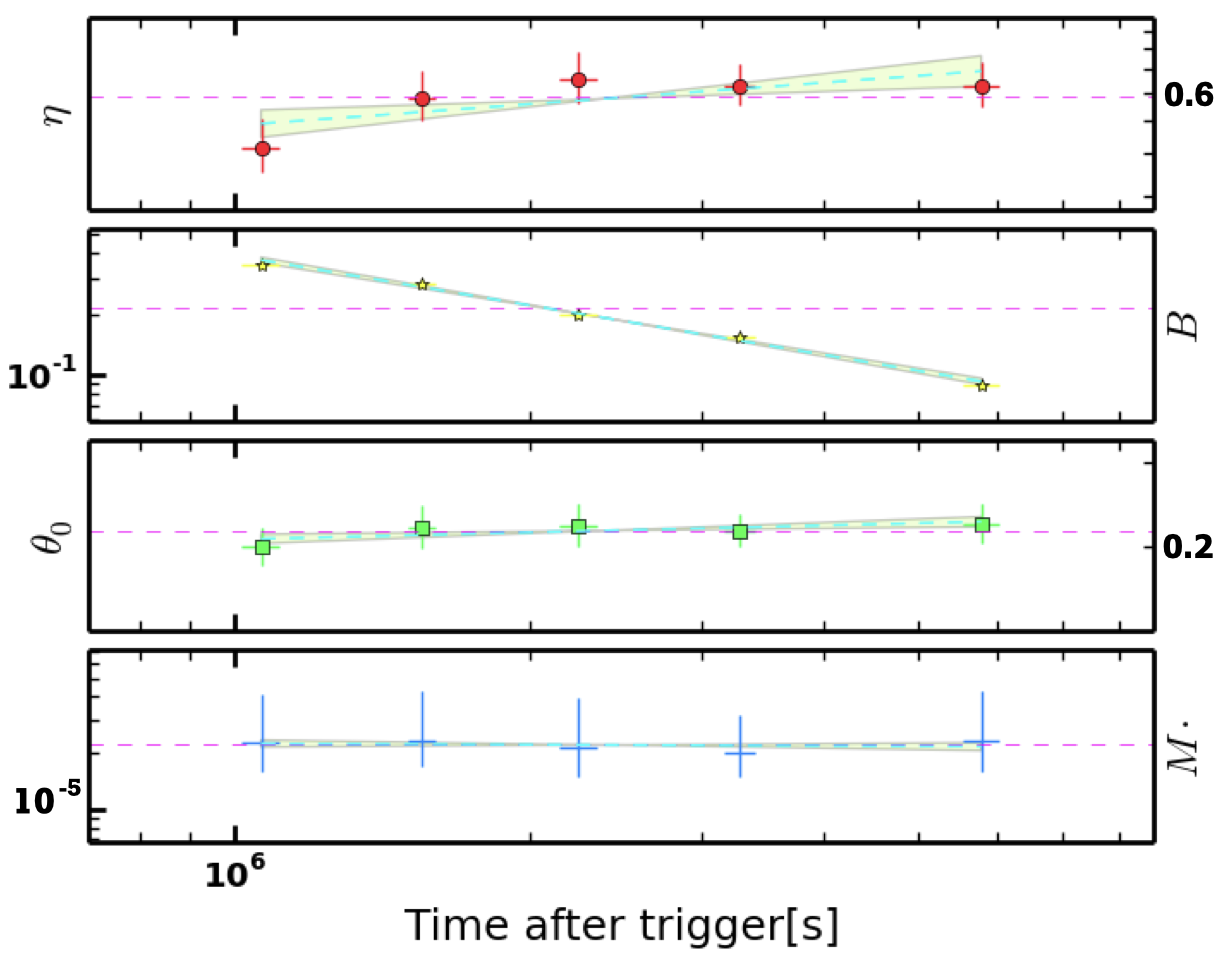}
\includegraphics[width=0.49\textwidth]{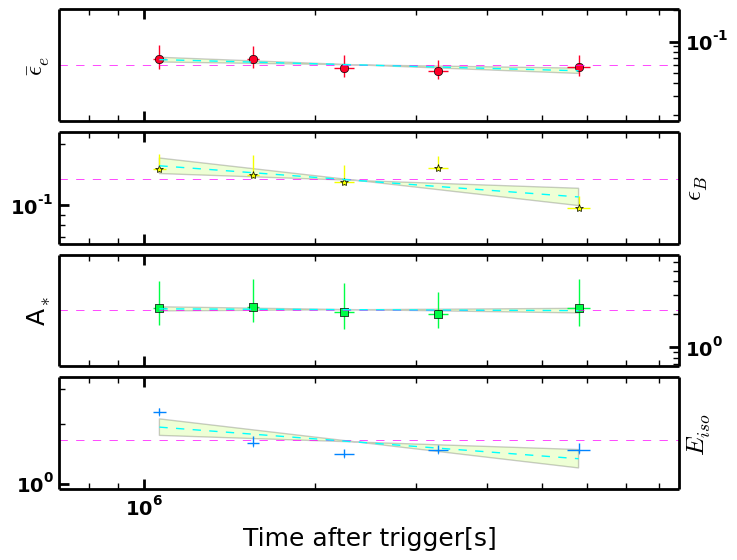}
\vspace{-0.3cm}
\caption{Slow cooling: Evolution of the derived  dynamical (top) and microphysical (bottom) parameters of the afterglow of GRB 100418A  without SSC emission. The blue dashed lines and shaded regions represent the results from the fit of the observed temporal evolution. The horizontal dashed purple lines show the average value for each parameter. \eiso\s is in units of $10^{52}$ erg.
\label{Fig:params_evol_grb100418A_slow}}
\end{figure}

\paragraph{Afterglow parameters} 
The analysis of the SEDs during the early fast cooling regime and the (later) slow cooling regime was done separately as the dependencies of the break frequencies on the parameters are different. 
All the results are reported in \tref{Table:params_grb100418a} and \tref{Table:params_sec_grb100418a}.

\tbf{Fast cooling:}  
During the early fast cooling phase (SED1-SED3) inverse Compton scattering is expected to play an important role in the cooling of the electrons.  
We test the strength of the inverse Compton scattering, but find \epsb\s $\gg1$, leaving the model with no physical meaning. 

We double-checked the fitting with the set of equations given by \cite{Granot+2000}, but confirm that no consistent solution is obtained. As mentioned in \cite{GS2002}, the emission during fast cooling in a wind environment is no longer dominated by the shock front but tends to come from smaller radii, and thus the break frequencies and corresponding flux densities are different by up to a factor of 3 between \cite{Granot+2000} and \cite{GS2002}. We therefore cannot deduce physical parameters from the 3 first epochs of evolution of the GRB 100418A afterglow.

\textbf{Slow cooling:} The last five epochs of the afterglow observations, i.e., SED4-SED8 are used in this case. The dynamical and microphysical parameters are in complete agreement with the theory. The inverse Compton scattering contribution was tested,
because the low value of \nuc\ and the large A$_*$ are indicative of SSC. However, we find no dominant contribution. 

All the values with and without SSC cooling are within the expected values from the theory. The average value for \epse\s is about 0.36 and for \eiso\s is about $2\times10^{52}$ erg. The relation \epse$/$\epsb\s is $<10$ which is in agreement with the SSC contribution being negligible during the slow cooling phase, and therefore is no longer included for the rest of the discussion. The value of \den\s (no SSC included) is of order unity,
and \epsb\s is about 0.1, which could require a larger value of $B$ in the shocked region. 
However, the temporal evolution of B (\al=-0.81\pmm0.05), as seen in \fref{Fig:params_evol_grb100418A_slow}, is as expected (\al=-3/4) of a magnetic field generated by shock compression of the seed magnetic field in the circumburst medium (CBM). Therefore, the difference in the expected values might just be related to the actual magnitude of $B_0$. If $B_0$ is of the order of a few mG, the value derived for \epsb\s is reproduced by theory.

\paragraph{Overall picture}

\begin{itemize}
\vspace{-0.15cm}
\item Cooling regime: fast cooling with \numm $>$\nuc\s up to $\sim$600 ks, and (rather late) transition to slow cooling with \nuc $>$\numm
\item the CBM profile corresponds to a stellar wind environment
\item Break frequency evolution: The shaded regions in \fref{Fig:frec_evol_grb100418A} for each frequency show the actual results of the fit of the temporal evolution. \nuc\s evolves with a slope of 0.57\pmm0.04, \numm\s has a slope -1.72\pmm0.06 and \nusa\s evolves with slope -0.56\pmm0.06. \nuc\s and \nusa\s are within $1\sigma$ uncertainty of the expected evolution of -0.5 and 0.6 respectively. \numm\s is $3.6\sigma$ away from the expected 1.5 value.
\item the plateau is compatible with continuous energy ejection
\item the observed break is due to both, the ceasing of the energy injection and the jet break of a uniform non-spreading jet with a collimation angle of about 0.22 rad
\item Energetics: \eiso\s has an average value $2\times10^{52}$ erg, and when compared with \eisog\s for this GRB, the required efficiency is about 6\%.
\item Emission mechanism: synchrotron with some SSC contribution possible
\end{itemize}

\paragraph{Comparison with literature:}

Our fitting results are consistent with those of \cite{marshall2011}, but they interpret the light curve break after one day as a consequence of the ceasing energy injection instead of a jet break, deduce a large beaming angle and very high beaming-corrected energy.

\cite{Laskar+2015} have analysed Swift/UVOT data and selected optical data published in GCNs. They provide solutions for 100418A for both, ISM and wind profile, and prefer the ISM solution because of the implied lower energy put during the energy injection phase. However, we note that this choice hinges on one single $R_c$-band data point, which suggests a very rapid optical rise during the first few hours. While our GROND data do not cover this early phase, the flat portion of the GROND light curve as well as the Swift/UVOT data suggest a much more modest flux increase. 
The parameters of our best-fit model are fully consistent with their wind solution.

\subsubsection{GRB 110715A}
\label{sect:theory_grb110715a}

\paragraph{Closure relations:}
\label{sect:closure_grb110715a}
The analysis of the first segment of the XRT observations ($t < 21.4$ ks) is in agreement with \numm\s < \nux\s < \nuc\s with an energy injection component in an ISM or \wind density profile, or in an ISM density profile without the energy injection component. However, the observations during this time interval might be altered by the contribution from SSC and therefore the closure relations could be modified, \ie \al\s is steeper when SSC is dominant. If SSC is included only \nuc\s < \nux\s is in agreement with the observations for either an ISM or a \wind density profile. A strong reason to have a SSC contribution is that it could explain the first break in the light curve, otherwise, the plateau phase would require a central engine that can "restart" itself after $10^4$ s. 
Therefore, the break would just be associated with the end of a dominant inverse Compton phase and the energy injection phase would just continue until the second break.

The plateau phase is in agreement with two scenarios: an energy injection phase where \nux\s < \nuc\ and q = -0.25\pmm0.10 for an ISM external medium and p = 3.10\pmm0.02, or alternatively \nux\s > \nuc\s with q = -0.36\pmm0.15 for p = 2.10\pmm0.02 in either a \wind or an ISM density profile. The second break in the X-ray light curve is associated with the end of the energy injection phase and/or with a jet break. Observations at the optical and X-ray frequencies have the same temporal slope during this last time interval, which is in agreement with both.

Three cases fit the data during this last time interval: first, for \nux\s < \nuc\s with the break associated only to the end of the energy injection phase in an ISM density profile. Second, where \nuc\s < \nux\s and the break is associated with both the end of an energy injection phase and a uniform non-spreading jet break in a \wind density profile. Third, \nuc\s < \nux\s with the break in the light curve associated uniquely to a non-spreading jet break but with an ongoing energy injection phase.

\begin{table}[ht]
\setlength\tabcolsep{5.0pt}
\caption{Derived microphysical and dynamical parameters for the GRB 110715A afterglow (see Tab. \ref{Table:params_grb100418a}).
  \label{Table:params_grb110715a}}
\vspace{-0.2cm}
\begin{tabular}{cccccc}
\toprule
SED & mid-time & \bepse$_{,-2}$& \epsb$_{,-3}$ & \den$_{,+1}$ & \eiso$_{,53}$ \\
     & [ks]   & & & & [erg] \\
\midrule
I  & 122.7    & $6.68^{+0.39}_{-0.27}$ & $1.53^{+0.11}_{-0.02}$ & $1.30^{+0.72}_{-0.31}$ & $1.22^{+0.16}_{-0.14}$ \\
II  & 173.2   & $7.26^{+0.36}_{-0.35}$ & $1.89^{+0.02}_{-0.01}$ & $0.98^{+0.51}_{-0.26}$ & $1.22^{+0.21}_{-0.12}$ \\
III & 254.5   & $7.97^{+0.43}_{-0.26}$ & $1.53^{+0.02}_{-0.02}$ & $1.10^{+0.57}_{-0.26}$ & $1.12^{+0.32}_{-0.23}$ \\
IV & 344.9   & $7.62^{+0.12}_{-0.10}$ & $1.62^{+0.02}_{-0.01}$ & $1.07^{+0.10}_{-0.07}$ & $1.36^{+0.21}_{-0.24}$ \\
V  & 1014.2 & $7.73^{+0.25}_{-0.19}$ & $1.81^{+0.01}_{-0.02}$ & $1.08^{+0.41}_{-0.18}$ & $1.34^{+0.38}_{-0.23}$ \\
VI & 1513.8 & $8.60^{+0.24}_{-0.26}$ & $1.74^{+0.01}_{-0.01}$ & $1.12^{+0.57}_{-0.29}$ & $1.17^{+0.27}_{-0.30}$ \\
\bottomrule
\end{tabular}
\end{table}

Including the sub-mm and radio data, we first recall that below \numm\s the flux at sub-mm and radio wavelengths would evolve with the same slope in an ISM density profile. 
Since this is not observed, any scenario where the CBM is homogenous can be discarded. The evolution of the frequencies as shown in Fig. \ref{Fig:GRB110715A_freq_evol}
(\nusa: \al = -0.72\pmm0.10; \nuc: \al = 0.56\pmm0.10) 
supports slow cooling and a \wind density profile. Therefore, the plateau phase can only be explained by a \wind density profile when q = -0.36\pmm0.15 and \nuc\s < \nux\s. As no SED evolution is detected in the XRT or optical/NIR bands, this implies that the observations during the pre- and post- plateau phase must be in the same spectral regime. Therefore, the pre-plateau phase is best explained by an IC contribution in a spectral regime where \nuc\s < \nux\s.
Finally, the post-plateau phase observations can be described by a spectral regime in \nuc\s < \nux\s (in a \wind CBM) and the break associated with a uniform non-spreading jet with or without the end of the energy injection phase.

In this last scenario, however, 
having a source that provides such long energy injection and without a sign of it at least in the sub-mm region is non-standard. Therefore, the only credible scenario is the association of the break in the light curve after the plateau phase with the end of the energy injection phase together with a uniform non-spreading jet 
in a \wind density profile for \nuc\s < \nux. A change in the flux evolution due to the non-spreading jet break would then take place compared to the normal evolution. The change is (k-3)/(4-k), \ie -3/4 for an ISM density profile and -1/2 for a \wind density profile. 
We therefore propose the stratification of shells as the favourable scenario for the energy injection phase with a strong contribution from IC during the early epochs.

\begin{table*}[th]
\caption{Energy efficiency, magnetic field magnitude, mass loss rate, and opening angle for the GRB 110715A afterglow. 
  \label{Table:params_sec_grb110715a}}
\vspace{-0.2cm}
\begin{tabular}{ccccccc}
\toprule
SED & mid-time [ks] & \thh[rad] & $\eta$ & $B$ [G] & \mloss$_{,-4}$ & \ejet$_{,51}$ [erg] \\
\midrule
I  & 122.7    & $0.18^{+0.02}_{-0.02}$ & $0.19^{+0.02}_{-0.02}$ & $0.48^{+0.03}_{-0.02}$ & $1.30^{+1.08}_{-0.47}$ & $2.45^{+0.17}_{-0.13}$ \\
II  & 173.2   & $0.17^{+0.01}_{-0.01}$ & $0.19^{+0.02}_{-0.01}$ & $0.33^{+0.01}_{-0.01}$ & $0.98^{+0.76}_{-0.39}$ & $2.12^{+0.20}_{-0.15}$ \\
III & 254.5   & $0.18^{+0.02}_{-0.01}$ & $0.21^{+0.02}_{-0.02}$ & $0.25^{+0.01}_{-0.01}$ & $1.10^{+0.86}_{-0.39}$ & $2.19^{+0.31}_{-0.22}$ \\
IV & 344.9   & $0.17^{+0.01}_{-0.01}$ & $0.18^{+0.01}_{-0.01}$ & $0.19^{+0.01}_{-0.01}$ & $1.07^{+0.15}_{-0.01}$ & $2.30^{+0.21}_{-0.26}$  \\
V  & 1014.2 & $0.17^{+0.02}_{-0.01}$ & $0.18^{+0.02}_{-0.02}$ & $0.09^{+0.01}_{-0.01}$ & $1.08^{+0.60}_{-0.27}$ & $2.29^{+0.30}_{-0.28}$ \\
VI & 1513.8 & $0.18^{+0.01}_{-0.01}$ & $0.20^{+0.01}_{-0.01}$ & $0.07^{+0.01}_{-0.01}$ & $1.13^{+0.87}_{-0.44}$ & $2.25^{+0.22}_{-0.33}$ \\
\bottomrule
\end{tabular}
\vspace{-0.1cm}
\tablefoot{The subscript denotes $C_x=C\times10^{-x}$. \mloss\s for a wind velocity of 1000 km. \ejet=\eiso$\times$\thh$^2$/2.
\eisog=$2.93^{+5.79}_{-2.81}\times10^{52}$ erg.}
\end{table*}

The radio and sub-mm observations have some discrepancies from the theoretical results. The flux from observations at 9.0 GHz and 5.5 GHz have an evolution with temporal slopes \al = -0.08\pmm0.11 and \al = -0.09\pmm0.07, respectively. In the case of a \wind density profile with a non-spreading jet break, the expected slope is \al = -1/2, which is 3.5$\sigma$ and 5.3$\sigma$ away from the observed \al\s at 9.0 GHz and 5.5 GHz, respectively. This could be associated to a strong interstellar scintillation effect, which is stronger at lower radio frequencies. Observations at 18 GHz are expected to have an initial slope of \al = -1/2 and then a decreasing flux with \al = 1/2. The observations are consistent with this within 2$\sigma$ uncertainty.

Finally, for observations at 44 GHz and 345 GHz, the flux is expected to decrease with \al=1/2. Observations at 345 GHz, and after the second epoch at 44 GHz show a decrease in flux with an \al\s of about 0.91\pmm0.12, although it is 3.4$\sigma$ away from the expected value, the difference might just be due to the low statistics in the sample. It is, however, not clear why the first two epochs, at a frequency of 44 GHz does not follow the expected values and are rapidly increasing with a slope of about -2. There is clearly an external effect that must be affecting the observations during these epochs, specially the first observation. If the flux at the first epoch is larger, then the rate of the increase in the flux would be slower, and it could be in agreement with the -0.5 if $\nu$<\nusa. This is a possible scenario where \nusa\s just crosses $\nu$ at 44 GHz as seen in the following section.

\paragraph{Afterglow parameters:}
\label{sect:micro_grb110715a}

From the analysis so far, we conclude that the best scenario describing the (late) observations is a uniform non-spreading jet expanding into a \wind density profile. The afterglow evolution went through an energy injection phase before the jet break. The power-law index $p$ of the non-thermal electron population is p = 2.10\pmm0.02. The cooling break is located below the NIR wavelengths, thus no spectral evolution in the optical/NIR or X-ray bands is observed. 

The measured break frequencies are used to derive the dynamical and microphysical parameters. The effect of energy injection is not taken into account, as it finishes before the start of the six analysed epochs. 
The results for all the parameters are reported in Tables \ref{Table:params_grb110715a}.

\begin{figure}[th]
\includegraphics[width=0.49\textwidth,height=8.5cm]{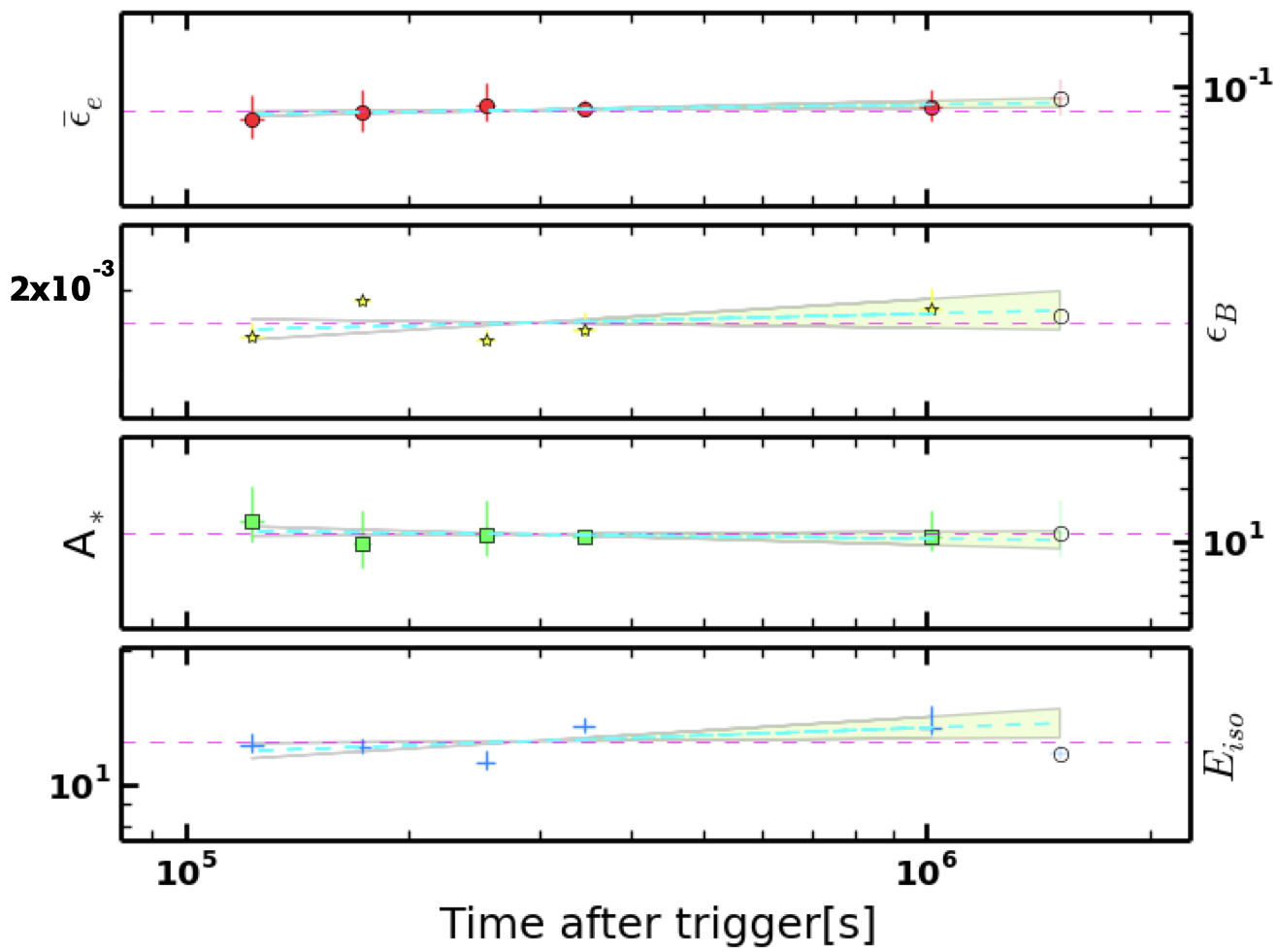}
\caption{Evolution of the derived microphysical and dynamical parameters of the afterglow of GRB 110715A, for the case without IC emission. The dashed-lines in cyan and shaded regions represent results from the fit of the observed temporal evolution (open data points were left out from the fit). The horizontal pink dashed-line shows the average value for each parameter. \eiso\s is in units of $10^{52}$ erg. 
\label{Fig:params_evol_grb110715A_all}}
\end{figure}

The results for the parameters are presented in \fref{Fig:params_evol_grb110715A_all}.
The top panel of \fref{Fig:params_evol_grb110715A_all} shows that all four parameters \epse, \epsb, \eiso\s and \den\s are constant in time, with measured formal slopes of 0.06\pmm0.04, 0.04\pmm0.06, -0.05\pmm0.07 and 0.06\pmm0.05, respectively, consistent with the standard model. The pink dashed line shows the average value for each parameter.

The ratio \epse/\epsb\s $>$590 implies that SSC could be important.  
However, including SSC in the derivation of the parameters, we obtain an unphysical \epsb\s $\approx$ 10.
We therefore assume that SSC was not relevant during the six analysed epochs presented here. 

The measured microphysical and dynamical parameters were used to derive the half-opening angle \thh\s of the collimated outflow, the magnitude of the magnetic field $B$ ($B = (32 \pi m_p c^2 n)^{1/2} \gamma$ \epsb$^{1/2}$; \citealt{1996ApJ...473..204S} 
) in the shock region, and the efficiency of conversion of the kinetic energy $\eta$.
Interestingly, $B$ is evolving with \al=-0.77\pmm0.04, 
consistent with theoretical evolution  of $-0.75$ (see sect. \ref{sect:microphys_par}). 
The values for the efficiency are of the order of 19\% which is just within the expected range of values 10\%--20\% \citep{Mochkovitch+1995, Kobayashi+1997}.  
The collimation angle \thh\s is about 0.17 rad, which implies a total energy in the jet after the beaming correction, of \ejet = 2.27$\times10^{51}$ erg.

\paragraph{Overall picture}

\begin{itemize}
\vspace{-0.15cm}
\item Cooling regime: slow cooling throughout
\item the CMB profile is a \wind density profile
\item Break frequency evolution: shown in \fref{Fig:GRB110715A_freq_evol} and consistent with
evolution in wind environment
\item the plateau seen in X-rays and optical is best explained with an energy injection phase with $q$ = -0.36\pmm0.15;
the preferred model for the energy injection component is a stratification of the mass shells.
\item the break at the end of the plateau phase coincides with a jet break of a uniform, non-spreading jet.
\item energetics: total energy in the outflow after the beaming correction is \ejet = 2.27$\times$10$^{51}$ erg
\item Emission mechanism: pure synchrotron
\end{itemize}

\paragraph{Comparison with literature:}
\cite{2017MNRAS.464.4624S} prefer an interpretation with $p=1.8$ and a wind termination shock with no energy injection, which (as explicitly mentioned) does not fit the flat part at 0.3 days. We explain this feature with energy injection.

\subsubsection{GRB 121024A}

GRB 121024A has been already analysed in a similar way \citep{Varela2016A&A...589A..37V}, and for the discussion in the following section, we just summarize the overall picture here.

\paragraph{Overall picture}

\begin{itemize}
\vspace{-0.15cm}
\item Cooling regime: slow cooling throughout
\item the CMB profile is a \wind density profile
\item hard electron spectral index $p=1.73\pm0.03$
\item jet break at 49.8$\pm$5.1 ks 
\item energetics: total energy in the outflow after the beaming correction is \ejet = 0.4$\times$10$^{51}$ erg
\item Emission mechanism: pure synchrotron
\end{itemize}

\subsubsection{GRB 130418A}
\label{sect:theory_grb130418a}

\paragraph{Closure relations:}
\label{sect:closure_grb130418a}

The evolution of the break frequencies (\fref{Fig:frec_evol_grb130418A}) allows us to unambiguously identify their physical origin (see Tab. \ref{Table:grb130418A_sedfit_freqs}).
In addition, we note that
(i) the spectral slope in the X-ray band \bx\s is flatter than the spectral slope in the optical/NIR bands \bo\s, implying SSC dominance;
(ii) the early decay is very slow, suggesting substantial energy injection;
(iii) the rising/falling sub-mm light curve indicates that \nusa\s is initially above 345 GHz and then moves towards lower values.

The combination of the temporal and SED information by means of the closure relations, and taking energy injection into account, lead to the following two options:
(i) The optical data  are consistent with either \nuc<\nuo\s and ISM or \wind density profile (and injection parameter q = 0.14\pmm0.10); or \nuo<\nuc\s and ISM (q = 0.09\pmm0.08);
(ii) The X-ray data suggest either \nux<\nuc\s and wind (q = 0.88\pmm0.16),  
or \nux<\nuc\s for either \wind or  ISM. 
Since \nuo\s should lie in the same segment as \nux\s in order to have a reasonable electron index $p$,
the above option 1) is preferred, i.e. the cooling frequency \nuc\s lies below the NIR band during all three epochs (see \tref{Table:grb130418A_sedfit_freqs} for the full parameters, and  \fref{Fig:GRB130418A_radio_SED}). Consequently, the electron index is $p$ = 2.32\pmm0.14 as derived from the SED above \nuc. The evolution of the break frequencies \nuc\s and \numm\s (\fref{Fig:frec_evol_grb130418A}) follows temporal slopes of \al=0.61\pmm0.03 (just marginally consistent with the expected 0.5), and  \al=-1.45\pmm0.06 (in perfect agreement with the theoretical value of \al=-1.5), respectively.
The injection parameter $q$ = 0.14\pmm0.10 is in agreement with both,
stratified mass shells, with parameter $s\sim4.2$ in a \wind density profile, and a magnetar model with an emission dominated by a Poynting flux that requires $q\sim0$.

\begin{table}[hb]
\caption{Derived microphysical and dynamical parameters for the GRB 130418A afterglow
(see Tab. \ref{Table:params_grb100418a}).
}
\label{Table:params_grb130418a}
\vspace{-0.2cm}
\begin{tabular}{cccccc}
\toprule
SED & mid-time  & \bepse$_{,-1}$& \epsb$_{,-5}$ & \den$_{,+1}$ & \eiso$_{,51}$  \\
    & [ks] & & & & [erg] \\
\midrule
I   & 288.1  & $0.96^{+0.07}_{-0.08}$  &  $7.67^{+1.25}_{-0.22}$ &  $4.5^{+2.1}_{-1.4}$ &  $7.4^{+0.3}_{-0.4}$ \\
II  & 415.7  & $1.07^{+0.21}_{-0.16}$  &  $6.5^{+1.9}_{-0.1}$ &  $5.3^{+1.7}_{-0.9}$ &  $7.8^{+0.3}_{-0.2}$ \\
III & 106.8  & $>$0.86          &  $<$8.40         &  $>$3.90        &  $<$7.8  \\
\bottomrule
\end{tabular}
\end{table}

\begin{table*}[ht]
\caption{Energy efficiency, magnetic field magnitude, mass loss rate, and opening angle for the GRB 130418A afterglow.  
\label{Table:params_sec_grb130418a}}
\vspace{-0.2cm}
\begin{tabular}{ccccccc}
\toprule
SED & mid-time [ks] & \thh$_{-1}$ [rad] & $\eta$ & $B$ & \mloss$_{,-4}$ & \ejet$_{,51}$ [erg] \\
\midrule
I   & 288.1  & $4.52^{+0.49}_{-0.43}$ & $0.35^{+0.05}_{-0.04}$ & $1.68^{+0.11}_{-0.09}$ & $4.48^{+2.81}_{-1.57}$ & $1.15^{+0.38}_{-0.41}$ \\
II  & 415.7  & $4.67^{+0.44}_{-0.38}$ & $0.33^{+0.03}_{-0.05}$ & $1.31^{+0.09}_{-0.07}$ & $5.34^{+2.47}_{-1.15}$ & $1.27^{+0.39}_{-0.23}$ \\
III & 106.8  & $>$4.27   	     & $>$0.33   & $>$0.69    & $>$3.90        & $<$1.09  \\
\bottomrule
\end{tabular}
\vspace{-0.1cm}
\tablefoot{The subscript of each quantity are $C_x = C \times 10^{-x}$. Mass loss rate for a wind velocity of 1000 km. \ejet=\eiso$\times$\thh$^2$/2.}
\end{table*}

IC lowers the initial value of \nuc\s by a factor of (1+Y)$^{-2}$ and changes the observed flux evolution to \al=1.39 when it is the dominant cooling effect. It also flattens the spectral slope above \nuc\s with an expected \be=1/3, which is in complete agreement with the observations. The first break in the optical light curve is therefore associated to the end of an energy injection phase. The second break is an achromatic break consistent with a uniform non-spreading jet. The sub-mm and radio data confirm that \nuc<\nuo\s and the evolution of the jet is in a \wind density profile. 
The density normalisation \den\s (47) is quite possible for a Wolf-Rayet star,
and in perfect agreement with the  \den\s $>$10 requirement for IC scattering being 
the dominant cooling process \citep{sari...esin2001}.

\begin{figure}[ht]
\includegraphics[width=0.49\textwidth]{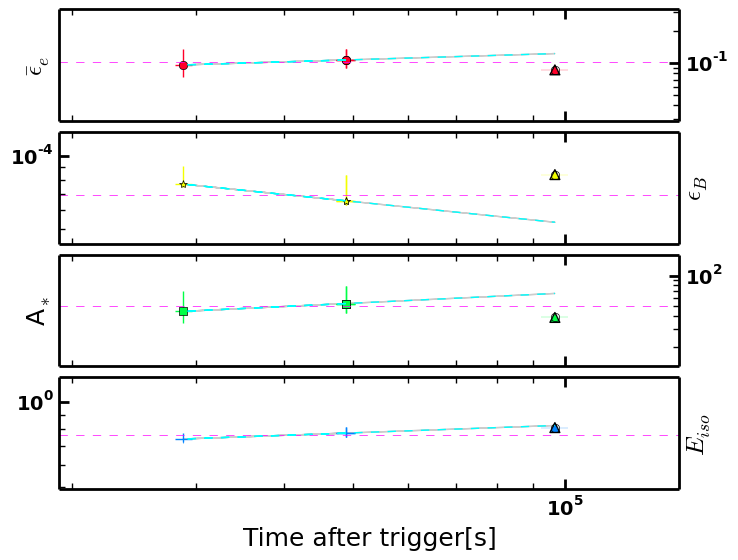}
\caption{Evolution of the derived microphysical and dynamical parameters of the afterglow of GRB 130418A. without IC.
The dashed lines in cyan
represent the results from the fit of the observed temporal evolution, although this is only the connection of two points, since the last epoch only has limits. When considering the uncertainties, all parameters are consistent with having no evolution with time. The horizontal dashed purple lines shows the average value for each parameter. \eiso\s is in units of $10^{52}$ erg.}
\end{figure}

\begin{figure}[ht]
\includegraphics[width=0.48\textwidth]{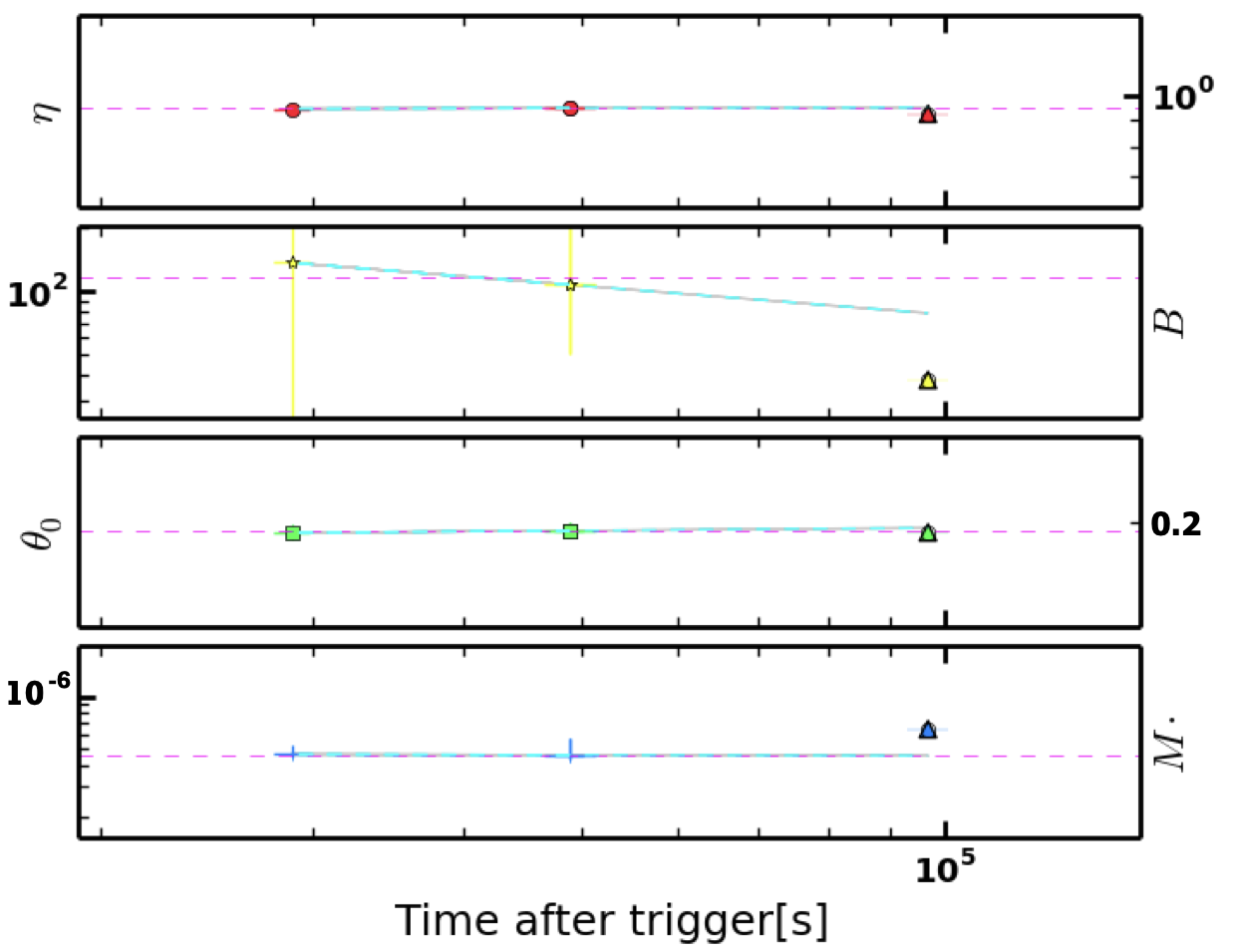}
\includegraphics[width=0.48\textwidth]{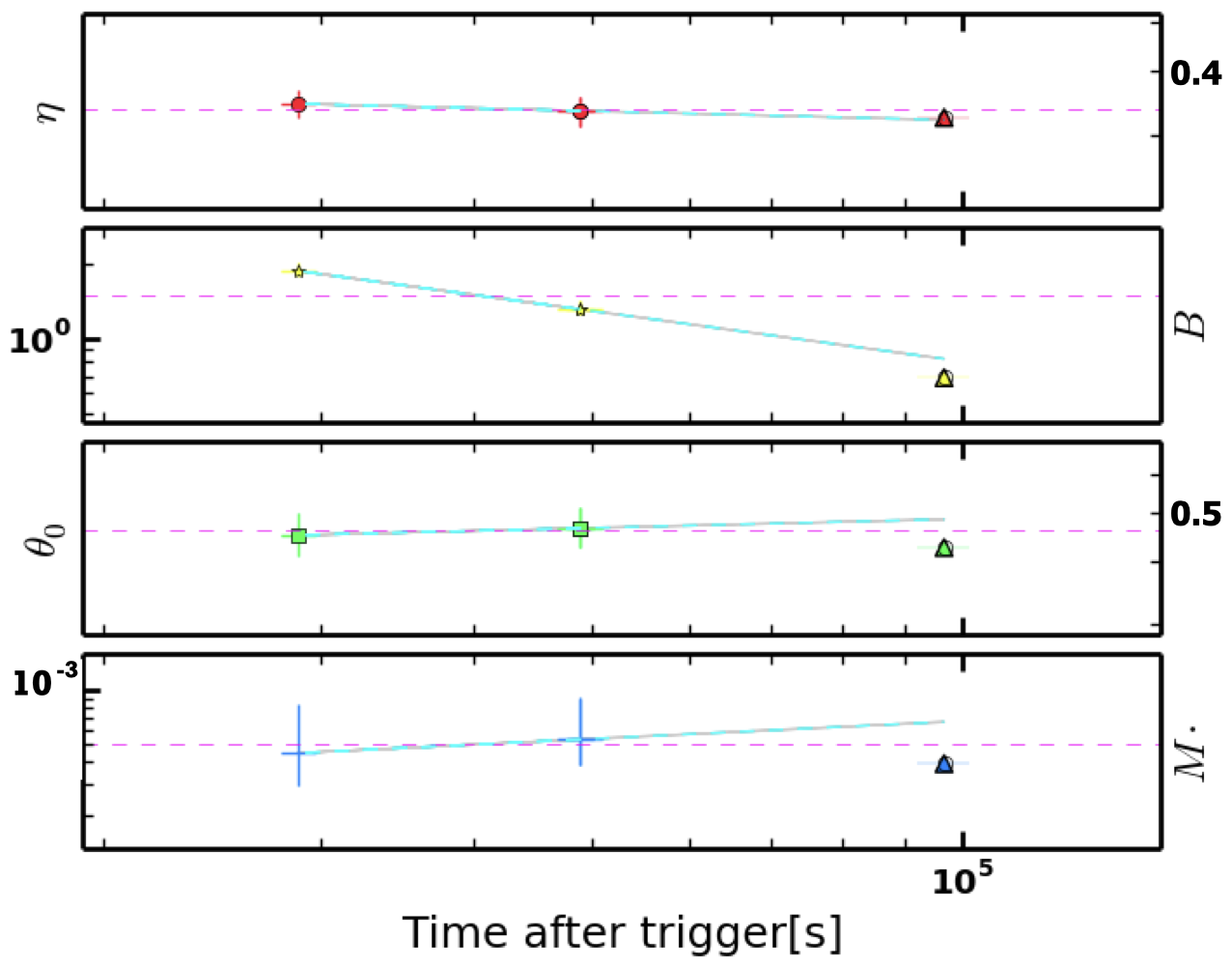}\\
\caption{Evolution of the energy efficiency $\eta$, magnetic field $B$, mass loss rate $\dot{M}$, opening angle $\theta_0$ derived from the measured microphysical and dynamical parameters of the afterglow of GRB 130418A with (top) and without (bottom) IC included. The dashed lines  represent the results from the fit of the observed temporal evolution. Dashed pink lines show the average of each parameter.}
\label{Fig:params_evol_sec_grb130418A}
\end{figure}

\paragraph{Afterglow parameters:}
\label{sect:micro_grb130418a}
Using the measured break frequencies (\tref{Table:grb130418A_sedfit_freqs})
we derived the microphysical and dynamical parameters, \ie \epsb, \epse, \eiso, \den\ (\tref{Table:params_grb130418a}). The energy injection phase ended by the time of the first break in the light curve at \tba=18.8\pmm3.5 ks and the break of the non-spreading jet is at \tbb=61.7\pmm8.1 ks. Therefore, the effect of the energy injection is not included in the derivation of the parameters, but the effect of the geometrical jet is included as a renormalisation of the peak flux to account for the difference with the expected spherical flux. 
When the derivation of the microphysical and dynamical parameters is performed with the SSC effect included, \epsb\s is of order $10^3$, which physically is not possible.
Without SSC, \epse\s and \epsb\s are less than unity, consistent with theory. The value for \epsb\s implies a large seed magnetic field in the CBM of order of mG (see sect. \ref{sect:microphys_par}). However, as expected by theory and needed by the early time observations, SSC was a dominant effect during the first stages of the afterglow evolution, with clearly
observed signatures in GRB 130418A: \nusa\s has larger values than usually expected, i.e., above sub-mm frequencies rather than being closer to radio frequencies, and the cooling break \nuc\ has lower values than commonly observed.

\paragraph{Overall picture}

\begin{itemize}
\vspace{-0.15cm}
\item Cooling regime: slow cooling throughout
\item the CMB profile is a \wind density profile
\item Break frequency evolution: shown in \fref{Fig:frec_evol_grb130418A} and consistent with evolution in wind environment
\item mild energy injection leading to modest decline rather than a plateau
\item the second break is consistent with a jet break, leading to \thh $\approx$ 0.45 rad
\item energetics: the measured isotropic energy \eiso\s  is 7.70$\times10^{51}$ erg, and the total energy in the outflow after the beaming correction is \ejet = 1.17$\times$10$^{51}$ erg.
\item Emission mechanism: synchrotron with clear SSC signature in early afterglow phase
\end{itemize}

\paragraph{Comparison with literature}
No results are published in refereed journals.

\begin{table*}[ht]
\caption{Main features for the four studied GRB afterglows.
  \label{Table:summarygrbs}}
\vspace*{-0.22cm}
\begin{tabular}{c|l|l|l|l|l}
\toprule
GRB  & \multicolumn{1}{c|}{X-ray} & \multicolumn{1}{c|}{GROND} &\multicolumn{1}{c|}{ Sub-mm} & \multicolumn{1}{c|}{Radio} & \multicolumn{1}{c}{Notes}\\
\midrule
\multirow{4}{*}{100418A} & \ Steep decay & \ 8 epochs & \ SMA: 3 det. & \ VLA: 7 det.&  \ Non-spreading JB \\
 & \ Plateau phase & \ Plateau phase  & \ PdBI: 7 det. & \ ATCA: 6 det. & \ Fast/slow cooling \\
 & \ EI, JB & \ No SED evol. & \ Evol. of $\alpha$& \ ISS & \ Wind\\
 & \ No SED evol. & \ EI, JB & &  & \\
\midrule
\multirow{3}{*}{110715A} & \ 2 breaks & \ 6 epochs & \ APEX: 1 det.  & \ ATCA: 22 det, 1 UL &\ EI \\
 & \ Plateau phase & & \  ALMA: 1 det. & \ ISS &\ Non-spreading JB \\
 & \ EI & &  & &\ Wind \\
 & & &  & & \ Early IC \\
\midrule
\multirow{6}{*}{121024A} & \ Plateau phase. & \ 6 epochs.  & \ APEX: 2 UL & \ EVLA: 1 det & \ $1 < p <2$ \\
& \ EI & \ Plateau phase. & & \ CARMA: 1 det & \ EI before $t_b$, p>2 \\
& \ Achromatic break.  & \ EI & & & \ Jet break, $1<p<2$ \\  
& \ No SED evol. & \ Achromatic break & & & \ EI after \tb, $p>2$ \\
& \  JB & \ No SED evol. & & & \ Polarisation det.\\
&   & \ Jet break &  & & \ Wind\\
\midrule
\multirow{5}{*}{130418A} & \ $< 10^4$ ks &\ 7 epoch & \ SMA: 1 UL. & \ CARMA: 1 det.  & \ No SED evol.\\
 & \ Jet break & \ 2 breaks, jet & \ APEX: 2 det, 1 UL. & \ WSRT: 1 UL & \ Early IC \\
 & \ \al\s too steep for EI & \ EI & &  & \ EI \\
  & \ \be\s too flat vs \be\s opt. & \ Achromatic break & &  & \ Jet break \\
   & &  \ JB & &  & \ Wind\\
\bottomrule
\end{tabular}
\vspace{-0.1cm}
\tablefoot{The number of epochs corresponds to the number of time slices used in the SED analysis. 
EI: Energy injection. ISS: Interstellar scintillation. JB: Jet break. UL: Upper limit. Det: detection.}
\end{table*}

\begin{table*}[th]
\setlength\tabcolsep{5.5pt}
\caption{Parameter summary of the four GRBs.
\label{Table:params_basic_grbs_1}}
\vspace{-0.15cm}
\begin{tabular}{c|c|c|c|c|c|c|c|c}
\toprule
GRB & \dustg\s (mag) \tablefootmark{a} & \dusth\s (mag) \tablefootmark{b} & \gasg$_{,22}$\s (cm$^{-2}$) \tablefootmark{a} & \gash$_{,22}$\s (cm$^{-2}$) \tablefootmark{b} & z & $\rho$ & \eisog$_{,52}$\s (erg) & \ejet$_{,51}$\s (erg) \\
\midrule
GRB 100418A & 0.22 & 0.01$^{+0.03}_{-0.01}$ & 0.06 & 0.57$^{+0.09}_{-0.08}$ & 0.625 & wind & $0.10^{+0.06}_{-0.03}$ & $0.40^{+0.11}_{-0.08}$ \\
GRB 110715A & 1.82 & 0.05$^{+0.01}_{-0.01}$ & 0.43 & 0.16$^{+0.03}_{-0.04}$ & 0.820 & wind & $2.93^{+5.79}_{-2.81}$ & $2.27^{+0.26}_{-0.24}$ \\
GRB 121024A & 0.27 & 0.18$^{+0.04}_{-0.04}$ & 0.08 & 0.30$^{+0.46}_{-0.29}$ & 2.298 & wind & $8.40^{+2.60}_{-2.20}$ & \,\,$0.40^{+0.15}_{-0.21}$ \tablefootmark{c} \\
GRB 130418A & 0.09 & 0.00$^{+0.01}_{-0.01}$ & 0.03 & 0.08$^{+0.08}_{-0.08}$ & 1.218 & wind & $0.39^{+0.51}_{-0.36}$ & $1.17^{+0.39}_{-0.32}$ \\
\bottomrule
\end{tabular}
\tablefoottext{a}{The Galactic gas absorption and dust extinction values are taken from \citet{2011ApJ...737..103S}.}
  \tablefoottext{b}{The host magnitudes are derived from the analysis of the combined SED using optical/NIR and X-ray data.}
  \tablefoottext{c}{This value was derived using the DC formalism. The value for the GS formalism corresponds to 5$\times$10$^{49}$ erg.}
\end{table*}

\section{Discussion}
\label{chap:comp}

The basic features of the four GRBs analysed here are presented in \tref{Table:summarygrbs} and \tref{Table:params_basic_grbs_1},
and in the following are placed in the context  of the current state in GRB afterglow studies.

\subsection{Evolution of the break frequencies}

\begin{figure}[!t]              
\includegraphics[width=0.47\textwidth]{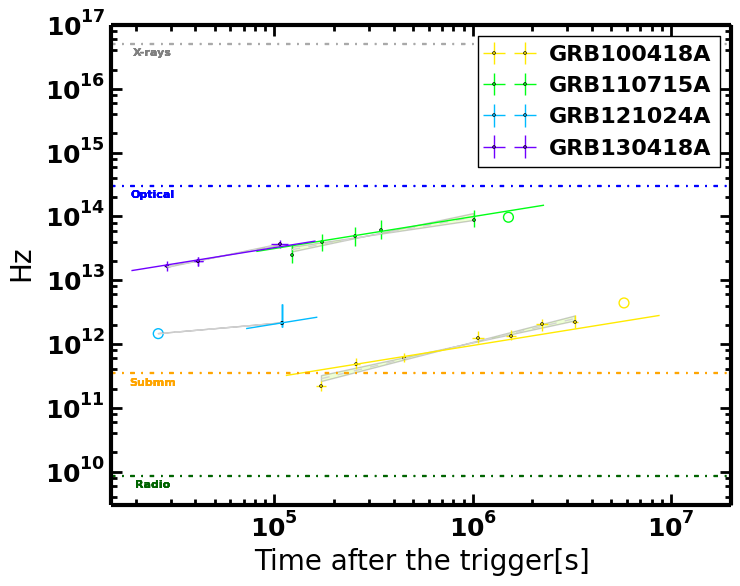}
\includegraphics[width=0.47\textwidth]{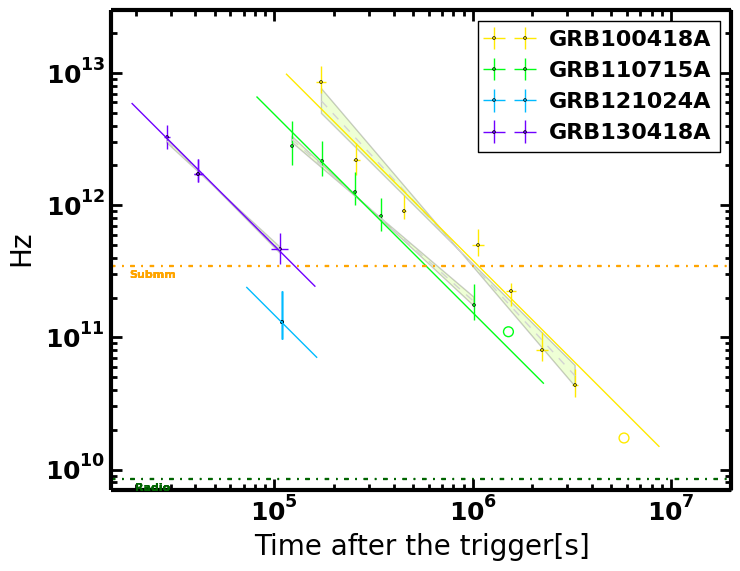}
\includegraphics[width=0.47\textwidth]{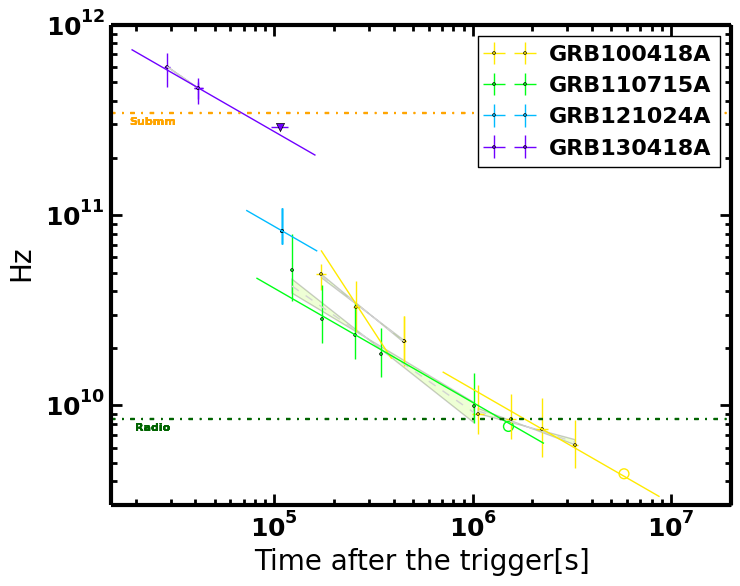}
\vspace{-0.22cm}
\caption{Evolution of the break frequencies \nuc\ (top), \numm\ (middle) and \nusa\ (bottom) for the afterglows of our four GRBs.
The solid lines corresponds to the expected evolution of each frequency from the standard afterglow theory. The 
shaded regions are the fits for each break frequency. The horizontal dashed lines mark the mid-frequency for the four main observing ranges, \ie X-rays, optical, sub-mm and radio.
\label{Fig:frec_evol_all}}
\end{figure}

The evolution of the break frequencies \nuc, \numm\s and \nusa\s for all our GRBs is composed in \fref{Fig:frec_evol_all} and \tref{Table:freq_evol_fit_all}. 
All breaks are evolving as predicted by the model during fast or slow cooling.
Effects such as ISS contribution to the radio observations or IC emission are included in a systematic way.
The impact of possible flares has been carefully considered which could give rise to some of the observed deviations (\eg GRB 110715A) due to the change in the temporal slopes and the flux values.

\begin{table}[ht]
\setlength\tabcolsep{3.2pt}
\caption{Temporal evolution of the measured break frequencies for our GRB afterglows (for 121024A we have only one measurement).
\label{Table:freq_evol_fit_all}}
\vspace{-0.2cm}
\begin{tabular}{c | c c | c c | c c }
\toprule
SED &  \nuc$_{\rm{T}}$ & \nuc$_{\rm{O}}$ & \numm$_{\rm{T}}$  & \numm$_{\rm{O}}$ & \nusa$_{\rm{T}}$ & \nusa$_{\rm{O}}$ \\
\midrule
100418A & 0.5 & 0.57\pmm0.04  & -1.5 & -1.72\pmm0.08 & -0.6 & -0.56\pmm0.06 \\
110715A & 0.5 & 0.56\pmm0.10  & -1.5 & -1.34\pmm0.06 & -0.6 & -0.72\pmm0.10 \\
130418A & 0.5 & 0.61\pmm0.03  & -1.5 & -1.45\pmm0.06 & -0.6 & -0.68\pmm0.08 \\
\bottomrule
\end{tabular}
\vspace{-0.1cm}
\tablefoot{The numbers  correspond to the temporal slope (\al) of a simple \pl\s fitting profile $\nu_i(t) \sim t^{-\alpha_i}$, with i = c, m and sa. The subscript $T$ corresponds to the theoretical value for a decelerating blast wave in a stellar-wind type environment, the subscript $O$ corresponds to the observed value.}
\end{table}

\subsection{Fireball parameters in context}

\subsubsection{GRBs with all basic fireball parameters measured}
\label{GRBsallpara}

For comparison purposes, we have collected those GRBs for which well-determined fireball parameters have been deduced without further assumptions (see Tab. \ref{Table:GRBwithparams} and appendix C for details). The list includes three GRBs (970508, 980703, 000926) for which different analyses have arrived at different sets of parameters, one of those even with drastically different $p$. We found nine GRBs with unique parameters: 3 with ISM (030329, 050904, 161219B), 5 with wind profile (060418, 090323, 090328, 140304, 181201A) preferred, and one with similarly good solutions for ISM or wind profile (140311A). All these are long-duration GRBs, as our four cases.
First,  we note that the addition of our 4 GRBs with well-determined fireball parameters is a worthy extension of the sample, and tips the balance towards the wind environment. More specific comparisons are made in the following sub-sections.

\subsubsection{Circumburst environment}
\label{CBMprofile}

The four GRB afterglows analysed here, and 5 of the 8 clear cases in the literature, are uniquely explained by a relativistic outflow expanding into a \wind density profile. In contrast, more than 50\% of the GRBs from samples based on X-ray and/or optical data sets alone are associated with an ISM density profile (\eg \citealt{Panaitescu2002, Schulze+2011, Gompertz+2018}). This previous dominance was surprising given the theoretical expectations for a wind-blown surrounding of massive stars
and considering the relation between GRBs and Type Ic supernovae
\citep{Hjorth+2003, 2003ApJ...591L..17S, Woosley+2006, 2014A&A...568A..19C, Klose+2019}.

There are several observational biases in pinpointing the external density profile. Firstly, wavelength coverage below \nuc{} is required, since
at $\nu_{\rm{obs}}$ > \nuc\s there is no distinction between ISM or \wind density profiles; only at $\nu_{\rm{obs}}$ < \nuc,  the closure relations allow us to identify the CBM profile. In the literature, X-ray samples show that for a large fraction (70-90\%) of the afterglows, \nux\s usually lies above \nuc\s\ \citep[\eg 22/31 GRBs \citealt{2007ApJ...655..989Z} and 280/300 GRBs;][]{2010ApJ...716L.135C}. Therefore, the CBM structure cannot be determined.

Optical samples, such as the one presented in \citet{2010ApJ...720.1513K}, suggest that less than 25\% of the afterglows (10/42) have \nuo>\nuc\s if p is assumed to be larger than 2. \citet{2009MNRAS.395..580C} and \citet{2006MNRAS.366.1357P} show that > 70\% of their samples (10 and 9 GRBs, respectively) have $\nu_{\rm{obs}}$ < \nuc\s. However, they did not associate the CBM with a \wind\s density profile, instead they show that 1 < k < 2, as expected for an inhomogeneous density profile. In addition to these samples, about 60\% of the afterglows in \cite{2011AA...526A..30G} and \cite{Schulze+2011} have a break between the optical and X-ray bands, \ie $\Delta$\be = 0.5 and/or $\Delta$\al = \pmm 0.25 (+ISM, -\wind).

In the radio band, the peak of the light curve can be used to infer the density profile, and recent estimates resulted in more than half of the sample being consistent with ISM, 20\% with wind, and the remaining 30\% with $0<k<2$ \citep{Zhang+2022ApJ927}.

A second problem is that, if closure relations are used, measurement errors on \al\s and \be\s may not allow a conclusive statement to be made. \citet{Schulze+2011} found that 22\% of their afterglow sample (6 out of 27) are related to a \wind\s density profile, but excluded one quarter of their originally selected GRBs due to inconclusive results.
The presence of breaks between \nuo\s and \nux\s may also be difficult to detect because the break is too close to an observed band, or due to the effects of \dusth\s and \gash.

Finally, assumptions may be needed, but even if these assumptions are very plausible, they may lead to different conclusions. An illustrative example of this is seen for GRB 970228, GRB 970508, GRB 980326 and GRB 980519. \citet{ChevalierLi2000} associated the four afterglows with a \wind density profile, but other authors identified an ISM profile as the preferred CBM for those GRBs (\eg \citealt{1997ApJ...488L.105V,1999ApJ...516..683F,1997Natur.387..876D,1998ApJ...500L.105G,1998ApJ...502L.123G,2000MNRAS.317..170W}). 
The differences in the approach are partly source-specific (e.g. the sudden optical rise in 970508 after 1 day), partly due to the non-availability of the wind model in the first years after the afterglow discovery.

Summarizing, the percentage of GRB afterglows associated with a \wind\s density profile has been increasing over the last years, but certainly deserves further investigation.

\begin{table*}[ht]
 \caption{Temporal slopes (\al) of the parameter change with time.
   \label{Table:par_evol_fit_all}}
 \vspace{-0.22cm}
\begin{tabular}{c c c c c c c c c }
\toprule
SED & \bepse & \epsb &  \den & \eiso & \thh & $\eta$ & $B$ & \mloss\\
\midrule
100418A & $\!$-0.14\pmm0.06 & 0.20\pmm0.11 & 0.02\pmm0.05 & 0.22\pmm0.12 & 0.05\pmm0.03 & 0.21\pmm0.12 & 0.81\pmm0.05 & 0.02\pmm0.05 \\
110715A & 0.06\pmm0.04 & 0.04\pmm0.06 & $\!$-0.05\pmm0.07 & 0.06\pmm0.05 & $\!$-0.03\pmm0.02 & 0.05\pmm0.03 & 0.78\pmm0.04 & 0.04\pmm0.07\\
130418A & 0.20\pmm0.04 & $\!$-0.29\pmm0.03 & 0.33\pmm0.04 & 0.09\pmm0.03 & 0.06\pmm0.03 & 0.06\pmm0.03 & 0.67\pmm0.04 & 0.33\pmm0.03\\
\bottomrule
\end{tabular}
\vspace{-0.1cm}
\tablefoot{Based on a simple \pl\s fitting profile $Q(t) \sim t^{-\alpha}$, where Q stands for the different microphysical and dynamical parameters.}
\end{table*}

\begin{figure}[ht]
\includegraphics[width=0.49\textwidth]{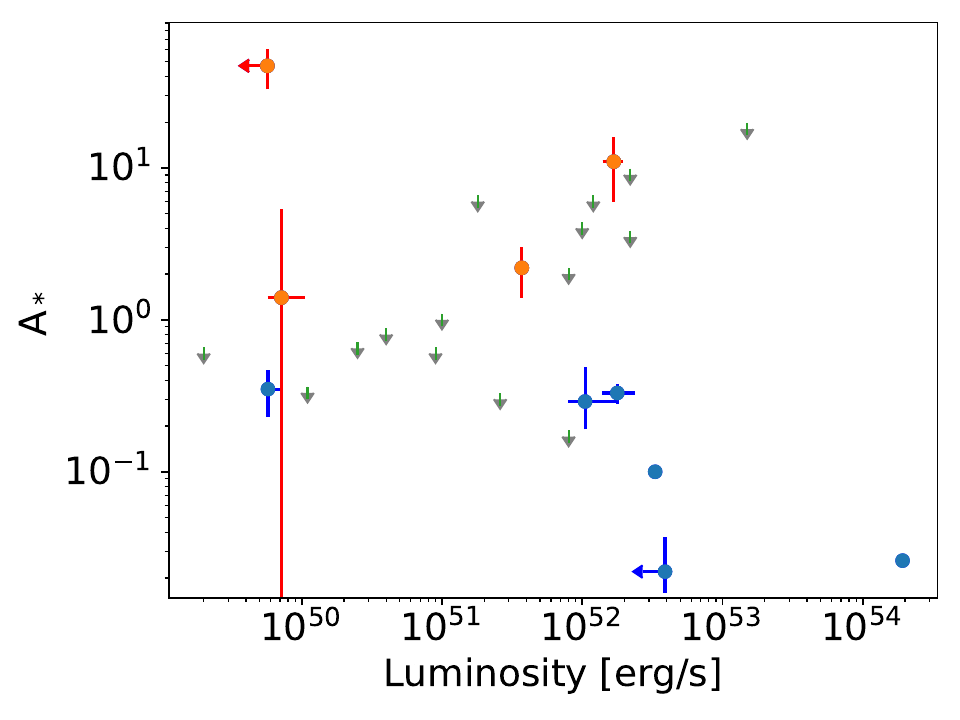}
\caption{Relation between wind density A$_*$ and the explosion luminosity (see Tab. \ref{Table:GRBwithparams}), with literature values (blue) and our four GRB afterglows in red. 
The gray arrows are limits as derived by \cite{HascoetBeloborodov+2014} for a GRB afterglow sample, with limits on the variability timescale and Lorentz factor. 
\label{Fig:E-wind}
}
\end{figure}

In a simplistic model, one could expect the wind density to scale with the mass of the Wolf-Rayet progenitor, and the latter with the GRB luminosity \citep{HascoetBeloborodov+2014}. However, neither our four GRBs nor the sample of Tab. \ref{Table:GRBwithparams} shows an obvious correlation (Fig. \ref{Fig:E-wind}), as already suggested by \cite{HascoetBeloborodov+2014}.

\subsubsection{Microphysical parameters \label{sect:microphys_par}}

The best fit temporal slopes for each of the derived quantities, assuming a single \pl\s model, are given in \tref{Table:par_evol_fit_all}.

\begin{figure}[th]
\includegraphics[width=0.49\textwidth]{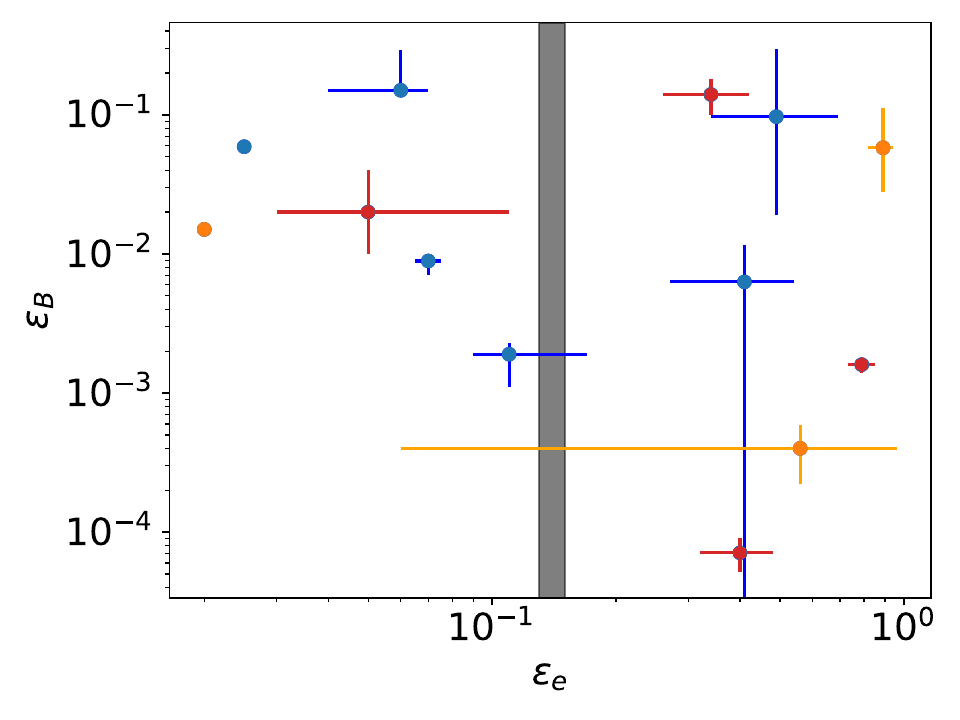}
\vspace{-0.6cm}
\caption{Collection of \epse\ and \epsb\ parameters as deduced from multi-wavelength modelling of afterglows (see Tab. \ref{Table:GRBwithparams}), with literature values for wind (blue) and ISM (orange) profile, and our four GRB afterglows in red (all wind). The shaded region is the narrow \epse\ distribution derived by  \cite{BeniaminiHorst2017}.
\label{Fig:epse-epsb}
}
\end{figure}

For the GRBs analysed here, \epse\s is constant throughout the time for all the afterglows.
When looking at the absolute values, including those from the literature (Fig. \ref{Table:GRBwithparams}), substantial scatter between the GRBs can be noticed (see Fig. \ref{Fig:epse-epsb}). This is in contrast to  \cite{BeniaminiHorst2017} who derived a narrow \epse\ distribution from the analysis of the peaks of 36 radio afterglows.

 \begin{figure*}[ht!]
\centering
\includegraphics[width=0.49\textwidth]{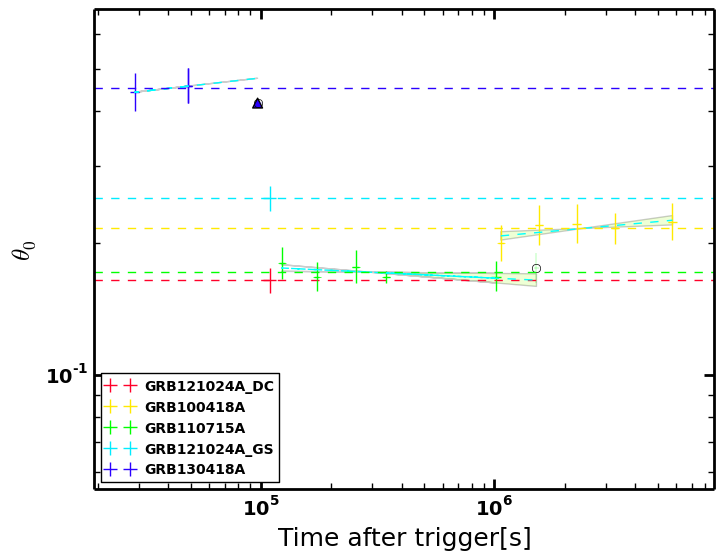}
\includegraphics[width=0.49\textwidth]{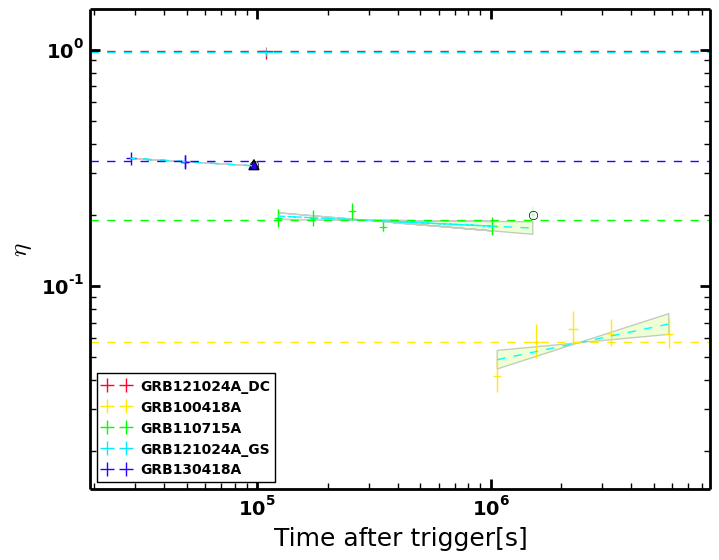}
\includegraphics[width=0.49\textwidth]{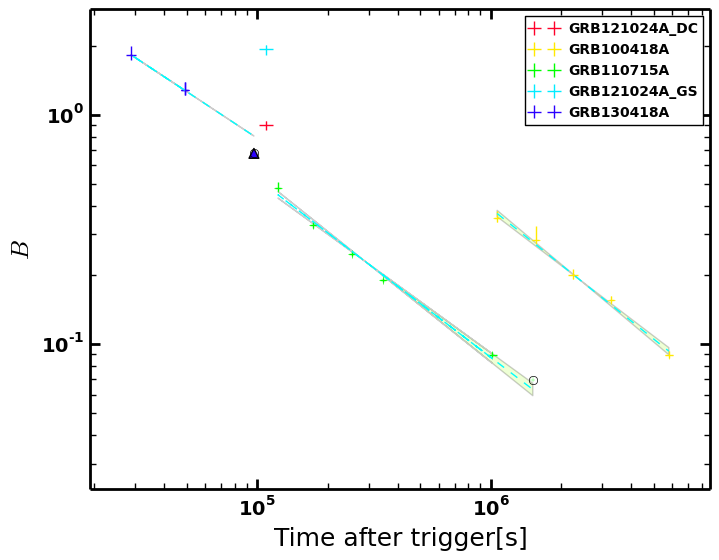}
\includegraphics[width=0.49\textwidth]{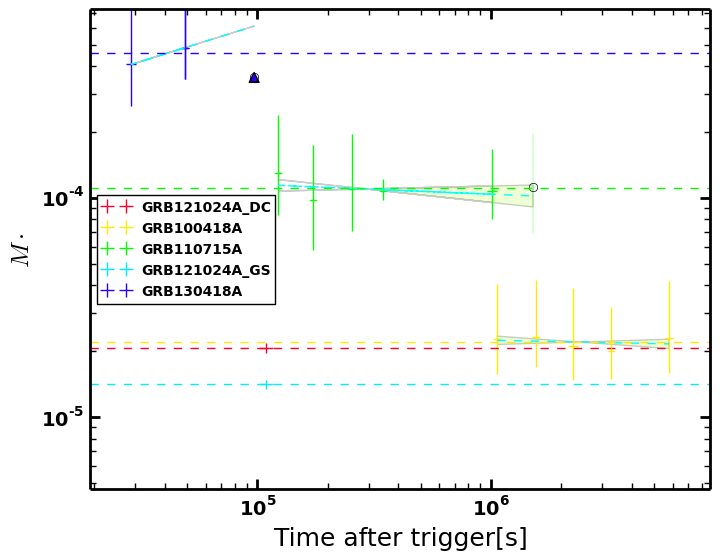}
\caption{Evolution of the secondary quantities according to the GRB afterglow standard model.
The dashed-lines show the average value of each parameter. The dotted lines represent the power-law fit to the data. 
$B$ has a slope $-3/4$ as expected from a magnetic field amplified by shock compression. \thh\s is in rad, $B$ is in G, and \mloss\s is in units of \msun\s yr$^{-1}$. 
\label{Fig:params_sec_evol_all}}
\end{figure*}

Interestingly, our tests for the temporal evolution of the microphysical parameters reveals a clear indication for an evolution of the magnetic field for GRBs 100418A and 130418A, see
\fref{Fig:params_sec_evol_all} and details in \tref{Table:par_evol_fit_all}. 
Already
in the early days of afterglow analysis, the effect of a non-constant
$\epsilon_B$ on the observables was tested \citep{2003ApJ...597..459Y}.
Though these authors found a degeneracy with different external medium types, models of varying $\epsilon_B$ and $\epsilon_e$ were advocated soon after \citep{Ioka+2006, FanPiran2006, 2023arXiv230208230H}.
Pretty clear observational evidence for $\epsilon_B$ rising with time
(as $t^{0.5}$) was demonstrated by \cite{Filgas+2011}, and \cite{2006MNRAS.366.1357P} and \cite{Kong+2010}
have also suggested a non-constant $\epsilon_B$ based on observational data.

We recall that two possibilities are discussed for the origin of the magnetic field in GRB outflows:: 
(i) the magnetic field is assumed to be generated by the shock from a seed that only needs to be very small in the CBM, or
(ii) the blast is merely shock-compressing an existing CBM magnetic field (which is usually assumed to be of order 10 $\mu$G), as inferred for some Fermi-detected GRBs \citep{2009MNRAS.400L..75K}.
The evolution that we find
follows the predictions for a magnetic field, which originates due to shock compression,
$t^{-3/(2(4-k))}$ \citep[$= t^{-3/4}$ for our case of $k=2$;][]{BM1976, 1979rpa..book.....R, 2011ApJ...734...77I}.

Our result  is interesting for two reasons: 
(1) It has no additional assumptions or linked parameters among the analysed epochs of each afterglow. This implies that the observed evolution relies completely on the derived parameters for each SED and actually tests the evolution of the magnetic field in the shocked region independently. 
(2) Our finding of magnetic field evolution suggests shock compression as its origin. The above option (i) has been preferred,
because the shock compression of the pre-existing field alone would lead to a negligible magnetic energy per particle \cite[e.g.][]{Gruzinov2001}, and previous afterglow studies suggested
that the derived magnetic fields are about two orders of magnitude higher than the values expected from compression of the intergalactic field 
\citep{2003ApJ...597..459Y, santana2014, 2016ApJ...820...94N}.
This is based on assumptions of a seed magnetic field similar to that of the Milky Way (6 $\mu$G near the Sun, up to 100 $\mu$G in some Galactic Center filaments), but there is no physical reason this applies. Based on the observation of higher magnetic fields in galaxies with large star formation rate \citep{2012SSRv..166..215B}, \cite{2014MNRAS.442.3147B} has argued that the strength of the global field might be correlated with the field in the vicinity of the GRB, and thus making shock compression a viable option.

A strong relation between \epse\s and the energy injection is expected. 
A strong energy injection affects the dynamics of the outflow and therefore the radiative processes. When the cooling process undergoes a radiative phase, \epse\s is expected to be close to one \citep{2006MNRAS.366.1357P}. Otherwise, if the cooling process is in an adiabatic regime, \epse\s is expected to be of order 0.1 or smaller \citep{Sari+1998}. 
At least qualitatively, we find that for the strongest injection phase, $q = -0.36$, the value of \epse\s is larger than the other cases of energy injection (0.3).

\subsubsection{Plateaus, energy injection and more}
\label{sect:plateaus_all}

The standard fireball model assumes an instantaneous energy injection. Observational data for many afterglows show instead that fast decays and/or plateau phases that do not follow the closure relations are commonly detected. For example, the plateau phase is part of the canonical X-ray light curve \citep{2006nousek,2006ApJ...642..354Z}
having been detected in more than 50\% of the X-ray afterglows.

In this study, three of the analysed GRBs have 
plateau phases up to 50--80 ks in the observer frame, with GRBs 100418A and 121024A also exhibiting clear optical plateaus. We have interpreted this above as prolonged energy injection (see \tref{Table:summarygrbs}), though this interpretation does not influence the main goal of the present work.
The injection parameters 
(\tref{Table:params_basic_grbs_2}), are consistent with  $q > 0$ 
(except for GRB 110715A, which is still consistent with $q > 0$ at 3$\sigma$, though).
The observed values for the injection parameters $q$ of our GRBs are consistent with either a mass stratification model or with a long-lived central engine with a relativistic reverse shock. The first scenario is preferred due to the lack of a convincing model for the long-lived central engine as well as the prediction for the reverse shock to dominate at early times and low frequencies, both in contrast to observations.
The third injection scenario, a Poynting flux dominated outflow, is discarded for GRB 100418A with high confidence (based on the $q$ value) and is unlikely for our other GRBs.

\begin{table}[ht]
 \caption{Spectral slopes \be\s, injection parameter $q$ and electron index
    $p$ for the analysed afterglows. \label{Table:params_basic_grbs_2}}
 \vspace{-0.22cm}
\begin{tabular}{cccc}
\toprule
GRB &  $\beta$ &  $q$ & $p$ \\
\midrule
121024A & 0.86\pmm0.02 & 0.52\pmm0.07     & 1.73\pmm0.03 \\
100418A & 1.11\pmm0.02 & o:0.23\pmm0.04  & 2.22\pmm0.04\\
          	       & 		         & x:0.00\pmm0.05  &  \\
110715A & 1.05\pmm0.01 & -0.36\pmm0.15 & 2.10\pmm0.02\\
130418A & x:0.58\pmm0.11 & x:0.88\pmm0.16 & 2.32\pmm0.14 \\
 		       & o:1.16\pmm0.07 & o:0.14\pmm0.10 \\
\bottomrule
\end{tabular}
\vspace{-0.1cm}
\tablefoot{o: optical, x: X-ray bands. In the cases of GRBs 100418A and 130418A, the final $q$ values correspond to the optical one. Details on the difference between the optical and X-ray values are given in the text.}
\end{table}

Besides prolonged energy injection, alternative interpretations for the common plateaus exist. One of these alternatives is forward shock emission from the edges of jet cores \citep{Beniamini+2020}, with two options: (i) de-beamed emission from the core coming gradually into view,  or (ii) from material
travelling close to the line of sight that has not yet decelerated (this option requires a wind-like environment). Due to the strong dependence of the luminosity on the viewing
angle, both interpretations can reproduce the large span of observed
plateau durations and luminosities,
and naturally predict a correlation between $E_\gamma, iso$, the duration of the plateau and the luminosity at the end of the plateau which is very close to the observed one. In our case, the optical emission for GRB 100418A is clearly above $\nu_m$ and $\nu_c$ during the plateau, thus both options  imply an optical decay slope of 1.1, as opposed to the observed 0.36$\pm$0.04. For GRB 130418A our slope errors are larger and imply consistency, while for GRB 121024A the interpretation is ambiguous given the hard $p$.

Another alternative interpretation is a forward shock expanding into a wind environment of particularly low density (factor 100 below typical Wolf-Rayet winds) with an unusually low Lorentz factor \citep[of order a few tens;][]{Dereli-Begue+2022}. This scenario is not applicable to our GRBs for two reasons: first, we derive the wind normalisation $A_*$ from our data, and these are in the canonical range; second, the peak of the forward shock is constrained by public optical data (not used in our work) to be smaller than 100\,s, 100\,s, 60\,s and 300\,s, much smaller than $>$10$^4$\,s as expected for small $A_*$ and $\Gamma$.

\subsubsection{Dynamics: \eiso}

Energy injection can have an important effect on the estimated total energy. Three of the four GRB afterglows analysed here have \eiso\s of order $10^{52}$ erg or larger.
GRB 110715A has the largest \eiso, and also  the largest $q$, i.e. the strongest energy injection.
While the exact time when the deceleration ($t_{\rm{dec}}$) phase starts is not known, it was assumed it to be the start of the plateau phase. Combining the time of the end of the plateau phase ($t_{\rm{inj}}$) with T$_{90}$ (as a proxy of $t_{\rm{dec}}$),
we derive the ratio \eiso($t_{\rm{dec}}$)/ \eiso($t_{\rm{inj}}$) = $(t_{\rm{dec}}/t_{\rm{inj}})^{1-q}$ of 0.03, 0.3, 0.5 and 0.4 for GRB 100418A, 110715A, 121024A and 130418A, respectively. These ratios imply that 
only in the case of GRB 100418A can we assume that the initial energy is negligible (as usually assumed in the energy injection scenario), while it should not be neglected for the other GRBs. Furthermore, these ratios imply that all the efficiencies $\eta$ are larger than the ones presented in 
\fref{Fig:params_sec_evol_all}
that were derived using \eiso\s after the energy injection phase. Actually, the values for $\eta$ would be 80\% for GRB 100418A and 130418A, 50\% for GRB 110715A and about 95\% for GRB 121024A. Similar values for the ratio \eiso($t_{\rm{dec}}$)/ \eiso($t_{\rm{inj}}$) have been found by \citet{Panaitescu2005}. The increment in $\eta$ was also observed in the analysis of the 31 afterglows using X-ray data presented by \citet{2007ApJ...655..989Z}, rising up to
about 90\%. However, this is just a qualitative statement, as the ratios and change in $\eta$ are highly dependent on the deceleration time and duration of the plateau phase.

\subsection{Synchrotron-self Compton Emission}

Among our four GRBs, 130418A shows a clear signature of SSC emission by exhibiting \bo>\bx\s, which is not possible in the standard synchrotron framework. Also, \bx\s is close to the expected \be=1/3 for an SED dominated by SSC and so does the X-ray temporal slope \aax\s = 1.26 that is consistent with an SSC dominated light curve when \nux\s > \nuc. Interestingly, with  \den\s = 45 it also matches the argument of \citet{sari...esin2001} that a large value of \den\s (of order 10) is required to be able to detect the SSC emission directly.
These findings are very similar to GRB 000926 \citep{2001ApJ...559..123H} which was among the first GRBs with direct SSC emission detection.

We note in passing that hard GeV spectra seen with Fermi/LAT in some GRBs indicate evidence for inverse-Compton emission \citep{2017ApJ...837...13P}.

\setlist{}

\section{Conclusions}

The analysis of the broadband multi-epoch data of four GRB afterglows presented here 
allows us to determine the external medium profile without ambiguity, as well as all the microphysical and dynamical parameters. 
Three important results can be highlighted: 

First, all our four GRBs exploded into a CMB profile corresponding to a wind-blown environment. This association is consistent with the collapsar model and the GRB-SN relation. 
Due to the similar X-ray evolution of ISM and wind interaction, radio observations are needed to infer the CMB profile.
Combined with a critical assessment of the literature, our finding suggests a larger percentage of GRBs associated with a \wind density profile than has been previously reported.

Secondly, the well-sampled multi-wavelength data allow us to extensively test, for the first time, the temporal evolution of all synchrotron break frequencies, and consequently test for the constancy of spectral slopes between the break frequencies, and the temporal behaviour of the main physical parameters of the fireball model.

The third important result is related to the magnetic field in the shocked region. For the first time, the evolution of the magnetic field strength in the shock is derived from afterglow data,
and is found to be in agreement with the prediction of the magnetic field originating from shock amplification of the CBM magnetic field. This supports shock compression as a natural and probable origin of the shocked magnetic field. Additionally, based on the values for \epsb, the seed magnetic field in the CBM region is about 10 mG in our cases. We note that \cite{Lemoine+2013} described the magnetization as the partial decay of the micro-turbulence generated in
the shock precursor, and inferred such decay for four GRBs which showed extended emission at $>$100 MeV. 

These findings have been substantially helped by well-sampled broad-band coverage of the afterglow emission from radio to X-rays. In particular, light curves in the radio and sub-mm range were key factors in deciding between different scenarios. One may suspect that when assumptions are made on certain parameters instead of constraining observations, the previous frequent inferences of an ISM density profile could have been premature.

\begin{acknowledgement}
We thank the referee for helpful comments.
Part of the funding for GROND (both hardware as well as personnel) was generously granted from the Leibniz-Prize to Prof. G. Hasinger (DFG grant HA 1850/28-1). KV and JG are grateful for APEX support by K. Menten, A. Weiss, and F. Bertoldi. APEX is operated by the Max-Planck Institut f\"ur Radioastronomie, the European Southern Observatory, and the Onsala Space Observatory. 
This work made use of data supplied by the UK Swift Science Data Centre at the University of Leicester.
\end{acknowledgement}

\bibliographystyle{aa}
\bibliography{main.bib}

\begin{appendix}

\onecolumn

\section{Finding charts and photometric comparison stars}

\begin{figure*}[!ht]
\includegraphics[width=0.335\textwidth]{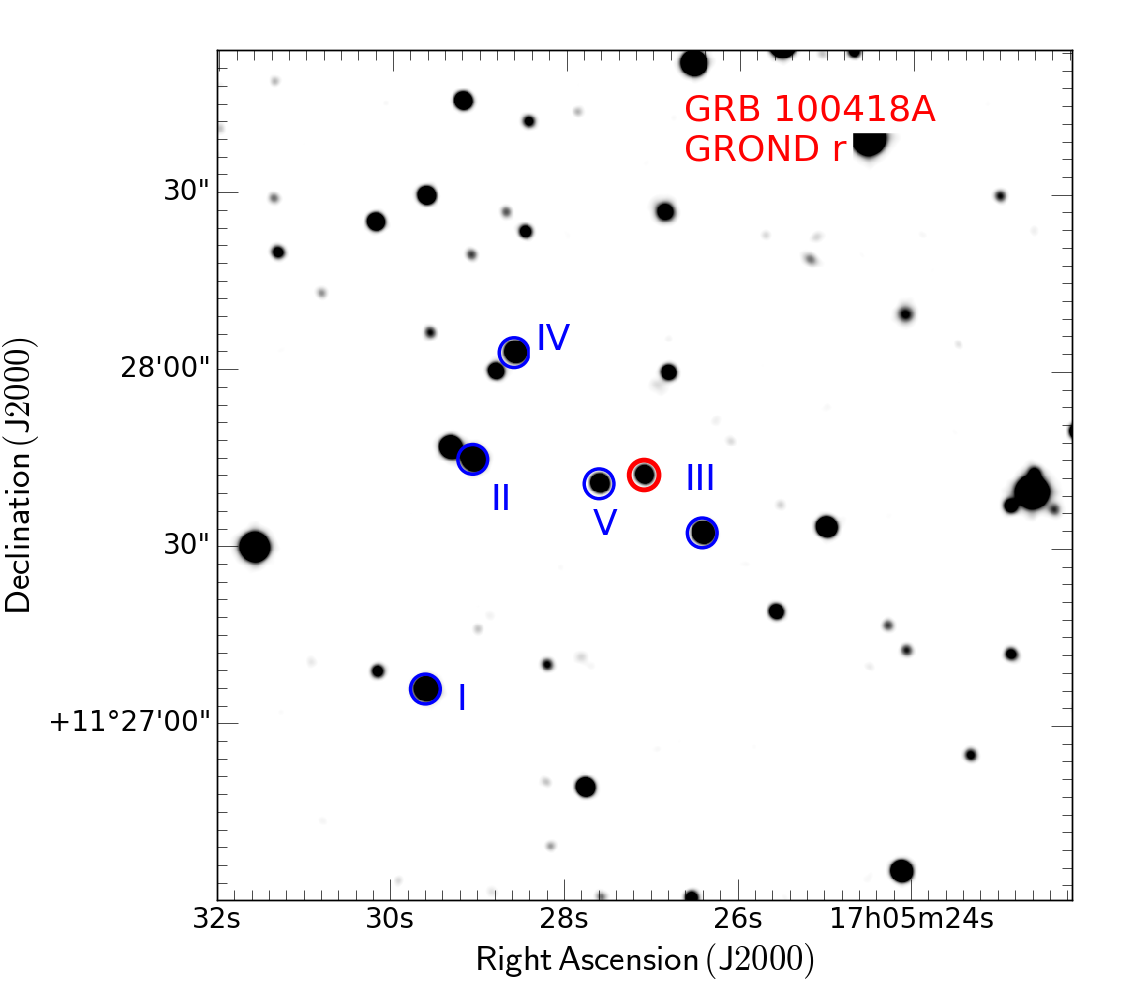}
\includegraphics[width=0.335\textwidth]{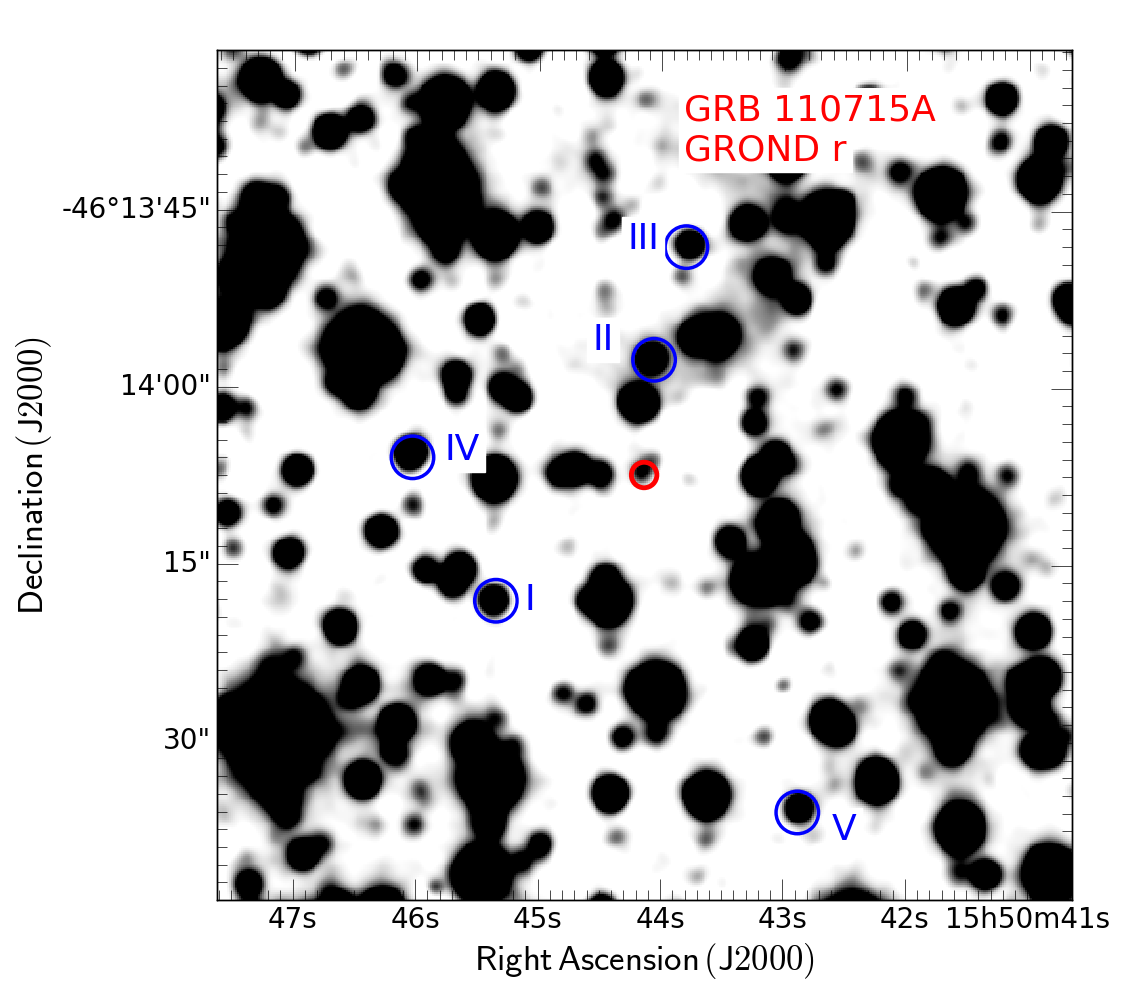}
\includegraphics[width=0.335\textwidth]{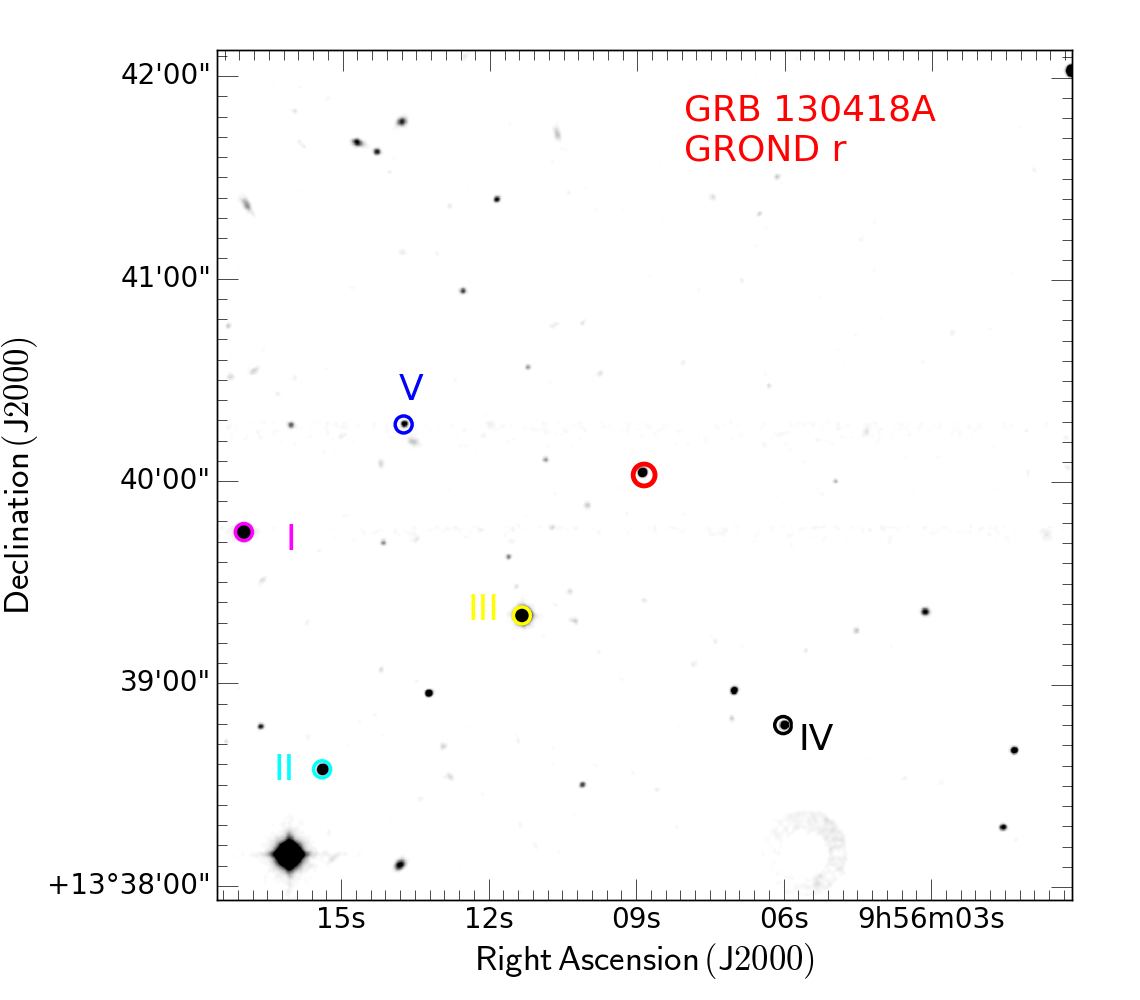}
\caption{GROND $r'$-band finding charts for GRB 100418 (left), 110715A (middle) and 130418A (right). The secondary reference stars (labelled with roman numbers) are reported in Tables \ref{Table:refstars_grb100418A}, \ref{Table:refstars_grb110715A} and  \ref{Table:refstars_grb130418A}, respectively. North is up and East to the left.}
\label{Fig:3FC}
\end{figure*}

\bigskip

\begin{table*}[!ht]
\setlength\tabcolsep{4.5pt}
\caption{Secondary stars of photometric calibration for GRB 100418A (left panel of Fig. \ref{Fig:3FC}). }
\label{Table:refstars_grb100418A}
\begin{tabular}{l*{9}{c}}
\toprule
Star & RA, Decl. (J2000.0) & $g'(\rm{mag}_{\rm{AB}})$ &$r'(\rm{mag}_{\rm{AB}})$ &$i'(\rm{mag}_{\rm{AB}})$ & $z'(\rm{mag}_{\rm{AB}})$ & $J(\rm{mag}_{\rm{Vega}})$ & $H(\rm{mag}_{\rm{Vega}})$ & $K_{\rm{s}} (\rm{mag}_{\rm{Vega}})$ \\
\midrule
I    & 17:05:29.60 +11:27:05.9 & 17.43\pmm0.05 & 17.05\pmm0.05 & 17.01\pmm0.05 & 16.98\pmm0.06 & 15.83\pmm0.09 & 15.59\pmm0.10 & ---  \\
II   & 17:05:29.06 +11:27:44.8 & 18.24\pmm0.05 & 16.99\pmm0.05 & 16.60\pmm0.05 & 16.39\pmm0.06 & 14.93\pmm0.09  & 14.26\pmm0.08  & 16.27\pmm0.39 \\
III  & 17:05:26.42 +11:27:32.5 &18.24\pmm0.05 & 17.68\pmm0.05 & 17.57\pmm0.06 & 17.49\pmm0.07 & 16.27\pmm0.07  & 15.89\pmm0.10 & --- \\
IV  & 17:05:28.59 +11:28:02.9 &19.05\pmm0.06 & 17.72\pmm0.06 & 17.26\pmm0.06 & 17.02\pmm0.06 & 15.51\pmm0.09  & 14.89\pmm0.08 & 17.10\pmm0.57  \\
V  & 17:05:27.61 +11:27:40.8 &18.89\pmm0.06 & 18.48\pmm0.07 & 18.42\pmm0.07 &18.38\pmm 0.07 & 17.18\pmm0.09  & 15.71\pmm0.08 &15.34\pmm 0.14  \\
\bottomrule
\end{tabular}
\end{table*}

\begin{table*}[!ht]
\setlength\tabcolsep{4.5pt}
\caption{Secondary stars of photometric calibration for GRB 110715A (middle panel of Fig. \ref{Fig:3FC}). }
\label{Table:refstars_grb110715A}
\begin{tabular}{l*{9}{c}}
\toprule
Star      & RA, Decl. (J2000.0) & $g'(\rm{mag}_{\rm{AB}})$ &$r'(\rm{mag}_{\rm{AB}})$  &$i'(\rm{mag}_{\rm{AB}})$  & $z'(\rm{mag}_{\rm{AB}})$  & $J(\rm{mag}_{\rm{Vega}})$  & $H(\rm{mag}_{\rm{Vega}})$  & $K_{\rm{s}} (\rm{mag}_{\rm{Vega}})$ \\
\midrule
I	& 15:50:45.34, -46:14:18.0	& 21.89\pmm0.04 & 20.18\pmm0.05  & 19.37\pmm0.05 & 18.98\pmm0.07 & 16.41\pmm0.16  & 15.92\pmm0.09 &  15.85\pmm0.07  \\ 
II 	& 15:50:44.05, -46:13:57.5	& 21.75\pmm0.04 & 20.53\pmm0.04  & 20.13\pmm0.04 & 19.73\pmm0.07 & 17.18\pmm0.12  & 16.42\pmm0.12 & ---   \\
III	& 15:50:43.79, -46:13:48.0	& 22.07\pmm0.05 & 20.90\pmm0.04  & 20.36\pmm0.04 & 19.97\pmm0.07  & 18.85\pmm0.36 & ---  & ---\\
IV	& 15:50:46.02, -46:14:05.8	& 22.05\pmm0.06 & 20.19\pmm0.05  & 21.07\pmm0.07 & 20.35\pmm0.08  & 17.87\pmm0.14 &  15.71\pmm0.08 & 15.64\pmm0.08 \\
V	& 15:50:42.87, -46:14:35.8	& 22.28\pmm0.07 & 20.88\pmm0.06  & 20.32\pmm0.07 & 19.86\pmm0.08  & 17.28\pmm0.16 & ---  & --- \\
\bottomrule
\end{tabular}
\end{table*}

\begin{table*}[!ht]
\setlength\tabcolsep{4.5pt}
\caption{Secondary stars of photometric calibration for GRB 130418A (right panel of Fig. \ref{Fig:3FC}). }
\label{Table:refstars_grb130418A}
\begin{tabular}{l*{9}{c}}
\toprule
Star & RA, Decl. (J2000.0) & $g'(\rm{mag}_{\rm{AB}})$ &$r'(\rm{mag}_{\rm{AB}})$ &$i'(\rm{mag}_{\rm{AB}})$ & $z'(\rm{mag}_{\rm{AB}})$ & $J(\rm{mag}_{\rm{Vega}})$ & $H(\rm{mag}_{\rm{Vega}})$ & $K_{\rm{s}} (\rm{mag}_{\rm{Vega}})$ \\
\midrule
I    & 9:56:16.99, +13:39:44.9& 17.15\pmm0.04 & 16.02\pmm0.05 & 15.59\pmm0.04 & 15.49\pmm0.05 & 13.90\pmm0.12 & 13.47\pmm0.11 & 13.42\pmm0.09 \\
II   & 9:56:15.39, +13:38:34.6 & 18.26\pmm0.04 & 17.55\pmm0.05 & 17.32\pmm0.05 &  17.30\pmm0.05 & 15.88\pmm0.17  & 15.62\pmm0.14  &  15.62\pmm0.10 \\
III  & 9:56:11.40, +13:39:20.3 & 15.64\pmm0.04 & 14.96\pmm0.06 & 14.73\pmm0.05 & 14.72\pmm0.05 & 13.34\pmm0.17 &  12.98\pmm0.12 & 13.10\pmm0.10  \\
IV  & 9:56:06.01, +13:38:47.9 & 20.24\pmm0.06 & 18.87\pmm0.05 & 18.28\pmm0.07 & 18.12\pmm0.06 & 16.57\pmm0.16  & 16.16\pmm0.11 & ---  \\
V  & 9:56:13.74, +13:40:16.9 & --- & 20.24\pmm0.06 & 19.09\pmm0.07 & 18.71\pmm0.07 &  16.95\pmm0.13 & 16.50\pmm0.18 & ---  \\
\bottomrule
\end{tabular}
\end{table*}

\newpage

\section{Observations and data analysis}

\subsection{GRB 100418A}
\label{sect:observations_grb100418a}

\paragraph{\swift}  
The \swift\ Burst Alert Telescope triggered and located GRB 100418A on 2010 April 18 at $T_0$ = 21:10:08 UT \citep{Marshall2010_GCN}. \swift\ slewed immediately to the position of the burst, with the observations starting 79 s after the trigger with the X-ray Telescope. The afterglow was located at RA, Decl. (J2000) = 17:05:27.24, 11:27:42.7 with an uncertainty of 3\farcs1. The observations started in Windowed Timing (WT) mode until $T_0$+174 seconds followed by Photon Counting (PC) mode observations up to $T_0$+3 Ms \citep{marshall2011}. The \swift/XRT light curve and spectral data in the 0.3--10 keV energy range were obtained from the XRT repository \citep{2007AA469Evans,2009MNRAS.397.1177Evans}. The Ultraviolet/Optical Telescope \citep[UVOT;][]{2005SSRv..120...95R} observed the afterglow during the same time interval as \tit{Swift}/XRT. The analysis of the white filter data of the first 150 seconds located the source at RA, Decl. (J2000.0)=17:05:26.96, 11:27:41.9 with an uncertainty of 1\farcs0 \citep{Marshall2010_GCN}. The observations show an initial plateau phase followed by a normal decay after $T_0$+50 ks, with a host of magnitude 22.7 in the white band \citep{Marshall2010_GCN_2}.
 
\paragraph{Optical/NIR}
GROND observations in the wavelength range from 400--2400 nm ($g'r'i'z'JHK_{\rm{s}}$) were targeted on the Swift/UVOT detected afterglow \citep{Marshall2010_GCN}. The observations started on April 19, 2010 at 4:50 UT \citep{Filgas2010_GCN} and continued for 6 hours during the first night. The afterglow was detected in all 7 bands at the position RA, Decl. (J2000) = 17:05:27.09, 11:27:42.3 with an uncertainty of 0\farcs4 in each coordinate (Fig. \ref{Fig:3FC}). VLT/X-shooter spectroscopy of this optical afterglow revealed a redshift of z=0.623 \citep{Antonelli_2010GCN10620}. The GROND observations of the field of GRB 100418A continued on the 2nd, 3rd, 4th, 6th and 23rd night after the burst. 
The data were corrected for Galactic foreground reddening \ebv$=0.07$ mag \citep{2011ApJ...737..103S}, corresponding to an extinction of \av$=0.22$ mag for $R_{\rm{v}}=3.08$.

\begin{table*}[ht]
\centering
\caption{Observed GROND magnitudes of the GRB 100418A afterglow for the seven highlighted epochs in light blue and light red in \fref{Fig:LC_opt_fit_100418A}.
\label{Table:grb100418A_mags_opt}}
\vspace{-0.2cm}
\begin{tabular}{l*{9}{c}}
\toprule
SED & mid-time [ks] & $g'(\rm{m}_{\rm{AB}})$ &$r'(\rm{m}_{\rm{AB}})$  &$i'(\rm{m}_{\rm{AB}})$  & $z'(\rm{m}_{\rm{AB}})$  & $J(\rm{m}_{\rm{Vega}})$  & $H(\rm{m}_{\rm{Vega}})$  & $K_{\rm{s}} (\rm{m}_{\rm{Vega}})$ \\
\midrule
I*  & 27.7   & 18.99\pmm0.05 & 18.64\pmm0.05 & 18.33\pmm0.07 & 18.08\pmm0.07 & 17.63\pmm0.07 & 17.26\pmm0.08 & 17.18\pmm0.14 \\
II* & 40.2   & 19.11\pmm0.05 & 18.48\pmm0.07 & 18.77\pmm0.06 & 18.24\pmm0.07 & 17.81\pmm0.09 & 17.49\pmm0.10 & 17.15\pmm 0.12 \\
\midrule
I   & 130.9  & 20.20\pmm0.06 & 19.87\pmm0.06 & 19.56\pmm0.07 & 19.36\pmm0.07 & 18.93\pmm0.09 & 18.66\pmm0.12 & 18.34\pmm0.11 \\
II  & 202.1  & 20.92\pmm0.06 & 20.60\pmm0.06 & 20.27\pmm0.06 & 20.01\pmm0.07 & 19.66\pmm0.09 & 19.51\pmm0.11 & 19.12\pmm0.16 \\
III & 217.8  & 21.07\pmm0.07 & 20.73\pmm0.07 & 20.36\pmm0.06 & 20.22\pmm0.08 & 19.85\pmm0.10 & 19.55\pmm0.13 & 19.22\pmm0.18 \\
IV &  296.8 & 21.34\pmm0.06 & 21.13\pmm0.07 & 20.72\pmm0.07 & 20.59\pmm0.07 & 20.07\pmm0.11 & 19.91\pmm0.16 & 19.72\pmm0.14 \\
V  &  476.4 & 21.96\pmm0.05 & 21.60\pmm0.04 & 21.36\pmm0.06 & 21.09\pmm0.08 & 20.53\pmm0.17 & 20.18\pmm0.16 & 19.98 UL \\
\midrule 
host & -- & 22.82\pmm0.06 & 22.36\pmm0.06 & 22.25\pmm0.07 & 22.14\pmm0.07 & 21.95\pmm0.08 & 21.70\pmm0.18 & 21.68\pmm0.25 \\
\bottomrule
\end{tabular}
\vspace{-0.1cm}
\tablefoot{Two epochs during the energy injection phase and five epochs after the break in the light curve. The host contribution was subtracted for each band. The Galactic foreground extinction is \dustg$=0.22$ mag.}
\end{table*}

\begin{table*}[!ht]
\centering
\caption{Observed magnitudes of the GRB 100418A afterglow during the epochs for the broad-band analysis (as highlighted in \fref{Fig:LC_opt_fit_100418A}). 
\label{Table:grbmags_sedr_GRB100418A}}
\vspace{-0.2cm}
\begin{tabular}{l*{9}{c}}
\toprule
SED & mid-time [ks] & $g'(\rm{m}_{\rm{AB}})$ &$r'(\rm{m}_{\rm{AB}})$  &$i'(\rm{m}_{\rm{AB}})$  & $z'(\rm{m}_{\rm{AB}})$  & $J(\rm{m}_{\rm{Vega}})$  & $H(\rm{m}_{\rm{Vega}})$  & $K_{\rm{s}} (\rm{m}_{\rm{Vega}})$ \\
\midrule
I     & 173   & 20.59\pmm0.05 & 20.26\pmm0.04 & 19.95\pmm0.06 & 19.72\pmm0.08 & 19.31\pmm0.13 & 18.99\pmm0.15 & 18.67\pmm0.24 \\
II    & 259   & 21.21\pmm0.04 & 20.88\pmm0.04 & 20.57\pmm0.06 & 20.34\pmm0.08 & 19.92\pmm0.14 & 19.61\pmm0.15 & 19.28\pmm0.23  \\
III   & 450   & 21.84\pmm0.05 & 21.51\pmm0.04 & 21.31\pmm0.06 & 21.18\pmm0.07 & 20.55\pmm0.13 & 19.89\pmm0.15 & >20.12 \\
IV   & 1065 & 23.35\pmm0.05 & 23.02\pmm0.05 & 22.71\pmm0.07 & 22.48\pmm0.08 & 22.06\pmm0.14 & 21.75\pmm0.14 & >21.42 \\
V    & 1555 & 23.92\pmm0.05 & 23.59\pmm0.05 & 23.28\pmm0.06 & 23.05\pmm0.09 & 22.63\pmm0.14 & 22.33\pmm0.15 & >21.99 \\
VI   & 2246 & 24.48\pmm0.06 & 24.15\pmm0.04 & 23.84\pmm0.07 & 23.61\pmm0.08 & 23.18\pmm0.16 & 22.88\pmm0.18 & >22.56 \\
VII  & 3283 & 25.05\pmm0.05 & 24.72\pmm0.06 & 24.42\pmm0.07 & 24.18\pmm0.09 & 23.76\pmm0.16 & 23.46\pmm0.17 & >23.13 \\
VIII & 5788 & 25.91\pmm0.07 & 25.58\pmm0.06 & 25.27\pmm0.08 & 25.05\pmm0.10 & 24.62\pmm0.15 & 24.32\pmm0.17 & >23.99 \\
\bottomrule
\end{tabular}
\vspace{-0.1cm}
\tablefoot{The host contribution was subtracted and the magnitudes are corrected for \dustg=0.22 mag.}
\vspace{-0.5cm}
\end{table*}

\paragraph{Sub-millimeter}
The optical counterpart of GRB 100418A was followed-up in the sub-mm wavelength range using the Submillimeter Array SMA and the Plateau de Bure interferometer PdBI over several days. \\
\tbf{SMA:} observations of the afterglow of GRB 100418A started on April 19th 2010 at 13:00 UT, 16 hours after the trigger \citep{2010GCN..10630...1M}. The observations were performed at a frequency of 340 GHz with an initial detection of the counterpart with a flux of 13.40\pmm1.60 mJy \citep{2012A&A...538A..44D}.  Follow-up observations were performed during the following 4 nights until the source was not detected any more (see \tref{Table:grb100418A_mags_radio}) down to a 3$\sigma$ upper limit of $<$4.2 mJy. \\
\tbf{PdBI:} observations started on April 19th 2010, 1.26 days after the trigger, and continued for 2 months until the source was not detected any more after 69 days down to a 3$\sigma$ upper limit of $<$0.57 mJy \citep{2012A&A...538A..44D}. The observations were performed at three different bands: 86.7 GHz, 103.0 GHz and 106.0 GHz with an initial ($T_o$ = 109 ks) detection at 103.0 GHz with a flux of 6.57\pmm0.07 mJy before our chosen epochs (see \tref{Table:grb100418A_mags_radio}).

\begin{table}[th]
\centering
\setlength\tabcolsep{7pt}
\caption{Sub-mm and radio fluxes of GRB 100418A.
   \label{Table:grb100418A_mags_radio}}
\vspace{-0.2cm}
\begin{tabular}{l*{8}{c}}
\toprule
SED & hmid-time  & SMA [mJy] & PdBI [mJy] & PdBI [mJy] & ATCA [mJy] & VLA [mJy]  & ATCA [mJy] \\
 & [ks] &  345 GHz &  103/106 GHz &  86.7 GHz  &  9 GHz  &  8.46 GHz  &  5.5 GHz  \\
\midrule
I     & 173   & 5.10 \pmm0.90 & -- 		       & --			 & 1.27\pmm0.09 & -- 		      & 0.86\pmm0.12 \\
II    & 259   & 5.40 \pmm1.10 & 3.43\pmm1.00 & -- 			 & --  		   & 0.46\pmm0.02 & -- \\
III   & 450   & 4.20 UL 	     & --      		       & 3.70\pmm0.07 & --  		   & 0.29\pmm0.02 & -- \\
IV   & 1065 & -- 		     & 1.13\pmm0.12 & -- 			 & --  	    	   & 0.52\pmm0.02 & -- \\
V    & 1555 & -- 	             & -- 		       & 1.14\pmm0.05 & -- 			   & 0.54\pmm0.02 & -- \\
VI   & 2246 & -- 		     & -- 		       & 1.18\pmm0.09 & -- 			   & 0.85\pmm0.03 & -- \\
VII  & 3283 & -- 		     & 0.61\pmm0.13 & -- 			 & 1.39\pmm0.18 & 1.02 \pmm0.06 & 0.90\pmm0.08 \\
VIII & 5788 & -- 		     & -- 		       & 0.55\pmm0.18 & 1.60\pmm0.20 & 0.82 \pmm0.06 & 1.27\pmm0.12 \\
\bottomrule
\end{tabular}
\vspace{-0.1cm}
\tablefoot{The epochs correspond to the eight highlighted epochs in \fref{Fig:LC_radio_fit_100418A}.}
\vspace{-0.3cm}
\end{table}

\paragraph{Radio}
We also use published data from the Very Large Array (VLA), the Australian Telescope Compact Array (ATCA) and the Westerbork Synthesis Radio Telescope (WSRT) \citep{2012ApJ...746..156C, 2012A&A...538A..44D, 2013ApJ...779..105M}. 
\textbf{ATCA} follow-up observation began on April 20th. The afterglow was followed for three epochs on the 2nd, 38th and 67th day after the trigger in two different bands, 5.5 GHz and 9.0 GHz (see \tref{Table:grb100418A_mags_radio}). 
WSRT observed also on April 20th, for a duration of $\sim$ 8 hours and detected a radio counterpart with a flux density of 369\pmm29 $\mu$Jy at a frequency of 4.8 GHz \citep{2010GCN.10647....1V}. 
The source was monitored at 8.46 GHz  using the \textbf{VLA} \citep{2013ApJ...779..105M} between 2 and 157 days after the trigger. 
It was also observed at frequencies of 4.95 GHz, 4.9 GHz and 7.9 GHz (see \tref{Table:grb100418A_mags_radio}).

\subsection{GRB 110715A}
\label{sect:observations_grb110715a}

\paragraph{\swift}  
On 2011 July 15 at $T_0$ = 13:13:50 UT \citep{2011GCNR..340....1S} the \swift Burst Alert Telescope triggered on and located GRB 110715A. \swift slewed immediately to the position of the burst, and the observations started 90 s after the trigger. The afterglow was located at RA, Decl. (J2000) = 15:50:44.07, -46:14:09.0 with an uncertainty of 2\farcs2 \citep{2011GCN..12161...1E}. 
The first few thousand seconds are covered in Windowed Timing (WT) mode, during which the flux decayed with a temporal slope \al\s of about 0.5. Photon Counting (PC) mode observations continued until $T_0$ + $1$ Ms, with two observed breaks in the light curve. The \tit{Swift}/XRT light curve and spectral data were obtained from the XRT repository \citep{2007AA469Evans,2009MNRAS.397.1177Evans}. UVOT observed the afterglow in the same time interval and located it at RA, Decl. (J2000.0) = 15:50:44.09, -46:14:06.5 with an uncertainty of 0\farcs6 \citep{2011GCN..12162...1K}. The observations show an initial decay phase up to $T_0+22$ ks, followed by a plateau phase up to $T_0+50$ ks and a final decay phase.  

\paragraph{Optical/NIR}
GROND observations of the afterglow
started on July 18, 2011 at 00:35 UT and continued for the next 2 hours during the first night. The afterglow was detected in all 7 bands \citep{2010GCN..10577...1U} at position RA, Decl. (J2000) = 15:50:44.10, -46:14:06.2 with an uncertainty of 0\farcs4 in each coordinate (\fref{Fig:3FC}). VLT/X-shooter spectroscopy identified the redshift as z = 0.82 \citep{Piranomonte_2011GCN12164}.
Observations continued on the 2nd, 4th, 6th and 8th night after the burst. 
The data were corrected for the Galactic foreground reddening of \ebv=0.59 mag \citep{2011ApJ...737..103S}, corresponding to an extinction of \dustg=1.82 mag for $R_{\rm{v}}=3.08$. 

\paragraph{Submillimeter}
The afterglow of GRB 110715A was followed up in the sub-mm wavelength range using the LABOCA bolometer camera \citep{refId0} in the Atacama Path Experiment Telescope (APEX) and with the antennas of the ALMA array. It was observed at a mid-frequency of 345 GHz with both instruments, with one epoch taken with each. See \tref{Table:grb110715A_mags1} and \fref{Fig:LC_radio_fit_110715A}.
\textbf{LABOCA} started observations on July 16 at 23:21 UT, observed for about 1.47 hours, and detected the source with a flux of 11.0\pmm2.3 mJy \citep{2011GCN..12168...1D}.
\textbf{ALMA} observed the source 2.5 days after the detection by \swift. The source was detected with a flux of 4.9\pmm0.60 mJy \citep{2012A&A...538A..44D}.

\paragraph{Radio: ATCA}
Radio observations were performed with ATCA, starting on July 18 at 12.2 UT \citep{2011GCN..12171...1H} and continued for more than 2.5 months
at four different frequencies (5.5, 9.0, 18.0, and 44.0 GHz). 
The observations at 9.0 GHz resulted in four detections and one upper limit. 
Details on the fluxes are given in \tref{Table:grb110715A_mags1} and \fref{Fig:LC_radio_fit_110715A} \citep{2012ApJ...746..156C}.

\begin{table*}[bh]
\centering
\caption{Observed GROND magnitudes of the GRB 110715A afterglow for the epochs used in the SED analysis. \label{Table:grbmags_sedradio_GRB110715A}}
\vspace{-0.2cm}
\begin{tabular}{l*{9}{c}}
\toprule
SED & mid-time [ks] & $g'(\rm{m}_{\rm{AB}})$ &$r'(\rm{m}_{\rm{AB}})$  &$i'(\rm{m}_{\rm{AB}})$  & $z'(\rm{m}_{\rm{AB}})$  & $J(\rm{m}_{\rm{Vega}})$  & $H(\rm{m}_{\rm{Vega}})$  & $K_{\rm{s}} (\rm{m}_{\rm{Vega}})$ \\
\midrule
I  & 122.7    & 21.11\pmm0.04 & 19.96\pmm0.05 & 19.24\pmm0.04 & 18.78\pmm0.04 & 18.11\pmm0.05 & 17.56\pmm0.05 & 17.26\pmm0.05 \\
II  & 173.2   & 21.67\pmm0.05 & 20.53\pmm0.04 & 19.79\pmm0.04 & 19.34\pmm0.04 & 18.67\pmm0.04 & 18.12\pmm0.04 & 17.82\pmm0.04 \\
III & 254.5   & 22.29\pmm0.04 & 21.15\pmm0.04 & 20.42\pmm0.04 & 19.96\pmm0.04 & 19.31\pmm0.04 & 18.74\pmm0.04 & 18.45\pmm0.04 \\
IV & 344.9   & 22.79\pmm0.06 & 21.65\pmm0.06 & 20.92\pmm0.06 & 20.46\pmm0.06 & 19.79\pmm0.06 & 19.24\pmm0.06 & 18.94\pmm0.06 \\
V  & 1014.2 & 24.54\pmm0.05 & 23.41\pmm0.04 & 23.41\pmm0.08 & 22.22\pmm0.06 & 21.56\pmm0.22 & 20.99\pmm0.21 & 20.69\pmm0.32 \\
VI & 1513.8 & 25.19\pmm0.08 & 24.06\pmm0.08 & 23.33\pmm0.09 & 22.87\pmm0.20 & 22.21\pmm0.21 & 21.64\pmm0.34 & 21.35\pmm0.31 \\
\bottomrule
\end{tabular}
\vspace{-0.1cm}
\tablefoot{The Galactic foreground extinction is \dustg=1.82 mag.}
\vspace{-0.7cm}
\end{table*}

\begin{table*}[th]
\centering
\caption{Sub-mm and radio fluxes of GRB 110715A.
\label{Table:grb110715A_mags1}}
\vspace{-0.2cm}
\begin{tabular}{l*{8}{c}}
\toprule
SED & mid-time  & APEX [mJy] & ALMA [mJy] & ATCA [mJy]  & ATCA [mJy] & ATCA [mJy]  & ATCA [mJy] \\
 & [ks] &  345 GHz &  345 GHz &  5.5 GHz  &  9 GHz  &  18. GHz  &  44.0 GHz  \\
\midrule
I   & 123   & 11.0\pmm2.0  & -- 			   & -- 		    & --  		        	& -- 		        	  & -- 		    \\
II  & 173   & --                    & -- 			   & -- 		    & --  		        	& -- 			  & 0.51\pmm0.24 \\
III & 254   & -- 		       	& -- 			   & -- 		    & --  		     	& -- 			  & 2.05\pmm0.66 \\
IV & 345   & -- 			& 4.90\pmm0.60 & 0.53\pmm0.17 & 0.44\pmm0.13 	& 0.73\pmm0.22 & -- 		     \\
V  & 1014 & -- 			& --                      & 0.43\pmm0.13 & --		       	& 1.47\pmm0.44 & 1.89\pmm0.59 \\
VI & 1514 & -- 			& -- 			   & 0.58\pmm0.17 & 0.71\pmm0.17 	& 1.10\pmm0.33 & 1.18\pmm0.66 \\
\bottomrule
\end{tabular}
\vspace{-0.1cm}
\tablefoot{The epochs correspond to the six highlighted epochs in \fref{Fig:LC_radio_fit_110715A}. Radio observations include an additional systematic uncertainty of 30\% to take into account the effects of interstellar scintillation. }
\vspace{-0.5cm}
\end{table*}

\subsection{GRB 130418A}
\label{sect:observations_grb130418a}

\paragraph{\swift}  
On April 18th 2013 the \swift\s Burst Alert Telescope detected GRB 130418A \citep{2013GCN.14377....1D} at 19:00:53 UT. 
\swift\s slewed to the position of the GRB and started observations 129.7 seconds after the trigger. The X-ray afterglow was detected by \swift/XRT at a position RA, Decl.(J2000) = 09:56:9.05, 13:39:55.4 with an uncertainty of 5\farcs3. The observations were performed in Windowed Timing (WT) mode within the time interval from $T_0+136$ s to $T_0+353$ s, and continued in Photon Counting (PC) mode in the time interval from $T_0+3.6$ ks to $T_0+407$ ks. The \swift/XRT light curve and spectral data in the energy range from 0.3–10 keV were obtained from the XRT repository \citep{2007AA469Evans,2009MNRAS.397.1177Evans}. The afterglow was also detected by the Ultraviolet/Optical Telescope (UVOT), and localised at a position of RA, Decl.(J2000) = 09:56:08.88, 13:40:02.7 with an uncertainty of 0\farcs5 and a magnitude in the $white$ band of 14.85\pmm0.05 in the first 150 s of exposure \citep{2013GCN.14384....1K}.

\paragraph{Optical/NIR}
GROND observations of the field of GRB 130418A started on April 19, 2013 at 01:20:33 UT, 6.3 hours after the trigger \citep{2013GCN.14386....1N} and continued for the next three hours. The observations were performed simultaneous in 7 bands in a wavelength range from 400-2400 nm ($g'r'i'z'JHK_{\rm{s}}$). The optical counterpart was detected in all 7 bands at a position RA, Decl.(J2000) = 09:56:8.85, 13:40:02.0 with an uncertainty of 0\farcs4 in each coordinate (\fref{Fig:3FC}). The measured redshift is z = 1.218 \citep{UgartePostigo_2013GCN14380}. The GROND observations continued on the third night, and in February 2014 with deep observations of the field to determine the possible host contribution. 
The data were corrected for Galactic foreground reddening of \ebv = 0.03 mag \citep{2011ApJ...737..103S}, corresponding to an extinction of \dustg = 0.09 mag for $R_{\rm{v}}=3.08$.

\paragraph{Submillimeter}
\tbf{APEX}
The afterglow of GRB 130418A was observed using the LABOCA bolometer camera \citep{refId0}\footnote{Based on observations collected during ESO time at the Atacama Pathfinder Experiment (APEX) under proposal 091.D-0131.} 
starting on April 19th 2013 at 23:50 UT, about 22 hours after the trigger. The observations were taken in mapping mode and the reduction of the data was done using the Bolometer Array analysis software \citep[BoA,][]{2012SPIE.8452E..1TSchuller}. There is an initial detection of the source with a flux of about 40 mJy with a fast decay after just a couple of hours, with a subsequent faint detection at 17 mJy. It was followed up for one the night of April 20th with no detection and a 2$\sigma$ limit of 10 mJy.
\tbf{SMA}
sub-mm observations were also performed using the Submillimetre Array (SMA) at Mauna Kea at a central observing wavelength of 340 GHz. The observations were performed on April 19, 2013 at 06:30 UT for 1.25 hours. No source was detected at the GRB afterglow position down to a 3$\sigma$ limit of 14.5 mJy 
\citep{2013GCN.14400....1M}.
\tbf{CARMA}
Millimetre observations using the Combined Array for Research in Millimetre-Wave Astronomy (CARMA) started observations of the field of GRB 130418A at a frequency of 93 GHz at 02:50 UT on April 19, 2013 and continued during 0.5 hours. The millimetre counterpart was detected with a flux of 3 mJy \citep{2013GCN.14387....1P}.

\paragraph{Radio:}
Radio observations of the GRB 130418A field with the WSRT were taken between April 21, 2013 13.53 UT and April 22 01.49 UT. No radio counterpart was detected with a 3$\sigma$ limiting magnitude of 69 $\mu$Jy \citep{2013GCN.14434....1V}.

\begin{table*}[bh]
\centering
\caption{Observed GROND magnitudes of the GRB 130418A afterglow for the seven analysed epochs. \label{Table:GRONDmags_130418A}}
\vspace{-0.2cm}
\begin{tabular}{l*{9}{c}}
\toprule
SED & mid-time [ks] & $g'(\rm{m}_{\rm{AB}})$ &$r'(\rm{m}_{\rm{AB}})$  &$i'(\rm{m}_{\rm{AB}})$  & $z'(\rm{m}_{\rm{AB}})$  & $J(\rm{m}_{\rm{Vega}})$  & $H(\rm{m}_{\rm{Vega}})$  & $K_{\rm{s}} (\rm{m}_{\rm{Vega}})$ \\
\midrule
I    & 24.8  & 18.87\pmm0.06 & 18.54\pmm0.04 & 18.31\pmm0.04 & 18.02\pmm0.04 & 17.69\pmm0.14 & 17.34\pmm0.15 & 17.11\pmm0.16 \\
II   & 26.7  & 18.97\pmm0.05 & 18.66\pmm0.04 & 18.43\pmm0.05 & 18.09\pmm0.05 & 17.83\pmm0.10 & 17.43\pmm0.10 & 17.23\pmm0.12 \\
III  & 33.4  & 19.29\pmm0.04 & 18.94\pmm0.04 & 18.72\pmm0.05 & 18.41\pmm0.05 & 18.10\pmm0.07 & 17.70\pmm0.07 & 17.51\pmm0.12 \\
IV & 194.7 & 24.24\pmm0.50 & 23.54\pmm0.50 & 23.31\pmm0.50 & 23.29\pmm0.50 & 22.42\pmm0.50 & 21.67\pmm0.00 & 20.55\pmm0.00 \\
\midrule
Ir   & 28.8  & 19.03\pmm0.02 & 18.74\pmm0.07 & 18.49\pmm0.02 & 18.19\pmm0.04 & 17.85\pmm0.04 & 17.43\pmm0.05 & 17.29\pmm0.07 \\
IIr  & 41.5  & 19.41\pmm0.06 & 19.12\pmm0.06 & 18.97\pmm0.07 & 18.64\pmm0.06 & 18.29\pmm0.12 & 18.13\pmm0.13 & 17.93\pmm0.28 \\
IIIr & 96.8  & 20.93\pmm0.25 & 20.65\pmm0.21 & 20.42\pmm0.22 & 20.18\pmm0.32 & 19.91\pmm0.33 & 19.41\pmm0.40 & 19.21\pmm0.40 \\
\bottomrule
\end{tabular}
\vspace{-0.1cm}
\tablefoot{Three epochs during the energy injection phase and four epochs after the break in the light curve. The host contribution was subtracted for each band. The magnitudes are corrected for Galactic foreground extinction \dustg$=0.09$ mag.}
\vspace{-0.3cm}
\end{table*}

\section{Compilation of GRBs with complete fireball parameter measurements}

\subsection{Selection criteria}

Below, we compile all GRBs for which the afterglow modelling has allowed to deduce all 5 parameters (density structure, energy, \epse, \epsb, $p$) without further assumptions, such as, e.g., a canonical $p$, energy equipartition (\epse\ = \epsb\ = 1/3), or the location of \nusa. This implicitly requires a sixth parameter to be measured, the redshift (in order to deduce proper fluxes and source-frame frequencies). In reference to earlier compilations, such as \cite{santana2014} for \epse\ and \epsb, we comment on each GRB which we deem does not fulfill our selection criterion. We admit that this selection is subjective, because many of the well-observed afterglows show deviations from the canonical behaviour, and assumptions made on these additional features might strongly influence the interpretation of the whole afterglow event. Even on a more basic level, alternative assumptions for energy injection vs. e.g. jet structure already lead to systematic differences, such as preferences for the external density profile \citep{Panaitescu2005}. Yet, we have attempted to apply the same criteria.

The only systematic difference between our four GRBs and those compiled here is the fact that we derive the efficiency $\eta$ from the final set of fireball parameters, while in general it is assumed at the onset of the analysis.

\subsection{GRBs with all basic fireball parameters measured}

The resulting GRBs with a complete set of fireball model parameters is summarized in Tab. \ref{Table:GRBwithparams}. We note that pre-Swift GRBs typically have very poor X-ray spectra, implying that $p$ is only (loosely) constrained by the global model rather than the X-ray spectral slope and the knowledge of the location of the cooling break. A good example of this is GRB 980703 with two equally well fitting models but substantially different $p$.

\begin{table*}[ht]
  \setlength\tabcolsep{5.5pt}
  \renewcommand{\arraystretch}{1.2}
\caption{GRBs with measurements of all fireball parameters.
\label{Table:GRBwithparams}}
\vspace{-0.2cm}
\begin{tabular}{c c c c c c c c}
\toprule
  GRB  & $p$ & n/A$_*$ & \epse & \epsb & $\theta_J$ & E$_{\rm k, iso}$ & Refs.\\
       &    & $\!\!$(cm$^{-3}$) / (5$\times$10$^{11}$ g cm$^{-1}$)$\!\!$ & & & (degree) & (10$^{52}$ erg) & \\
\midrule 
~~~970508$^{(1)}$ & 2.12$^{+0.03}_{-0.008}$ & 0.20$^{+0.01}_{-0.02}$ (n) & 0.342$^{+0.09}_{-0.01}$ & 0.25$^{+0.006}_{-0.02}$ & 48$^{+2.0}_{-1.6}$ & 3.7$^{+0.1}_{-0.1}$ (E$_{\rm k,iso}^{\nu_c = \nu_m}$) & A \\
       & 2.39$^{+0.10}_{-0.12}$ & 0.63$^{+4.37}_{-0.50}$ (n) & 0.38$^{+0.30}_{-0.19}$ & 0.0032$^{+0.0468}_{-0.0030}$ & 42$^{+30}_{-10}$ & 3.4$^{+3.0}_{-2.6}$ (E$_{\rm k,iso}^{\nu_c = \nu_m}$) & B \\
       & 2.2 & 0.3 (A$_ *$)& 0.2 & 0.1 & -- & 0.3 & C \\
980703 & 3.08 & 7$\times 10^{-4}$ (n)  & 0.075  & 4.6$\times$10$^{-4}$ & 13.4 & 290 & D\\
            & 2.54 & 27.6 (n)   & 0.27  & 1.8$\times$10$^{-3}$ & 13.4 & 11.8 (E$_{\rm k,iso}^{\nu_c = \nu_m}$) & ~~~E$^{(2)}$\\
       & 2.11 & 1.42 (A$_*$)& 0.69 & 2.8$\times$10$^{-2}$ & 17.8 & 0.66 (E$_{\rm k,iso}^{\nu_c = \nu_m}$) & ~~~E$^{(2)}$\\
       & 2.05$^{+0.09}_{-0.09}$ & 0.38$^{+0.43}_{-0.21}$ (n)& 0.93$^{+0.33}_{-0.24}$ & 0.13$^{+0.19}_{-0.06}$ & 11.4$^{+2.4}_{-2.4}$ & 0.95$^{+0.53}_{-0.31}$ (E$_{\rm k,iso}^{\nu_c = \nu_m}$)& B\\
000926 & 2.40$^{+0.01}_{-0.02}$ & 22$\pm$5 (n) & 0.10$\pm$0.02 & 6.5$^{+1.5}_{-1.1} \times10^{-2}$ & 8.1$^{+0.5}_{-0.6}$ & 0.1 (E$_{\rm k,iso}$ sr$^{-1}$) & F, G \\
       & 2.43$\pm$0.06 & 27$\pm$3 (n) & 0.30$\pm$0.05 & 0.8$\pm0.3 \times10^{-2}$ & 7.8$\pm$0.2& 18$\pm$2 & H \\
       & 2.79$^{+0.05}_{-0.04}$ & 16$\pm$3 (n) & 0.15$\pm$0.01 & 2.2$^{+0.5}_{-0.6} \times10^{-2}$ & 9.3$^{+0.4}_{-0.2}$& 15 (E$_{\rm k,iso}^{\nu_c = \nu_m}$) & I \\
\midrule
~~~030329$^{(3)}$ & 2.12$\pm$0.05 & 8.6$^{+12.0}_{-5.0}$ (n) & 0.56$^{+0.4}_{-0.5}$ & 4.0$^{+1.9}_{-1.8}$ $\times10^{-4}$ & 6.2$^{+0.02}_{-0.03}$ & 0.14$^{+0.14}_{-0.08}$ & J\\
050904 & 2.14  & 680 (n) & 0.02 & 0.015 & 8 & 88 & K\\
060418& 1.97$^{+0.02}_{-0.04}$ & 0.35$\pm$0.12 (A$_*$) & 0.06$^{+0.01}_{-0.02}$ & 0.15$^{+0.14}_{-0.01}$ & 22.5$^{+0.9}_{-2.5}$ & 0.12$^{+0.03}_{-0.01}$  & L\\
~~~090323$^{(4)}$ & 2.71$\pm$0.02 & 0.10$\pm$0.01 (A$_*$) & 0.07$\pm$0.005 & 0.0089$^{+0.0007}_{-0.0018}$ & 2.8$^{+0.4}_{-0.1}$ & 116$^{+13}_{-9}$  &M\\
~~~090328$^{(5)}$ & 2.81$^{+0.14}_{-0.07}$ & 0.33$\pm$0.05 (A$_*$) & 0.11$^{+0.06}_{-0.01}$ & 0.0019$^{+0.0004}_{-0.0008}$ & 4.2$^{+1.3}_{-0.8}$ & 82$^{+28}_{-18}$  &M\\
100418 & 2.22$\pm$0.04 & 2.2$\pm$0.8 (A$_*$) & 0.34$\pm$0.08 &  0.14$\pm$0.04 & 12$\pm$6& 1.6$\pm$0.1 &this work$\!\!$\\
110715A&2.10$\pm$0.02 & 11$\pm$5 (A$_*$) & 0.79$\pm$0.06 & (1.6$\pm$0.2)$\times$10$^{-3}$  & 9.7$\pm$0.9 & 12$\pm$2 &this work$\!\!$\\
121024A& 1.73$\pm$0.03 & 1.4$^{+4.0}_{-1.4}$ (A$_*$) & 0.05$^{+0.06}_{-0.02}$ & 0.02$^{+0.02}_{-0.01}$ & 18$^{+4}_{-1}$ &0.15$^{+0.07}_{-0.03}$ & N\\
130418A& 2.32$\pm$0.14  & 47$\pm$14 (A$_*$) & 0.40$\pm$0.08 & (7.1$\pm$1.9)$\times$10$^{-5}$ & 2.6$\pm$0.4 & 0.77$\pm$0.05 &this work$\!\!$\\
140304A& 2.59  & 0.026 (A$_*$)& 0.025   & 0.059  & 1.13 &490 &O\\
~~~140311A$^{(6)}$ & 2.08$^{+0.01}_{-0.01}$ & 11.1$^{+9.1}_{-3.7}$ (n) & 0.60$\pm$0.10 & 0.22$^{+0.23}_{-0.14}$ & 4.1$\pm$0.3 & 8.7$^{+2.5}_{-1.5}$ & P\\
      & 2.07$^{+0.03}_{-0.02}$ & 0.29$^{+0.20}_{-0.10}$ (A*) & 0.49$^{0.20}_{0.15}$ & 0.097$^{+0.202}_{-0.078}$ & 2.9$\pm$0.2 & 12.5$^{+8.6}_{-3.0}$ & P\\
161219B& 2.079$^{0.009}_{0.006}$ & 3.2$^{+1.4}_{-1.2} \times10^{-4}$ (n) & 0.89$^{0.05}_{0.07}$ & 5.8$^{+5.4}_{-3.0} \times10^{-2}$  &13.5  &0.46$^{0.14}_{0.09}$ &Q\\
181201A& 2.11$\pm$0.01  & 0.022$^{+0.015}_{-0.006}$ (A$_*$)   & 0.41$^{+0.13}_{-0.14}$  & 6.3$^{+10.3}_{-5.3} \times10^{-3}$  & -- & 25.7$^{+9.0}_{-7.9}$  &R\\
\bottomrule
\end{tabular}
\tablebib{
(A)~\cite{2003ApJ...597..459Y};
(B) \cite{Aksulu+2020}, assumed ISM profile; their values have been adapted to be consistent with the usual assumption of 100\% electrons being accelerated;
(C) \cite{ChevalierLi2000}, no errors provided;
(D) \cite{2001ApJ...554..667P};
(E) \cite{2003ApJ...590..992F}, no errors provided;
(F) \cite{Panaitescu2002};
(G) \cite{Panaitescu2005};
(H) \cite{2001ApJ...559..123H};
(I) \cite{Yost2004PhD, 2003ApJ...597..459Y};
(J) \cite{2005AA...440..477R};
(K) \cite{Frail+2006}, no errors provided;
(L) \cite{Cenko2010};
(M) \cite{Cenko2011};
(N) \cite{Varela2016A&A...589A..37V}
(O) \cite{Laskar+2018ApJ859};
(P) \cite{Laskar+2018ApJ858};
(Q) \cite{Laskar+2018ApJ862};
(R) \cite{Laskar+2019}.
}

\smallskip

\setlength\parskip{2pt}
\noindent{$^{(1)}$~The \cite{Panaitescu2002} modelling results hinge on the interpretation of the  sudden re-brightening 
as an observer being initially outside the jet, and the related reddening as \nuc-passage, compared to the more canonical interpretation in terms of a late shell-collision \citep{Vlasis+2011}.
Similarly, also none of the other afterglow modelling attempts cover the clean afterglow\cite[e.g.][]{WijersGalama1999, Granot+1999, Frail+2000ApJ537}.
\cite{ChevalierLi2000} find consistency with a wind environment when fitting just the radio data and the $R$-band flux normalisation, but fix $p$.}

\noindent{$^{(2)}$ ISM and wind environment models explain the data equally well; \cite{2003ApJ...590..992F} prefer the ISM model due to the lower \epse.}

\noindent{$^{(3)}$ The values given are for the narrow jet of the two-component jet model; the wider jet explains the data at $>$1.5 days post-burst, and carries substantially more energy.}

\noindent{$^{(4)}$ The wind interpretation is consistent with \cite{McBreen+2010}, though they derive $\theta_J <$ 1\fdg1, and did not constrain the microphysical parameters. \cite{Lemoine+2013} assume the $>$100 MeV emission to be synchrotron, ignore spectral slopes, and use a two-zone model with decaying micro-turbulence to infer the fireball parameters, thus going substantially beyond the standard model.}

\noindent{$^{(5)}$ \cite{McBreen+2010} noted that modest host extinction of order 0.2 mag is consistent with the GROND data and reduces $p$ to $\approx$2.4; otherwise consistent jet angle and energetics, but again no microphysical parameters. \cite{Lemoine+2013} assume the $>$100 MeV emission to be synchrotron, ignore spectral slopes, and use a two-zone model with decaying micro-turbulence to infer the fireball parameters, thus going substantially beyond the standard model.}

\noindent{$^{(6)}$ \cite{Laskar+2018ApJ858} prefer the ISM solution with the argument that the wind solution overpredicts the early radio data. However, the wind solution fits the X-rays much better, making it a similarly viable option in our view.}
\bigskip
\end{table*}

\noindent GRBs with published fireball parameters, but with one or two parameters fixed at some assumed value, or other assumptions are listed with a summary of the offending detail(s):

\begin{itemize}[itemsep=2pt]
\item 971214: the ambiguity of the interpretation of the spectral break make it impossible for \cite{WijersGalama1999} to constrain the physical parameters
\item 980329: no redshift available; \cite{Yost+2002ApJ577} provide fits for z=1,2,3, and \cite{2003ApJ...597..459Y} assume z=2
\item 980519: no redshift available; \cite{Panaitescu2002} assume z=1 in their modelling
\item 980703: \cite{2003ApJ...597..459Y} find ISM and wind models fitting equally well, and reject the wind model based on some extreme parameter values
\item 990123: \cite{2001ApJ...554..667P} encounter inconsistencies re. the optical and radio decay slopes, and inclusion of energy injection or a structured jet do not solve these issues \citep{Panaitescu2005}.
\item 990510: \cite{2001ApJ...554..667P} find IC to dominate, affecting \nuc (which is not measured), and $p$ is inferred, as no X-ray spectral slope is used.
\item 991208: the \cite{Panaitescu2002} modelling hinges on the interpretation of the unusual flat radio light curve during the first 10 days; \cite{Galama+2003} find no explanation for the disparate slopes of the  radio and optical light curves, while \cite{LiChevalier2001} propose a non-standard electron energy distribution which itself is unconstrained due to the lack of X-ray observations.
\item 991216: \cite{2001ApJ...554..667P} find major discrepancies with the standard model, and suggest a solution with a broken power law electron distribution, having a hard $p$=1.2 at low energies. Energy injection is an alternative to this solution, but still does not reproduce the X-ray decline \citep{Panaitescu2005}.
\item 000301C: \cite{LiChevalier2001} apply a wind model with a (non-standard) electron energy distribution which steepens with energy, while \cite{Panaitescu2002} require a high-energy break in the electron distribution to describe the optical data. Inclusion of energy injection or a structured jet do not solve these issues \citep{Panaitescu2005}. \cite{Berger+2000} rules out a time-varying $p$.
\item 000418: the \cite{Panaitescu2002} analysis is ambiguous with respect to the CMB profile
\item 010222: \cite{Panaitescu2002} obtain no satisfactory fit and find inconsistencies for all assumptions, see also \cite{Panaitescu2005}.
\item 011211: The standard model does not fit satisfactorily, and for the structured outflow model with two more parameters the convergence of the numerical fitting is unclear \citep{Panaitescu2005} 
\item 020405: the interpretation of an early jet break hinges on two optical data points, and the strong and long-duration reverse shock\citep{Berger+2003ApJ587}
\item 020813: \cite{Panaitescu2005} find no unique solution, with fireball parameters varying drastically depending on the model.
\item 021004: \cite{Bjoernsson+2004} set $p$=2.2, and the jet break time is set from the polarisation variation, though it is not seen in the optical light curve.
\item 030226: The standard model does not fit satisfactorily, and for the structured outflow model with tregisterwo more parameters the convergence of the numerical fitting is unclear \citep{Panaitescu2005}
\item 050416A: With $\nu_a$ unconstrained, \cite{Soderberg+2007} assume \epse, \epsb $<$1/3
\item 050820A: \cite{Cenko2010} find two alternative solutions with different $p$.
\item 051022: Without optical detection, no distinction between ISM and wind environment is possible, and \epse\ and \epsb\ are not well constrained \citep{Rol+2007}.
\item 051221: With $\nu_a$ unconstrained, \cite{Soderberg+2006} assume \epse, \epsb $<$1/3
\item 070125: \cite{Chandra+2008} only find solutions (both wind and ISM) which have an isotropic-equivalent kinetic energy a factor 10 smaller than the isotropic $\gamma$-ray energy.
\item 080129: \cite{Gao2009} interpret the optical flare at 500 s after the GRB trigger with several assumptions to derive \epse\ and \epsb, but neither determine $p$ nor the external density profile. 
\item 080319B: even with \epsb\ fixed at equipartition,  \cite{Cenko2010} find two alternative solutions with different $p$, and \cite{Racusin+2008} invoke a double-jetted system resulting in \epsb\ a factor 100 smaller.
\item 080916C: \cite{Beniamini+2015} assume \epse = 0.1
\item 080928: The numerical fit of a model with three energy injections does not constrain the external density profile and returns a $p$ which is inconsistent with the X-ray and optical spectral slope \citep{Rossi+2011}.
\item 090423: With $\nu_a$ unconstrained, \epse, \epsb\ and $n$ have large uncertainties \citep{Chandra+2010}
\item 090510: \cite{Beniamini+2015} assume \epse = 0.1
\item 090902B: \cite{Cenko2011} find a better fit for an ISM density profile, but fix either \epsb\ (according to their Tab.10) or \epse\ (according to their text) at the equipartition value. \cite{Beniamini+2015} assume \epse = 0.1. \cite{Lemoine+2013} assume the $>$100 MeV emission to be synchrotron, ignore spectral slopes, and use a two-zone model with decaying micro-turbulence to infer the fireball parameters, thus going substantially beyond the standard model.
\item 090926A: \cite{Cenko2011} provide a solution for fixed \epse = 0.33, while \cite{Beniamini+2015} assume \epse = 0.1
\item 100414: \cite{Beniamini+2015} assume \epse = 0.1
\item 100901A: \cite{Laskar+2015} assume \epse, \epsb $<$1/3, and the interpretation hinges on the effects of the pan-chromatic re-brightening at 0.25d post-burst.
\item 110625A: \cite{Beniamini+2015} assume \epse = 0.1
\item 110731A: \cite{Beniamini+2015} assume \epse = 0.1. \cite{Lemoine+2013} assume the $>$100 MeV emission to be synchrotron, ignore spectral slopes, and use a two-zone model with decaying micro-turbulence to infer the fireball parameters, thus going substantially beyond the standard model.
\item 120326A: \cite{Laskar+2015} assume \epse, \epsb $<$1/3, and the interpretation hinges on the effects of the pan-chromatic re-brightening at 0.4d post-burst.
\item 120404A: \cite{Guidorzi+2014} interpret the pan-chromatic brightening at 0.03d as passage of $\nu_m$, implying a superfast ($t^{-12}$) passage, and do not fit the X-rays. \cite{Laskar+2015} assume \epse, \epsb $<$1/3, and the interpretation also hinges on the effects of the pan-chromatic re-brightening. 
\item 130427A: \cite{Beniamini+2015} assume \epse = 0.1; due to a strong reverse shock the self-absorption frequency could not be measured, and thus \epse, \epsb $<$1/3 is assumed, leaving\epsb\ with large uncertainties \citep{Laskar+2013, Perley+2014}.
\item 140713A: without the detection of an optical/NIR afterglow, the modelling of \cite{Higgins+2019} hinges on the association with the nearby host galaxy (and thus its redshift), and the assumption on the location of \nuc
\item 141121A: we admire the unique solution of \cite{Cucchiara+2015} for the complicated afterglow behaviour of this GRB, but note that a Newtonian reverse shock and the allowance of different \epsb\ in forward and reverse shock are beyond the "standard" afterglow model, leading to our decision to not include it.
\item 160131A: X-ray-to-optical and radio data separately provide good solutions, respectively, but with \epsb\ differing by a factor of 100, thus not allowing for a unique solution \citep{Marongiu+2022}.
\item 160509A: \nusa\ is not constrained, and the radio data can neither be fit with BOXFIT nor analytically. While \cite{Kangas+2020} provide a complete parameter set, the errors provided in a footnote suggest that neither $E$ nor $n$ are constrained. The independent measurement of extinction and IR fluxes by \cite{Kangas+2020} make the \cite{Laskar+2016} model incompatible with the IR data. 
\item 160625B: \nusa\ is not constrained and $p$ is inconsistent with the X-ray spectral slope \citep{Kangas+2020}.  \cite{Alexander+2017} assume \epse, \epsb $<$1/3.
\end{itemize}

\end{appendix}

\end{document}